\documentclass[%
 aip,
 amsmath,amssymb,
 reprint,%
]{revtex4-1}

\usepackage{graphicx}% Include figure files
\usepackage{dcolumn}% Align table columns on decimal point
\usepackage{bm}% bold math
\usepackage{color}
\usepackage[colorlinks,citecolor=blue]{hyperref}%\usepackage[mathlines]{lineno}% Enable numbering of text and display math
\newcommand{\gcc}{g$\,$cm$^{-3}$}

\begin{document}

\preprint{AIP/123-QED}

\title{Equation of State of Hot, Dense Magnesium Derived with First-Principles Computer Simulations }% Force line breaks with \\

\author{Felipe Gonz\'alez-Cataldo}
 \affiliation{Department of Earth and Planetary Science, University of California, Berkeley, CA 94720, USA}

\author{Fran\c{c}ois Soubiran}
  \affiliation{Department of Earth and Planetary Science, University of California, Berkeley, CA 94720, USA}
\affiliation{CEA, DAM, DIF, 91297 Arpajon, France}
\affiliation{\'Ecole Normale Sup\'erieure de Lyon, Universit\'e Lyon 1, Laboratoire de G\'eologie de Lyon, CNRS UMR 5276, 69364 Lyon Cedex 07, France}
\author{Burkhard Militzer}
 \email{militzer@berkeley.edu}
\affiliation{Department of Earth and Planetary Science, University of California, Berkeley, CA 94720, USA}
 \affiliation{Department of Astronomy, University of California, Berkeley, CA 94720, USA}

\begin{abstract}
Using two first-principles computer simulation techniques, path integral Monte-Carlo and density functional
theory molecular dynamics, we derive the equation of state of magnesium in the regime of warm dense matter, with densities ranging from 0.43 to 86.11~\gcc~and temperatures from 20,000 K to $5\times10^8$~K. These conditions are relevant for the interiors of giant planets and stars as well as for shock compression measurements and inertial confinement fusion experiments. We study ionization mechanisms and electronic structure of magnesium as a function of density and temperature. We show that the L shell electrons 2s and 2p energy bands merge at high density. This results into a gradual ionization of the L-shell with increasing density and temperature. In this regard, Mg differs from MgO, which is also reflected in the shape of its principal shock Hugoniot curve. For Mg, we predict a single broad pressure-temperature region where the shock compression ratio is approximately 4.9. Mg thus differs from Si and Al plasma that exhibit two well-separated compression maxima on the Hugoniot curve for L and K shell ionizations.  Finally we study multiple shocks and effects of preheat and precompression. 
\end{abstract}
\maketitle

\section{Introduction}

The physical properties of hot, dense plasmas have been studied with a variety of experimental and theoretical techniques.~\cite{Ebeling1991}
Understanding how dense plasmas behave is of importance for technologies based upon laser and particle beams, such as inertial confinement fusion (ICF),~\cite{Zhang2018,Betti2016,Seidl2009,Miyanishi2015} for the interpretation of high-velocity impact and shock wave experiments~\cite{Hammel2010,Millot2015,Kirsch2019}
as well as for the understanding of astrophysical processes.~\cite{Cotelo2011,Chabrier2002}
Warm dense matter (WDM) is a particularly challenging state of matter to study because
it is too dense to be described by plasma theory that is designed for weakly interacting particles,
but also too hot to be studied with most methods in condensed matter physics. Condensed matter theory is a well-established field that can accurately describe solids and liquids at moderate temperatures, but the treatment of high temperature conditions becomes increasingly difficult because many ground-state methods are not well suited to incorporate partially or completely ionized electronic orbitals that become relevant because of the thermal ionization. Developing a framework of theoretical methods or computer simulations that can consistently and accurately describe the low and high temperature regimes is, therefore, of high importance. The regime of WDM includes the deep interiors of planets in our solar system and that of exoplanets,~\cite{ExoplanetArchive,Guillot1999} where the equation of state (EOS) of materials in the regime of WDM is required to model their interior structure and the evolution.~\cite{Militzer2016b,Baraffe2014}

Magnesium (Mg) is of high importance in geophysics because as part of MgO and MgSiO$_3$ it belongs to the fundamental building materials in planetary formation.~\cite{Valencia2010,Bolis2016,Musella2019} The properties of these compounds in the WDM regime have recently been studied with first-principles simulations.~\cite{GonzalezMilitzer2019,Soubiran2019,GonzalezMilitzer2020} Shock compression experiments on MgO~\cite{Mcwilliams2012,Miyanishi2015,Bolis2016} and SiO$_2$,~\cite{Hicks2006,Millot2015} combined with first principle calculations, demonstrated that these mantle minerals
become electrically conducting in the fluid phase. Super-Earth planets can thus generate magnetic fields within their mantles.~\cite{Soubiran2018,Stixrude2020}

Considerable efforts have also been made to characterize the EOS of magnesium at high pressure. This includes the determination of the phase boundary between the hcp and bcc solid phases~\cite{Stinton2014} and the melting temperature with shock wave experiments,~\cite{Urtiew1977,Errandonea2010,Errandonea2001} which usually require models that link the Gruneisen parameter to the shock Hugoniot curve.~\cite{Qiang2002} Very recently, Beason {\it el al.}~\cite{Beason2020} again employed shock wave experiments to directly observe the hcp-bcc transition and melting along the principal Hugoniot curve. At the highest shock velocities,  the x-ray diffraction measurements indicate that the temperatures reached were sufficiently high to melt the sample, which  occurs around 63 GPa on the Hugoniot curve. These results also indicate that the hcp-bcc phase boundary intersects the Hugoniot above 27 GPa, fully transforming to bcc by 37 GPa.
%However, these experiments are well below the conditions of pressure and temperature explored in this work, and no experimental data are available in the K- %and L-shell ionization regimes yet. Therefore, more experiments are needed in order to validate our predictions.

%Molten Mg~\cite{Miki1998}.\\
%Hugoniot~\cite{Agarwal2017}. 
%Shock experiments in MgSiO3~\cite{Fratanduono2018}.

The EOS and thermodynamic properties of Mg have also been investigated with {\it ab initio} computer simulations~\cite{Sinko2009,Greeff1999,Khishchenko2004,Lomonosov2002} that have characterized the phase diagram.  Recent calculations of the melting curve have explored the reentrant melting phenomenon,~\cite{Hong2019} and Mehta {\it et al.}~\cite{Mehta2006} demonstrated that the choice of the pseudopotential in {\it ab initio} simulations has very little effect on the computed thermodynamic properties.

Path integral Monte Carlo (PIMC) methods have gained considerable interest as a state-of-the-art, stochastic first-principles technique to compute the properties of interacting quantum systems at finite temperature. This formalism results in a highly parallel implementation and an accurate description of the properties of materials at high temperature where the electrons are excited to a significant degree.~\cite{Mi06,Benedict2014,Driver2015,Hu2016,ZhangBN2019}
The application of the PIMC method to first and second-row elements has been possible due to the development of free-particle~\cite{Ce95,Ce96} and Hartree-Fock nodes.~\cite{MilitzerDriver2015} The latter approach enables one to efficiently incorporate localized electronic states into the nodal structure, which extends the applicability of the path integral formalism to heavier elements and lower temperatures.~\cite{ZhangSodium2017,Driver2018} Furthermore, PIMC treats all electrons explicitly and avoids the use of pseudopotentials. The PIMC simulation time scales as $1/T$, proportional to the length of the paths, which is efficient at high-temperature conditions, where most electrons including the K shell are excited. Predictions from PIMC simulations at intermediate temperatures have been shown to be in good agreement with predictions from density functional theory molecular dynamics (DFT-MD) simulations.~\cite{Mi09,ZhangCH2018}

Kohn-Sham DFT~\cite{Hohenberg1964,Kohn1965} is a first-principles simulation method that determines the ground state of quantum systems with high efficiency and reasonable accuracy, which has gained considerable use in computational materials science. The introduction of the Mermin
scheme~\cite{Mermin1965} enabled the inclusion of excited electronic
states, which extended the applicability range of the DFT
method to higher temperatures. The combination of this method with
molecular dynamics has been widely applied to compute the EOS of
condensed matter, warm dense matter (WDM), and some dense
plasmas.~\cite{Root2010,Wang2010,Mattsson2014,Zhang2018} Unless the number of partially occupied orbitals is impractically large, DFT is typically
the most suitable computational method to derive the EOS because it accounts for electronic shell and
bonding effects. The main source of uncertainty in DFT is the use of an approximate exchange-correlation (XC) functional. The errors resulting from the XC functional often cancel between different thermodynamic conditions. Furthermore this error may only be a small fraction of the internal energy, which besides pressure is the most relevant quantity for the EOS and the derivation of the shock Hugoniot curve.~\cite{Karasiev2016} However,
the range of validity of this assumption in the WDM regime remains to 
be verified for different classes of materials through the comparison 
with laboratory experiments and other computational techniques like PIMC simulations.

In this work, we combine the PIMC and DFT-MD simulation methods to study the
properties of magnesium in the regime of WDM. The combination of both
methods allows us to study a much wider temperature and density interval 
and furthermore to test the validity of the approximations in the
methods. We study the regimes of thermal and pressure ionization of
the electronic shells and provide an equation of state that spans a
wide range of temperatures and pressure.  We describe the electronic
properties of liquid Mg and show how the band gap between s and p
states changes upon compression, and provide a structural
characterization of the liquid.  Finally, we determine the shock Hugoniot curve and explore the effects of precompression.

\section{Simulations Methods}

\begin{figure}
    \centering
    \includegraphics[width=\columnwidth]{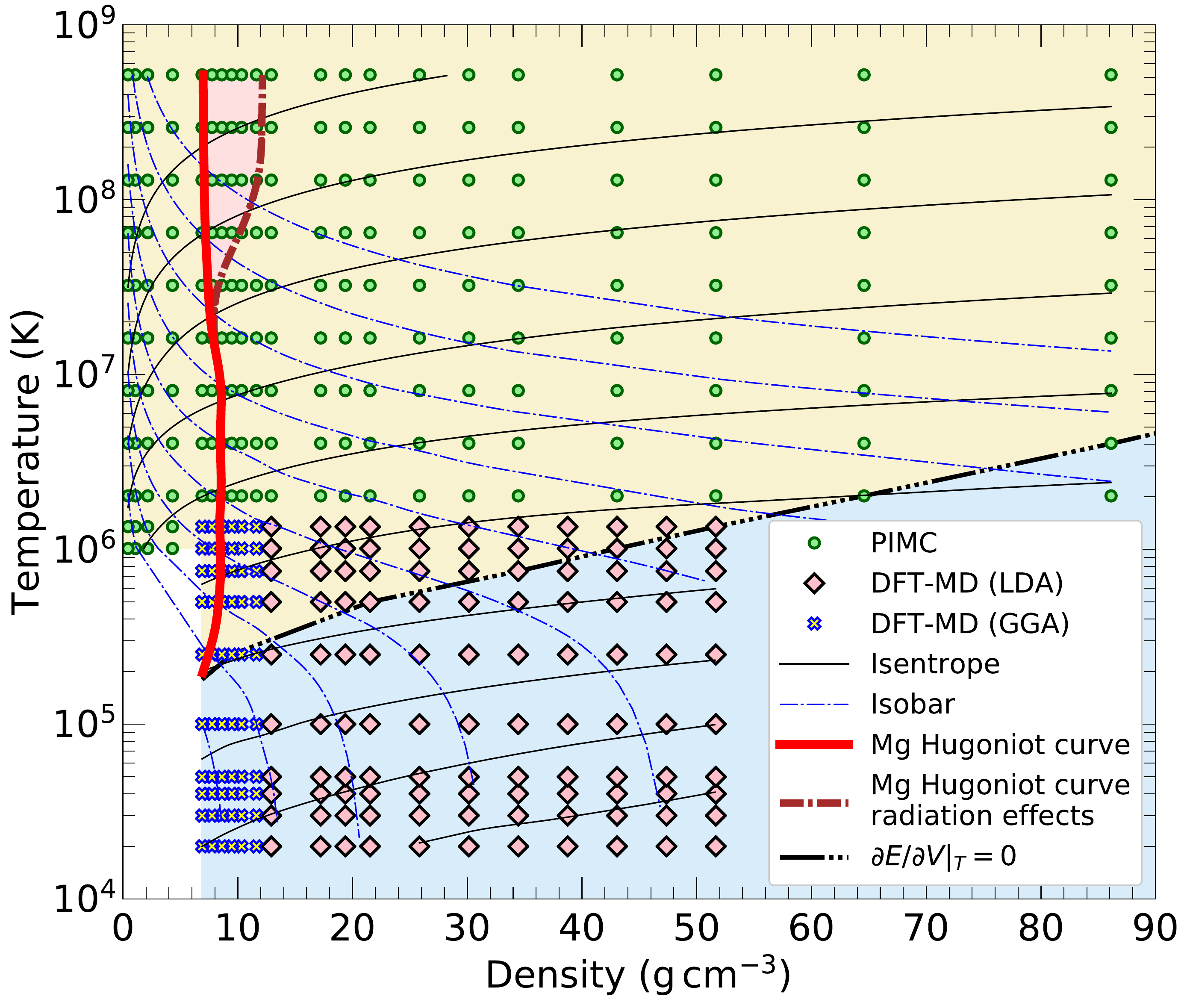}
    \caption{Temperature-density conditions of our PIMC and DFT-MD simulations along with
computed isobars, isentropes and shock Hugoniot curves that were derived, with and without radiation effects, for an initial density of
$\rho_0=1.73686577$ \gcc. The high-temperature region of thermal ionization is separated by the black dashed line from the regime of pressure ionization at lower temperatures.
\label{fig:Tvsrho}}
\end{figure}

We perform first-principles computer simulations of magnesium for a range of extreme density and temperature conditions that we illustrate in Fig.~\ref{fig:Tvsrho}. At high temperature, we employ PIMC simulations, while at lower temperatures we use standard Kohn-Sham DFT-MD calculations. 

\subsection{PIMC simulations}

The fundamental techniques for the PIMC simulations of bosonic systems
were developed in Ref.~\onlinecite{PC84} and reviewed in Ref.~\onlinecite{Ce95}.
Subsequently the algorithm was extended to fermionic systems by introducing the {\em restricted} paths approach.~\cite{Ce91,Ce92,Ce96}
The first results of this simulation method were
reported in the seminal work on liquid $^3$He~\cite{Ce92} and dense
hydrogen.~\cite{PC94} In subsequent articles, this method was applied to
study hydrogen,~\cite{Ma96,Mi99,MilitzerThesis,MC00,MC01,Mi01}
helium,~\cite{Mi06,Mi09,Mi09b} hydrogen-helium mixtures~\cite{Mi05}
and one-component plasmas.~\cite{JC96,MP04,MP05} In recent years, the
PIMC method was extended to simulate plasmas of various first-row
elements~\cite{Benedict2014,DriverNitrogen2016,Driver2017,ZhangCH2017,ZhangCH2018,Zhang2018}
and with the development of Hartree-Fock nodes, the simulations of
second-row elements became
possible.~\cite{MilitzerDriver2015,Hu2016,ZhangSodium2017,Driver2018}

The PIMC method is based on the thermal density matrix of a quantum
system, $\hat\rho=e^{-\beta \hat{\cal H}}$, that is expressed as a
product of higher-temperature matrices by means of the identity
$e^{-\beta \hat{\cal H}}=(e^{-\tau \hat{\cal H}})^M$, where
$M$ is an integer and $\tau\equiv\beta/M$ represents the time step of a path integral in
imaginary time. The path integral emerges when the operator $\hat\rho$
is evaluated in real space,
%This allows to simplify, in the path integral formulation, the action of the path $S[\mathbf R_t]$~\cite{MilitzerThesis,MC00}
%in the density matrix element
%which allows to write it as an integral over paths $\mathbf R_t$~\cite{MilitzerThesis,MC00},
\begin{equation}
\left<\mathbf R|\hat\rho| \mathbf R'\right>=\frac{1}{N!}\sum_{\mathcal P}(-1)^{\mathcal P}\oint_{\mathbf R\to\mathcal P\mathbf R'}\mathbf{dR}_t\, e^{-S[\mathbf R_t]}.
%=\sum_s e^{-\beta\epsilon_s}\Psi_s^*(\mathbf R)\Psi_s(\mathbf R'),
\label{PI}
\end{equation}
The sum includes all permutations, $\mathcal P$, of
$N$ identical fermions in order project out the antisymmetric
states.  For sufficiently small time steps, $\tau$, all many-body
correlation effects vanish and the action, $S[\mathbf R_t]$, can be
computed by solving a series of two-particle
problems.~\cite{PC84,Na95,BM2016} The advantage of this approach is that 
all many-body quantum correlations are recovered
through the integration over paths. The integration also enables
one to compute quantum mechanical expectation values of thermodynamic
observables, such as the kinetic and potential energies, pressure,
pair correlation functions and the momentum
distribution.~\cite{Ce95,Militzer2019} Most practical implementations of
the path integral techniques rely on Monte Carlo sampling techniques
because the integral has $D \times N \times M$ dimensions in
addition to sum over permutations ($D$ is the number of spatial dimensions). The method becomes
increasingly efficient at high temperature because the length
of the paths scales like $1/T$. In the limit of low temperature, where
few electronic excitations are present, the PIMC method becomes
computationally demanding and the Monte Carlo sampling can become inefficient.
Still, the PIMC method avoids any exchange-correlation approximation
and the calculation of single-particle eigenstates, which are embedded 
in all standard Kohn-Sham DFT calculations. 

The only uncontrolled approximation within fermionic PIMC calculations
is the use of the fixed-node approximation, which restricts the paths
in order to avoid the well-known fermion sign
problem.~\cite{Ce91,Ce92,Ce96} Addressing this problem in PIMC is
crucial, as it causes large fluctuations in computed averages due to
the cancellation of positive and negative permutations in
Eq.~\eqref{PI}. We solve the sign problem approximately by restricting
the paths to stay within nodes of a trial density matrix that we obtain 
from a Slater determinant of single-particle density matrices,
\begin{equation}
\rho_T({\bf R},{\bf R'};\beta)=\left|\left| \rho^{[1]}(r_{i},r'_{j};\beta) \right|\right|_{ij}\;,
\label{FP}
\end{equation}
that combines free and bound electronic states,~\cite{MilitzerDriver2015,Driver2018} 
\begin{eqnarray}
\label{rho1}
\rho^{[1]}(r,r';\beta) &=& \sum_{k} e^{-\beta E_k} \, \Psi_k(r) \, \Psi_k^*(r')\\
 &+& \sum_{I=1}^{N} \sum_{s=0}^{n} e^{-\beta E_s} \Psi_s(r-R_I) \Psi_s^*(r'-R_I)\;.
\quad.
\end{eqnarray}
The first sum includes all plane waves, $\Psi_k$ while the second represents $n$ 
bound states $\Psi_s$ with energy $E_s$ that are localized around all atoms $I$. 
Predictions from various slightly differing forms of this approach have been 
compared in Ref.~\onlinecite{ZhangSodium2017}.

The PIMC simulations were performed with the CUPID code.~\cite{MilitzerThesis} We used periodic boundary conditions and treated 8 Mg nuclei and 96 electrons explicitly as paths. We enforced the nodal constraint in small steps of imaginary time of $\tau=1/8192$ Ha$^{-1}$, while
the pair density matrices~\cite{Militzer2016} were evaluated in steps of 1/1024 Ha$^{-1}$. This
results in using between 1280 and 5 time slices for the temperature
range that was studied with PIMC simulations. These choices converged
the internal energy per atom to better than 1\%.  We have shown the
associated error is small for relevant systems at sufficiently high
temperatures.~\cite{Driver2012}
{\bf For example, in Ref.~\onlinecite{Driver2015}, pressure and internal energy from simulations
with 8 and 24 nuclei were shown to be in sufficiently good agreement. This convergence test underlines that, with simulations of 8 nuclei, we can obtain good thermodynamic average of the pressure and internal energy under conditions where their value is primarily controlled by the ionization of the electrons. }

\subsection{DFT-MD simulations}

Kohn-Sham (KS) DFT-MD,~\cite{Hohenberg1964,Kohn1965} on the other hand, is a method used to compute the properties of matter in the cold and warm dense matter regime. We thus used the DFT-MD code VASP \cite{Kresse1999} to perform simulations up to 2 million Kelvin to complement the PIMC calculations. We restricted our DFT-MD calculations to a range of densities from 6.89 to 51.67 \gcc (1.6- to 12-fold the reference density of $\rho^*=4.3055475$ \gcc). We used cubic simulation cells with periodic boundary conditions that, depending on the temperature, contained between 8 and 64 Mg atoms. It has been shown in previous work that such a small cell is not detrimental to the accuracy of the EOS data at high temperatures.~\cite{Driver2015b,ZhangCH2017,Driver2018,Soubiran2019} To keep the temperature constant in a given simulation, we use a Nos\'e thermostat.~\cite{Nose1984,Nose1991} The time step was adapted to the density and the temperature, ranging from  0.16 to 0.44 fs for simulation times from 1000 to $16\,000$ time steps to ensure a reliable estimation of the thermodynamic quantities.

Our DFT-MD calculations were performed within the Mermin scheme \cite{Mermin1965} and employed projector augmented wave (PAW) \cite{Blochl1994} pseudopotentials. From the available pseudopotentials in the VASP library, we chose a hard pseudopotential with a 1s$^2$ frozen core and a PAW sphere radius of 1.75 Bohr radii. To describe the exchange-correlation effects, we used the Perdew-Burke-Ernzerhof (PBE) \cite{PBE} functional for the lowest densities, as it has shown to give good results for MgO.~\cite{Soubiran2018,Soubiran2019} Since  the provided Mg PBE pseudopotential did not give proper results for high densities, we switched to the local density approximation (LDA). We obtain a very good agreement between both functionals at 4-fold the reference density $\rho^*$ (see section~\ref{sec:EOS}).
As shown in Ref.~\onlinecite{Mehta2006}, the choice of the pseudopotential in {\it ab initio} simulations of Mg has very little effects on the computed thermodynamic properties.
For very high temperatures, the Mermin approach requires computation of many excited states with low occupation numbers. That is why we computed up to 5000 bands, even when we employed reduced cell size of 8 atoms for temperatures above 10$^6$~K. This high number of bands ensured that every band with occupation fraction of 10$^{-5}$ or greater was included. For high temperature conditions, the energy cut-off for the plane wave basis set had to be increased up to 6000~eV. The error in the total energy due to this band cut-off is less than 0.1\%. We sampled the Brillouin zone with the $\Gamma$ point only, which was found to be sufficient for the convergence of the thermodynamic quantities under the conditions of interest.

\section{Results and discussion}

%%%%%%%%%%%%%%%%%%%%%%%%%%%%%%%%%%%%%%%%%%%%%%%%%%%%%%%%%%%%%%%%%%%%%%%%%%%%%%%%%%%%%%%%%%%%%%%%%%%
\subsection{Equation of state \label{sec:EOS}}

In order to make the internal energies of VASP DFT-MD simulations
compatible with the all-electron PIMC energies, we shifted
the energies generated with the LDA and GGA functionals by
$-$199.722498 and $-$200.011012~Ha per atom, respectively.
This shift was derived by
performing all-electron calculations for the isolated,
non-spin-polarized Mg atom with the OPIUM
code~\cite{OPIUM} and comparing the results with corresponding VASP
calculations.

\begin{figure}[!t]
    \centering
    \includegraphics[width=\columnwidth]{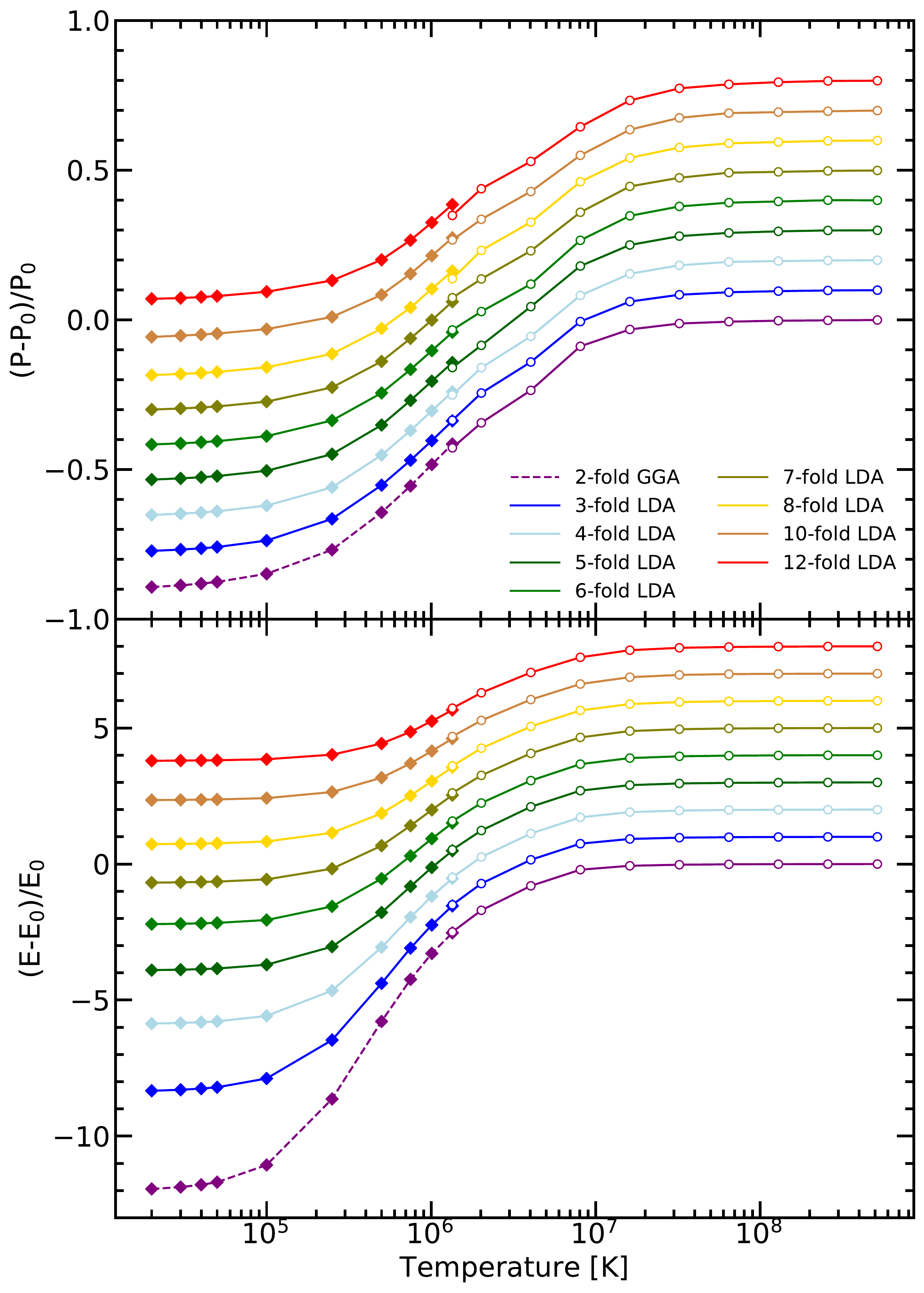}
    \caption{The temperature dependence of the excess pressure and internal energy relative to predictions of a noninteracting Fermi gas of electrons and classical nuclei. Every curve represents a different density, ranging from 2- to 12-fold the reference density of $\rho^*=4.305502$ \gcc. The open circles show the PIMC results at high temperature while the solid diamonds represent the DFT predictions. With the exception of the dashed curve for 2-fold $\rho^*$, the isochores have been shifted for clarity in steps of +0.01 and +1.0 in the upper and lower panels, respectively. }
    \label{fig:EOS}
\end{figure}

We show the pressure and energy of Mg as a function of temperature in Fig.~\ref{fig:EOS}, relative to an ideal Fermi gas of electrons and classical nuclei with pressure $P_0$ and internal energy $E_0$, in order to magnify the excess contributions that result from the particle interactions.
With increasing temperature, these contributions gradually decrease from the strongly interacting condensed matter regime, where chemical bonds and bound states dominate, to the weakly interacting, fully ionized plasma regime.

The DFT-MD results show good agreement between LDA and PBE calculations at 4-fold the reference density, $\rho^*$. No discontinuities are observed in the thermodynamic properties when the functional is changed. As temperature increases, DFT simulations become increasingly inefficient, as the number of partially occupied orbitals, that have to be explicitly computed, increases considerably. Around $10^6$ K, PIMC simulations are feasible, but also computationally demanding. However, they become more efficient at higher temperatures.  For Mg, we obtained good agreement between PIMC and DFT at 1,347,305~K, as we show in Fig.~\ref{fig:EOS}. Near this temperature, the relative difference in the pressure is less than 4.8\%, and the difference in the energy ranges from 2.5 to 7.1~Ha per atom. The largest energy differences occur at the highest densities, where the frozen cores of DFT pseudopotential overlap significantly. Overall, however, the agreement is more than satisfactory. 

\begin{figure}[!t]
    \centering
    \includegraphics[width=\columnwidth]{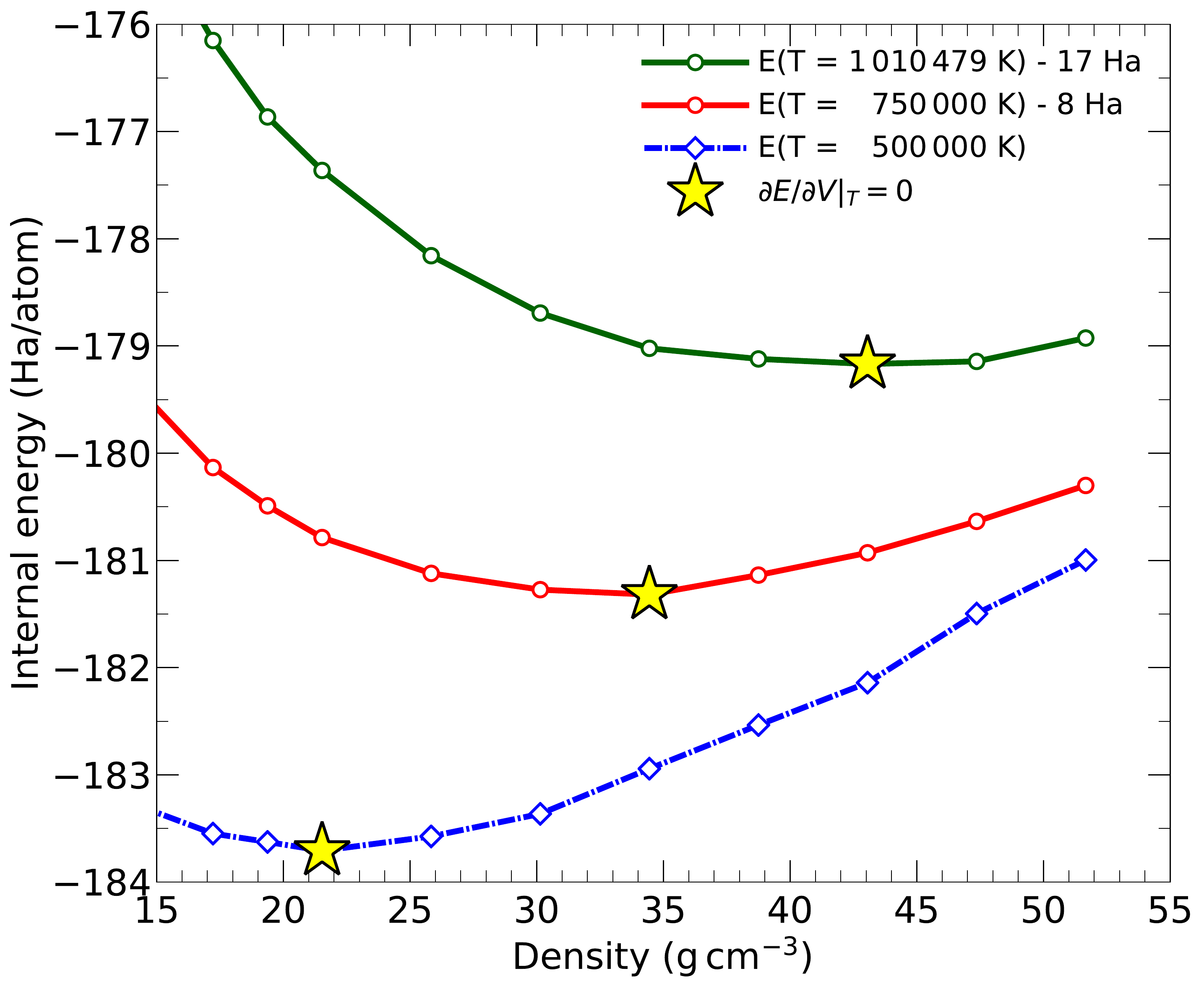}
    \caption{Internal energy vs. density curves for three temperatures, specified in the legend. The three stars mark the energy minima, $\partial E / \partial V|_T=0$, that we use in Figs.~\ref{fig:Tvsrho} and ~\ref{fig:PvsT}  to distinguish between the regimes of thermal and pressure ionization. The energies of the upper two curves have been shifted for clarity.}
    \label{fig:E_rho}
\end{figure}

In Fig.~\ref{fig:E_rho}, we show the internal energy, $E$, as
a function of density along three isotherms. We find that all three $E(\rho)_T$ curves have a minimum. With increasing temperature, the location of this minimum shifts towards higher densities. This minimum in the energy is related to the following condition for the thermal pressure coefficient, 
\begin{equation}
    \label{Eq:Emin}
    \beta_V \equiv \left(\frac{\partial P}{\partial T}\right)_V=\frac{P}{T},
\end{equation}
which is only satisfied if $\left.\frac{\partial E}{\partial \rho}\right|_T=0$.~\cite{GonzalezMilitzer2019}
At low density, the slope $\left.\frac{\partial E}{\partial \rho}\right|_T$ is negative because the system
is more ionized, as we will discuss in the next section.
At high density, the slope $\left.\frac{\partial E}{\partial \rho}\right|_T$ is positive for two possible reasons.
First, there is the confinement effect, which increases the kinetic energy of the free electrons and, second, the orbitals of the bound electrons hybridize and may even be pushed into the continuum of free electronic states, which is commonly referred to as pressure ionization. As previously,~\cite{GonzalezMilitzer2020} we use this energy minimum as a criterion to distinguish the thermal ionization regime from the pressure ionization regime.
\begin{figure}[!t]
    \centering
    \includegraphics[width=\columnwidth]{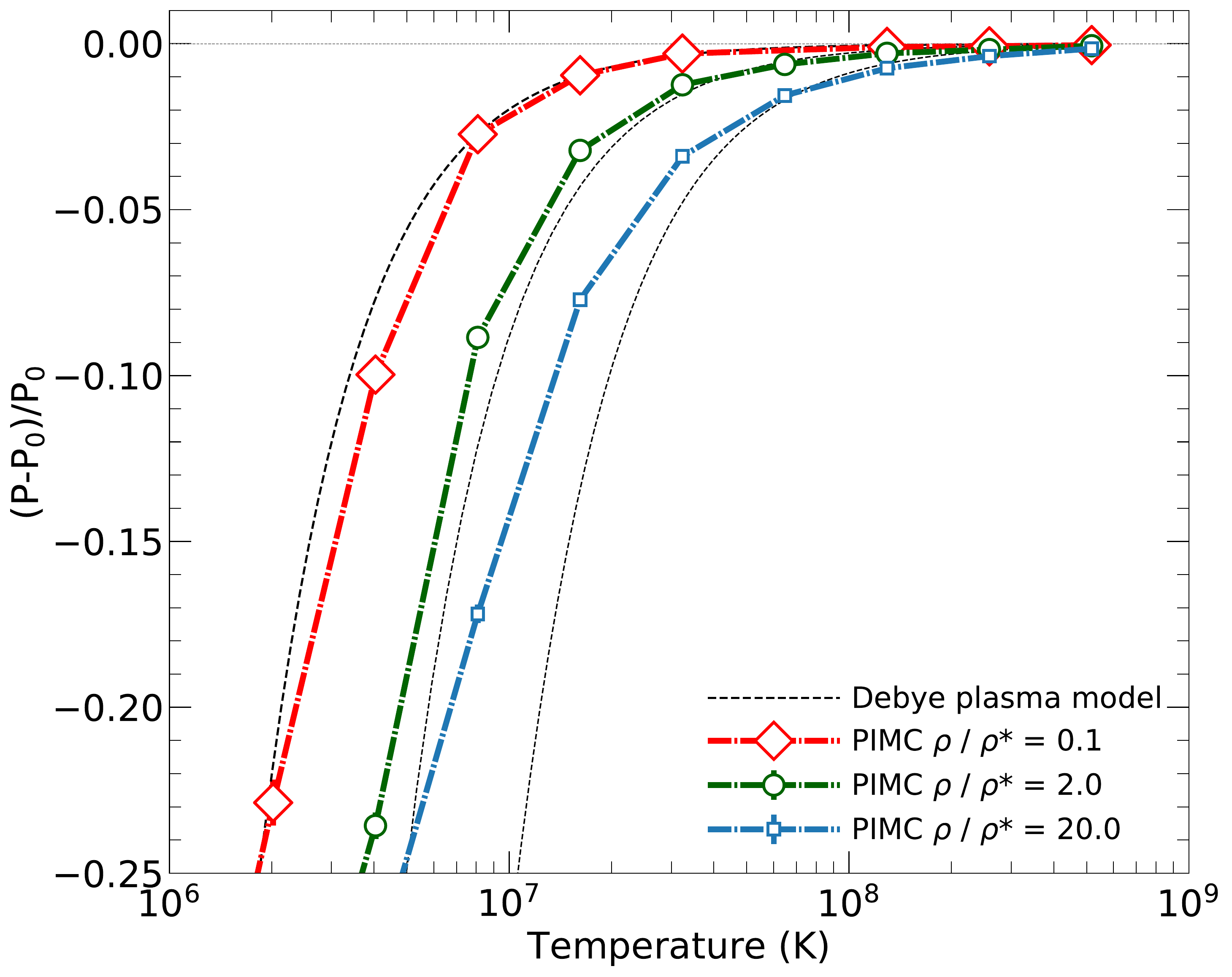}
    \caption{Excess pressure computed with PIMC simulations compared with the Debye plasma model for the three densities of 0.1, 2.0 and 20.0 $\times$ ($\rho^*=4.305502$ \gcc).}
    \label{fig:Mg_P_T_Debye}
\end{figure}

In Fig.~\ref{fig:Mg_P_T_Debye}, we show  that the PIMC results converge to predictions from the classical Debye-H\"uckel plasma model~\cite{Debye1923} in the limit of high temperature. For low densities, the agreement is reached at lower temperatures because there are more particles in the Debye sphere and the screening approximation is more accurate.~\cite{Mi09} As expected, the Debye-Hückel model becomes inadequate for lower temperatures ($T < 8\times  10^6$ K) since it does not include any bound electronic states. 
The temperature range from $2\times10^6$ to $1\times 10^7$ K encompasses significant portions of K shell ionization regime, which is precisely where the full rigor of PIMC simulations are needed to acquire an accurate EOS table.

\begin{figure}[!th]
    \centering
    \includegraphics[width=\columnwidth]{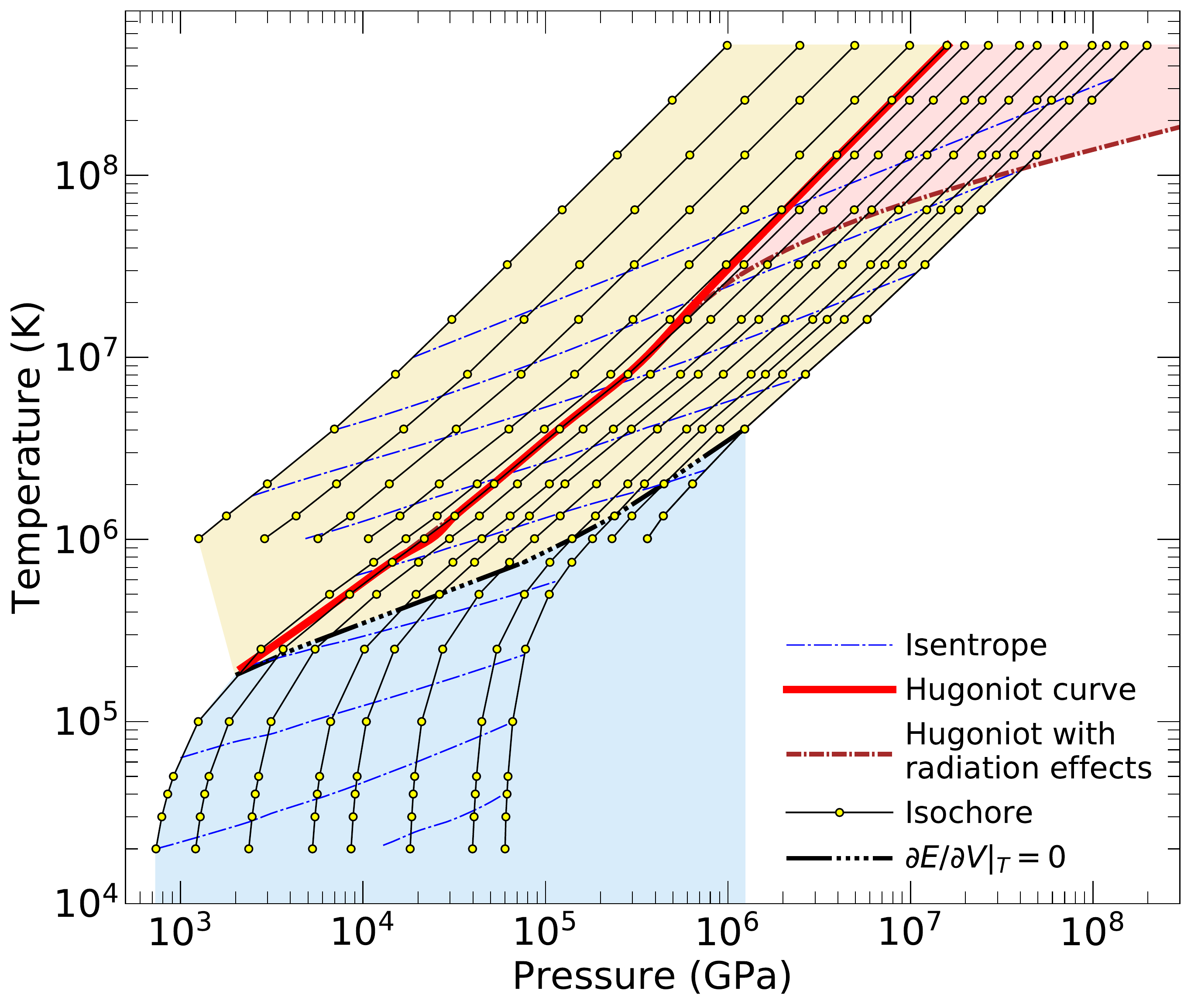}
    \caption{Temperature-pressure conditions for the PIMC and DFT-MD
    calculations along isochores corresponding to the densities of
    0.430550 (upper left curve) to 86.110045 \gcc~(lower right curve). The shock Hugoniot curves with and without radiation effects were included as well as a number of isentropes. As in Fig.~\ref{fig:Tvsrho}, the high-temperature region of thermal ionization is separated by the black dashed line from the regime of pressure ionization at lower temperatures.
  }\label{fig:PvsT} 
\end{figure}

In Fig.~\ref{fig:PvsT}, we show all EOS points that we computed in a pressure-temperature diagram. 
We include the principal shock Hugoniot curve that we have discuss in section~\ref{hug}.
Our entire EOS table is provided as supplementary material to facilitate a comparison with future experiments and as a benchmark for other faster and likely more approximate EOS methods. {\bf In this table, we provide the pressures and internal energies, as well as their one-sigma error bars, that were computed on a grid in temperature and density. The error bars were derived with the blocking method.~\cite{AT87} Their size is controlled by the length of the simulation and the number particles as well as by the temperature and density conditions that control the state of the material. In general, we find it easiest to obtain converged results for the lowest and the highest temperatures. At high temperature, the paths in the PIMC simulations are short, which makes it very efficient to move the electrons and nuclei in the system that is controlled by screening interactions. At low temperature, DFT-MD simulations are very efficient because very few excited states need to be included. More computer time needs to be invested to perform simulations in the regime of $T \sim 10^6$ K, which can be challenging to study with both methods.}

%%%%%%%%%%%%%%%%%%%%%%%%%%%%%%%%%%%%%%%%%%%%%%%%%%%%%%%%%%%%%%%%%%%%%%%%%%%%%%%%%%%%%%%%%%%%%%%%%%%
\subsection{Degree of Ionization}

In this and the two following sections, we report PIMC and DFT-MD results for the electronic structure of the magnesium plasma as a function of temperature and density. We study the ionization of the 1s orbital of Mg atoms as a function of temperature and explore the nucleus-electron pair correlation functions.  We also show how the electronic density of states (DOS) and the 2s-3p band gap, obtained from DFT-MD simulations, are affected by temperature, providing further insights into the temperature-density evolution of ionization effects, important for continuum lowering.~\cite{Vinko2014,Lin2017,Driver2018,Soubiran2019}

%\onecolumngrid
\begin{figure*}[!hbt]
    \centering
    \includegraphics[width=17.5cm]{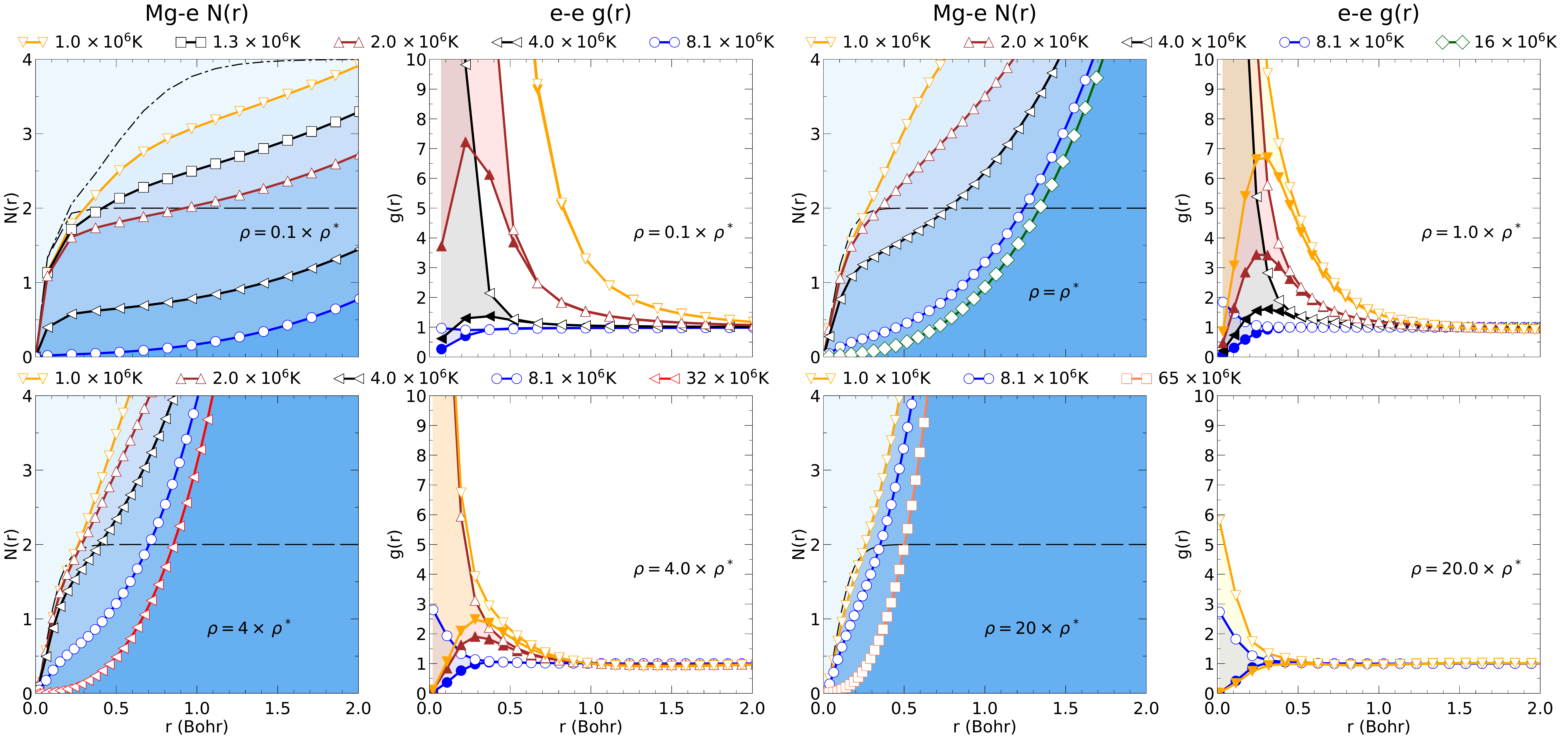}
    \caption{ Integrated nucleus-electron pair correlation functions, $N(r)$, and electron-electron pair correlation functions, $g(r)$, are shown for four densities: 0.1, 1.0, 4.0 and 20.0 $\times$ ($\rho^*=4.305502$ \gcc).  The temperatures, indicated above the panels, were chosen to show conditions where these functions show significant change. In columns 2 and 4, we plot the $g(r)$ functions for pair of electrons with parallel (filled symbols) and anti-parallel spins (open symbols). The $N(r)$ functions in columns 1 and 3 represent the average of number of electrons contained within a sphere of radius, $r$, around a given nucleus. All curves with symbols were derived from PIMC simulations. For comparison, we show the corresponding functions with thin dashed lines for isolated nuclei with doubly occupied 1s core states that we computed with the GAMESS software.~\cite{GAMESS} In the upper left diagram, the thin dash-dotted lines show the curve for doubly occupied 1s and 2s states. \label{fig:N(r)}
  }
\end{figure*}
%\twocolumngrid

In PIMC simulations, a measure of the degree of ionization can be
obtained from the integrated nucleus-electron pair correlation
function, $N(r)$, given by  
\begin{equation}\label{eq:N(r)}
N(r)= \left<\frac{1}{N_I}\sum_{e,I}\Theta(r -\|\vec r_e-\vec r_I \|)\right>,
\end{equation}
where $N(r)$ represents the average number of electrons within a sphere of
radius $r$ around a given nucleus of atom of type $I$.  The summation includes
all electron-nucleus pairs and $\Theta$ represents the Heaviside function.
Fig.~\ref{fig:N(r)} shows the integrated nucleus-electron pair correlation
function for temperatures from $1\times 10^6$ K to $65\times 10^6$ K and
densities from 0.431 \gcc~(0.1$\times \rho^*$) to 86.11 \gcc~(20$\times \rho^*$), where $\rho^*=4.305502$ \gcc~is a reference density that we chose for convenience. For comparison, the $N(r)$ functions of an isolated Mg nucleus with a
doubly occupied 1s orbital is included. Unless the 1s state
is ionized, its contribution will dominate the $N(r)$ function at small
radii of $ r < 0.2 $ Bohr radii. For larger radii, other
electronic shells and electron located near 
neighboring nuclei contribute also. Still, this is the most
direct approach available to compare the degree of 1s ionization of the nuclei.~\cite{MilitzerDriver2015}

For $\rho=0.1 \times \rho^*$, there is partial ionization of the 1s state of the Mg nuclei at $1.0\times 10^6$ K already (see top left panel of Fig.~\ref{fig:N(r)}). Ionization at this temperature has been observed in other single-component plasmas, such as carbon, oxygen, and silicon~\cite{Driver2017b,Driver2015b} at similar conditions.
In contrast, when Mg is bonded to other chemical species at a similar density, such as in MgSiO$_3$ or in MgO, partial ionization of K shell of Mg nuclei typically does not occur below $2\times10^6$ K.~\cite{GonzalezMilitzer2019,Soubiran2019} However, for temperatures above $4\times10^6$ K at this density, the $N(r)$ profile around the Mg nuclei, hence the degree of ionization, is very similar in pure Mg, MgO, and MgSiO$_3$. We conclude that the ionization onset of Mg 1s states occurs at lower temperature for pure Mg than it does in MgO, and MgSiO$_3$ plasma, where oxygen species provide additional electrons that can be ionized more easily. 

A comparison of the $N(r)$ functions in the upper panels of Fig.~\ref{fig:N(r)} shows that the degree of 1s ionization is reduced when the density is increased from 0.1 to 1.0$\times \rho^*$. Even less ionization is observed at higher densities of 4.0 and 20.0$\times \rho^*$, as lower panels show. The degree of 1s ionization is consistently reduced with increasing density when the results are compared for the same temperature. Most noticeable are the changes in the $N(r)$ function for a temperature of $8.1 \times 10^6$ K.
For $\rho=0.1 \times \rho^*$, the 1s states of the Mg nuclei are essentially fully ionized while there is a substantial 1s occupation for a density of $20.0 \times \rho^*$. Fig.~\ref{fig:N(r)} also illustrates that temperatures above $32\times10^6$ K are  sufficient to fully ionize the system. In this case, the system  behaves similar to an ideal gas and the pressure and energy scale linearly with temperature. Both depend weakly on density at these temperatures (see Figs.~\ref{fig:EOS},~\ref{fig:Mg_P_T_Debye}, and~\ref{fig:PvsT}).

In Fig.~\ref{fig:N(r)}, we also show the electron-electron pair
correlation functions, $g(r)$, that we derived from our all-electron
PIMC simulations. Without Coulomb interactions, pairs of electrons with opposite spin would be uncorrelated ($g(r)=1$ for all $r$). Also for sufficiently large separations, any pair of electrons is uncorrelated. However, for small separations, the pair correlation function of electrons with alike spin drops to zero, because of Pauli exclusion. This also remains true in systems with Coulomb interaction where the electrons are strongly attracted to the nuclei, as we show in Fig.~\ref{fig:N(r)}. 
When temperature increases for a given density, the pair correlation functions
decrease and approach eventually to 1, which shows that the kinetic energy may dominate over the Coulomb repulsion.  An exception are same-spin electron. Their $g(r)$ will go to zero for small $r$ at any temperature. At low temperature and low density, there is a very high correlation for both
parallel and anti-parallel spin electrons, which is caused by both types of electrons occupying bound states of a given nucleus.

%%%%%%%%%%%%%%%%%%%%%%%%%%%%%%%%%%%%%%%%%%%%%%%%%%%%%%%%%%%%%%%%%%%%%%%%%%%%%%%%%%%%%%%%%%
\subsection{Electronic Density of States}
\begin{figure}[!t]
    \centering
    \includegraphics[width=\columnwidth]{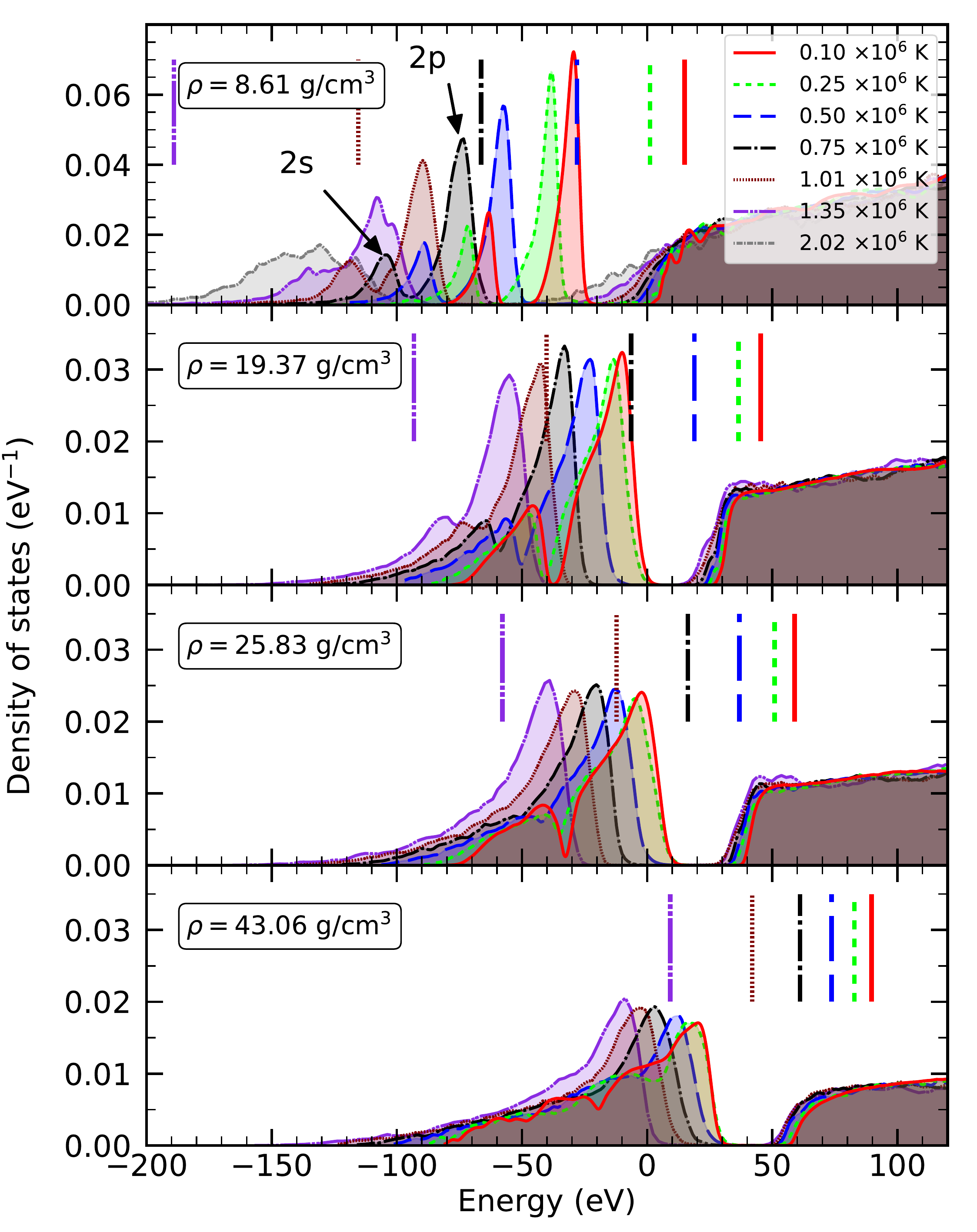}
    \caption{Electronic density of states derived from DFT-MD simulations is shown for different temperatures and densities. The vertical bars indicate the value of the average chemical potential. At low density, separate 2s and 2p peaks can be identified. With increasing density and temperature, these peaks broaden and eventually merge. A gap between 2p and 3s bands is seen for all conditions (see Fig.~\ref{fig:conduction_band}). The 3s states are always part of a broad conduction band.}
    \label{fig:DOS}
\end{figure}

We also studied the electronic density of states (DOS) of liquid Mg
through the analysis of the eigenenergies provided by Kohn-Sham DFT. 
With a  Brillouin zone sampled by the Gamma-point, we obtained smooth DOS curves by averaging over the MD-simulation snapshots and applying a Gaussian smearing of 0.1 eV to the band energies. The DOS at every snapshot was aligned at its respective Fermi energy, and then we averaged all of them together. The average Fermi energy was then subtracted out and the integrated DOS was normalized to 1.

Our average DOS functions, shown in Fig.~\ref{fig:DOS}, display two distinctive peaks at each temperature, representing bound 2s and 2p electrons of the L-shell, followed by the 2p-3s valence band gap and a continuum of conducting states, generated by the M-shell (3s) electrons.
Since the pseudopotential of our DFT-MD simulations has a frozen 1s core, these states do not appear in the DOS plots.
At low density, there is also a gap between the 2s and 2p peaks, which is present up to temperatures of $250\,000$~K. For higher temperatures, thermal excitations fill in the gap in between these two peaks.

Contrary to MgO,~\cite{Soubiran2019} where the band gap with the continuum closes completely due to the hybridization of oxygen and magnesium atomic orbitals, we observe that in pure Mg, the 2p-3s band gap does not disappear with compression.
However, as the density increases and the atomic orbitals start overlapping, the gap between the 2s and 2p bands does close with compression, causing these bands to merge and the height of the DOS peaks to decrease, as more electrons are promoted to the continuum.
The broadening of the peaks indicates that the electronic states are less localized because the overlap between atomic orbitals becomes more significant. The increasing occupation of continuum states also increases the internal energy of the system, which may also trigger the effect of pressure ionization that we illustrate in Fig.~\ref{fig:E_rho}. 

We use vertical lines in Fig.~\ref{fig:DOS} to mark the value of the chemical potential (or average Fermi energy). At low temperatures, it is located in the conduction band but it shifts toward lower energies as temperature increases. At $500\times10^3$~K and $8.61$ \gcc, the Fermi energy lies in the middle of the band gap, which would correspond to an insulator-like behavior if the smearing effects were not present. In fact, the Fermi smearing at this temperature is large enough (43.1 eV) to allow partial occupations in the conduction band, which implies a high electrical conductivity.
If the temperature is increased to $750\times10^3$~K at this density, the Fermi energy reaches the 2p band, which implies that the occupied DOS decreases, which indicates that there is partial ionization of the 2p electrons at these conditions. % so our results have important consequences for the ionization of Mg at extreme conditions.

%%%%%%%%%%%%%%%%%%%%%%%%%%%%%%%%%%%%%%%%%%%%%%%%%%%%%%%%%%%%%%%%%%%%%%%%%%%%%%%%%%%%%%%%%%
%In order to understand how band gap affects the degree of ionization...

\begin{figure}[!t]
    \centering
    \includegraphics[width=\columnwidth]{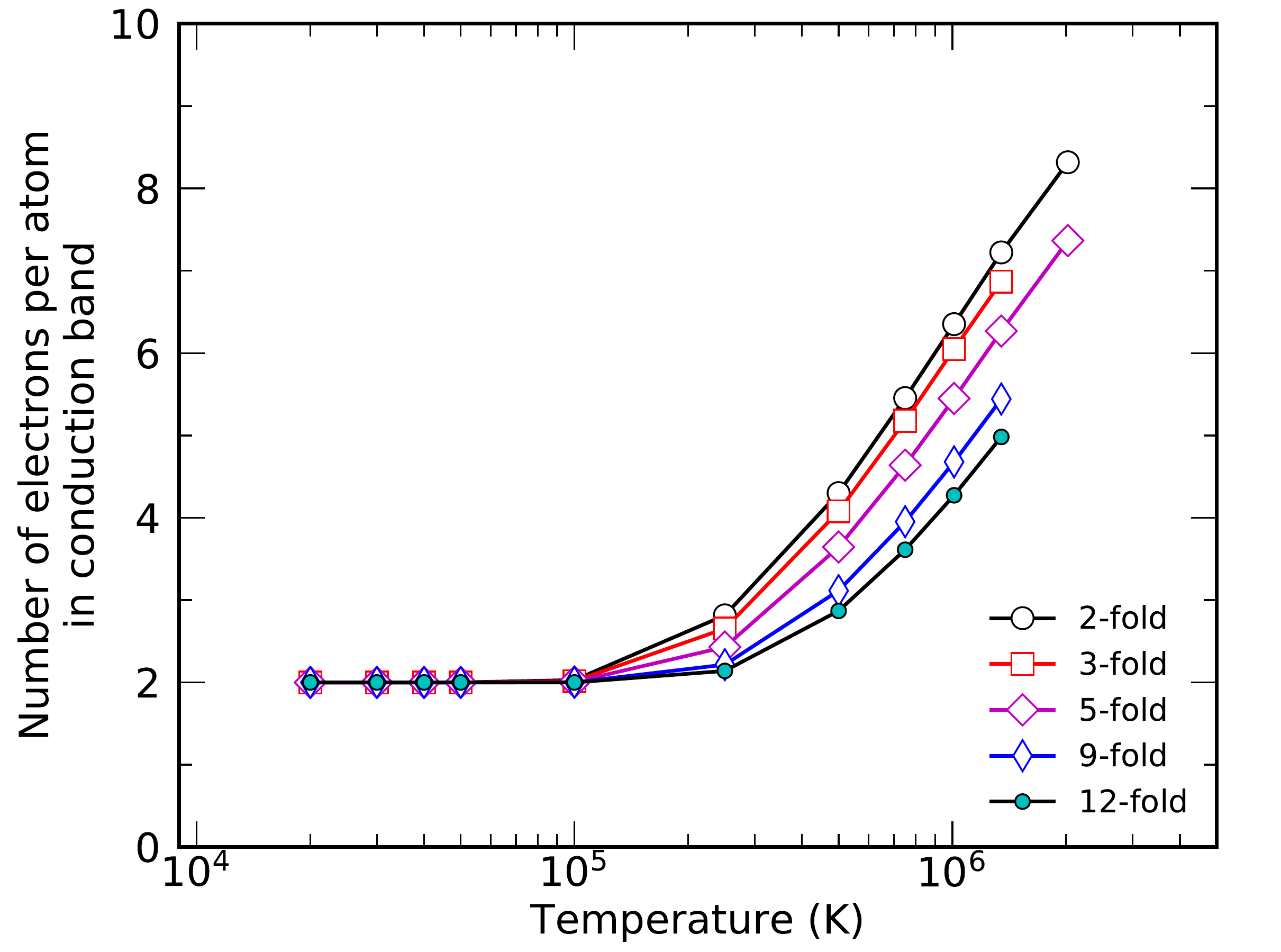}
    \includegraphics[width=\columnwidth]{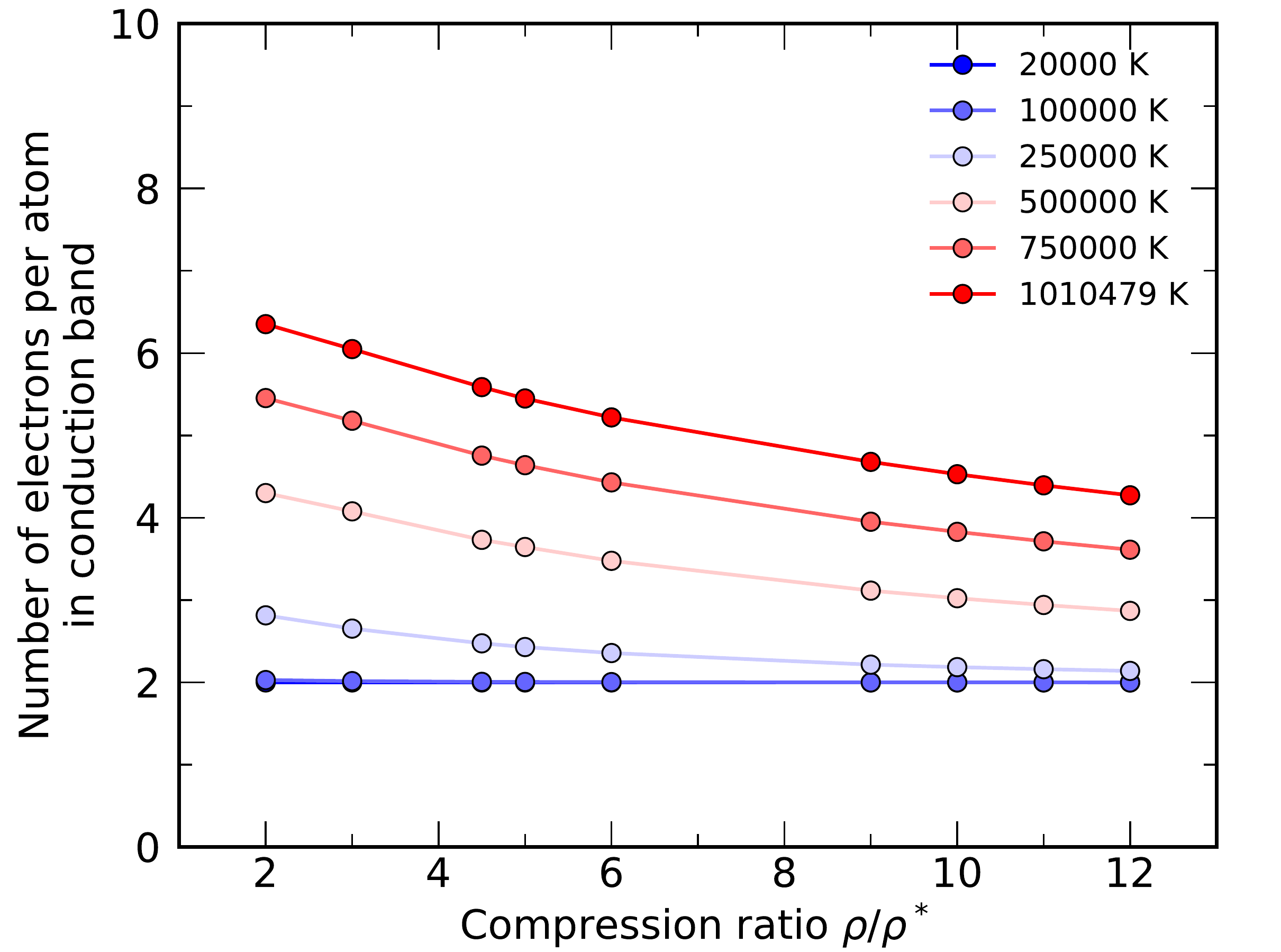}
    \caption{Average number of electrons per atom in the conduction band (3s) as a function of temperature and density. This number increases with increasing temperature and decreasing density as the average degree of ionization rises. Density axis has been normalized to a reference density of $\rho^*=4.3055475$ g cm$^{-3} $. }
    \label{fig:conduction_band}
\end{figure}

In Fig.~\ref{fig:conduction_band}, we plot the average number of electrons that have been promoted to the continuum. We sum up all band occupations excluding the lowest $N_e/2-N_I$ bands. $N_e$ and $N_I$ are the number of electrons and ions in the cell, respectively. $N_I$ is subtracted so that the 3s electrons are part of the continuum in this definition. The resulting number of electrons per atom in the continuum is equivalent, over the range of conditions explored here, to the average ionic charge $\left<Z\right>$ that has been used in Ref.~\onlinecite{Driver2018}. In Fig.~\ref{fig:conduction_band}, we show that there are only 2 electrons per atom in the conduction band for temperatures up to $10^5$ K, which means that no ionization below this temperature is expected at any of the densities under consideration. For these lower temperatures, the Fermi energy always lies in the conduction band, as we showed in Fig.~\ref{fig:DOS}.
At $2.5\times10^5$ K and above, the number of electrons in the conduction band is substantially larger than 2, which means that the degree of ionization increases. This number, and hence the degree of ionization, increases with decreasing density, and the number difference between densities is more notorious at higher temperatures.
\begin{figure}[!t]
    \centering
    \includegraphics[width=\columnwidth]{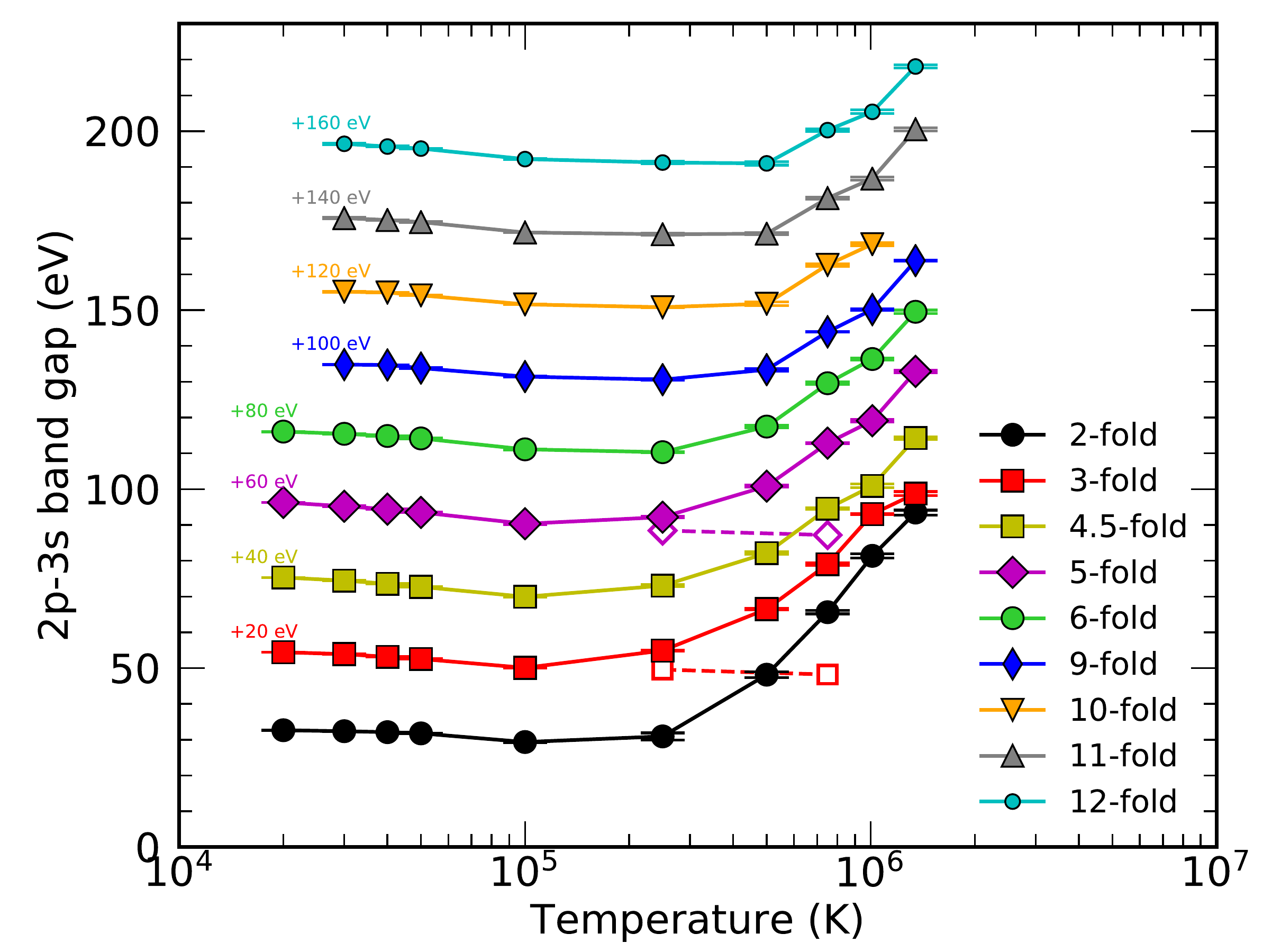}
    \includegraphics[width=\columnwidth]{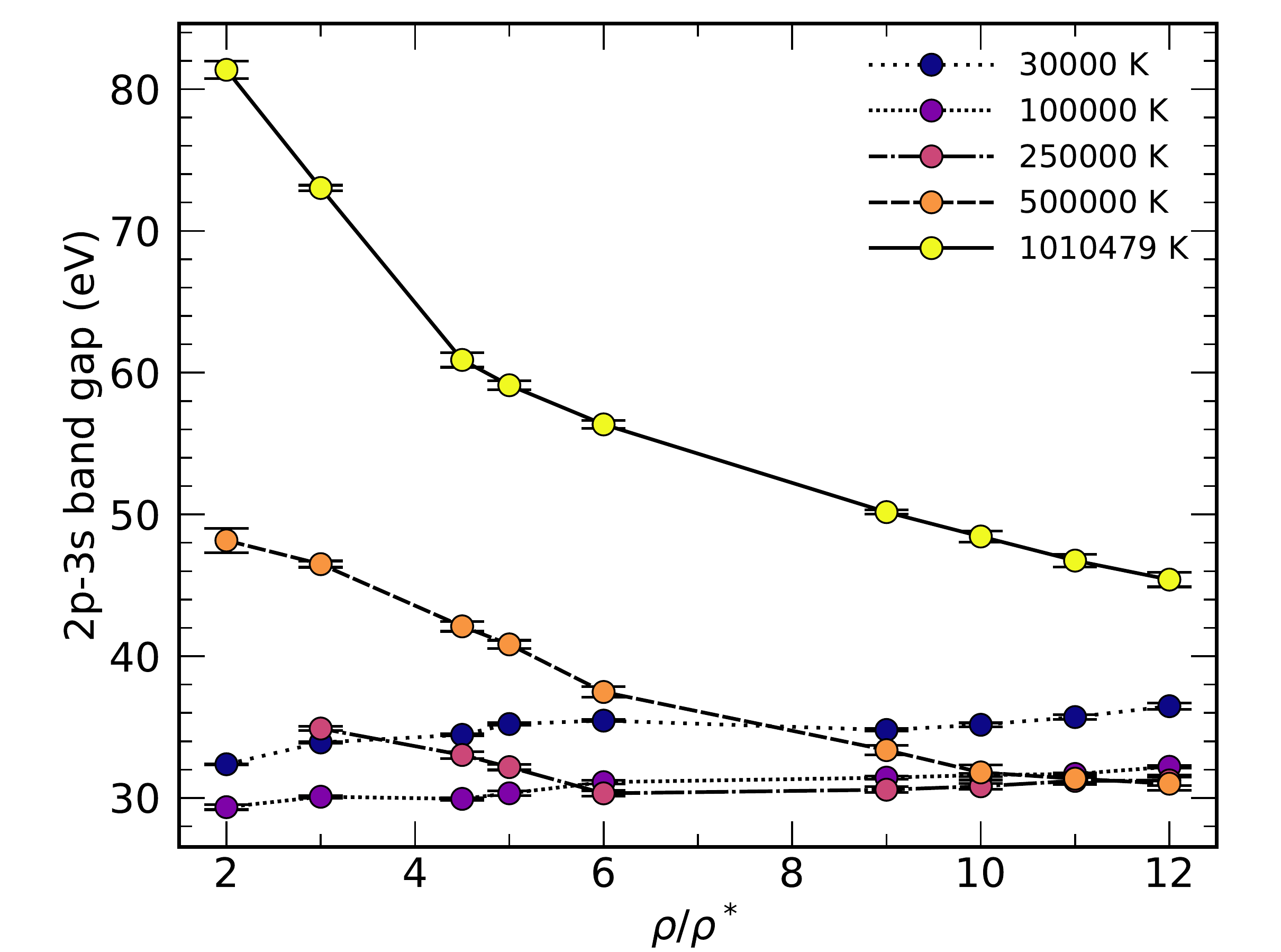}
    \caption{Gap between 2p and 3s bands derived from DFT-MD simulations is plotted as function of temperature and density. For clarity, the curves in the upper panel have been shifted vertically by the specified amounts. The open symbols show gaps from simulations at $\rho/\rho^*=3$ and 5 without electronic excitations, which we identified as the primary cause for the gap to increase with increasing temperature and decreasing density (see Fig.~\ref{fig:conduction_band}). In the upper panel, every curve represents a different density, normalized to the reference density of $\rho^*=4.3055475$ g cm$^{-3} $, which also defines horizontal axis of the lower plot. }
    \label{fig:gap}
\end{figure}

In Fig.~\ref{fig:gap}, we show how the 2p-3s band gap observed in Fig.~\ref{fig:DOS} depends on density and temperature. In the low-temperature regime up to 250,000 K, we see that there is a slight decrease in the  band gap with increasing temperature at fixed density that we attribute to the collisions between the nuclei that disorder the local electronic structure. Around 250,000 K, the band gap attains a minimum and then increases rapidly with temperature, because the degree of ionization increases, as we observed in Fig.~\ref{fig:conduction_band}.
In the upper panel of Fig.~\ref{fig:gap}, we find that the gap minimum is shifted towards higher temperature as the density is increased, because it is more difficult to ionize the system at such conditions. 
Along the isotherms (lower panel), the band gap decreases with density if the temperature is sufficiently high but it remains almost constant at low temperatures. Over the density interval from 2 to 12 $\times \rho^*$, the band gap changes by less than 10 eV for temperatures below $2.5\times10^5$ K, which represents a change of only 32\%. However, as a function of temperature, the band gap at a given density can change by more than 70 eV, which represents an increase of three times its value at low temperatures.  Thus, the valence band gap is more affected by temperature than by compression.

%%%%%%%%%%%%%%%%%%%%%%%%%%%%%%%%%%%%%%%%%%%%%%%%%%%%%%%%%%%%%%%%%%%%%%%%%%%%%%%%%%%%%%%%%%

To explain why this transition occurs, we notice that
at temperatures below $2.5\times10^5 $ K, there is not enough thermal excitation of the 2p states to promote them to the continuum. The Fermi energy is located in the conduction band and the energy difference is too high with respect to the 2p states. The bound states are thus not ionized under these conditions.
But above 250$\,$000 K there is significant thermal excitation of the 2p states, giving rise to ionization. In more ionized systems, fewer electrons screen the charges of the nuclei. The eigenenergies of the bound states thus decrease because of the lower effective nuclear charge. As a consequence, the band gap between the 2p states and the continuum increases.

For temperatures higher than 250$\,$000 K, a density increase forces the 2p states to recombine and the screening is therefore increased, which results in a decrease of the gap. The temperature of 250$\,$000 K is a turning point because the band gap, which is typically about 30 eV, corresponds to a temperature of 330$\,$000 K. Therefore, it is expected to have a significant thermal ionization of the 2p levels above this temperature.
We observe for instance at 8.61~\gcc~(2 $\times \rho^*$), that the Fermi energy leaves the conduction band around $2.5\times10^5 $ K, approaching the 2p band with increasing temperature (see Fig.~\ref{fig:DOS}). The occupation of the 2s and 2p bands thus decreases. At 750$\,$000~K, the Fermi energy is right above the merged 2s-2p peaks, which corresponds to the point where the Hugoniot curve has its first peak, as we will discuss in section~\ref{hug}. By 1.0$\times10^6$ K the ionization is such that the Fermi energy lies in between the 2s and 2p peaks and then goes over the 2s peak at about 1.3$\times10^6$ K. Although a band gap still exists at these temperatures, the thermal excitations have ionized the 2s and 2p levels and promoted almost all L shell electrons to the partially occupied states of the conduction band.

%This coincides with the point at which the band gap and number of electrons in the conduction band start increasing (see Figs.~\ref{fig:gap} and~\ref{fig:conduction_band}). We can conclude that, as temperature increases, the DOS peaks broaden, decreasing the availability of localized states, while the decreasing band gap and Fermi energy compete and do not promote electrons to the conduction band until the Fermi energy reaches the bottom of the conduction band. This occurs around 250 000 K for low densities, but higher temperatures are required at higher densities. After this point, the band gap starts increasing with temperature, but the Fermi energy keeps decreasing, which allows more electrons in the conduction band.

This picture is consistent with the ionization observed at much higher temperatures in our PIMC simulations (see Fig.~\ref{fig:N(r)}), where the integrated nuclear-electron pair correlation function $N(r)$ always increased upon compression, reducing the number of electrons in the conduction band and, hence, increasing the number of electron in the bounded 1s state. Therefore, we can conclude that at $2.5\times10^5$ K, the hybridized 2s and 2p bands start contributing to the conduction band, while the 1s electrons do so at $4.0\times10^6$ K.

%%%%%%%%%%%%%%%%%%%%%%%%%%%%%%%%%%%%%%%%%%%%%%%%%%%%%%%%%%%%%%%%%%%%%%%%%%%%%%%%%%%%%%%%%%%%%%%%%%%

\subsection{Structure of the fluid}
%\onecolumngrid
%\twocolumngrid

In order to characterize the structure of the fluid, we analyzed the trajectories of nuclei obtained from the DFT-MD simulations a function of the density and temperature. With the radial pair correlation function, $g_{\alpha\beta}(r)$, we can measure the local atomic coordination. This function can be interpreted as the probability
of finding an particle of type $\alpha$ at distance $r$ from a particle of type $\beta$.
The nuclear pair-correlation function is defined as,
\begin{equation}
g_{\alpha\beta}(r)= \frac{V}{{4\pi r^2N_\alpha N_\beta}}\left<\sum_{i=1}^{N_\alpha}\sum_{j\neq i}^{N_\beta}\delta\left(r-\|\vec r_{ij}\|\right)\right>,
\end{equation}
where $N_\alpha$ and $N_\beta$ are the total number of nuclei of type $\alpha$ and $\beta$,
respectively. $V$ is the cell volume, and  $\vec r_{ij}=\vec r_i-\vec r_j$
the separation between nuclei $i$ and $j$.

\begin{figure}[!hbt]
    \centering
    \includegraphics[width=9cm]{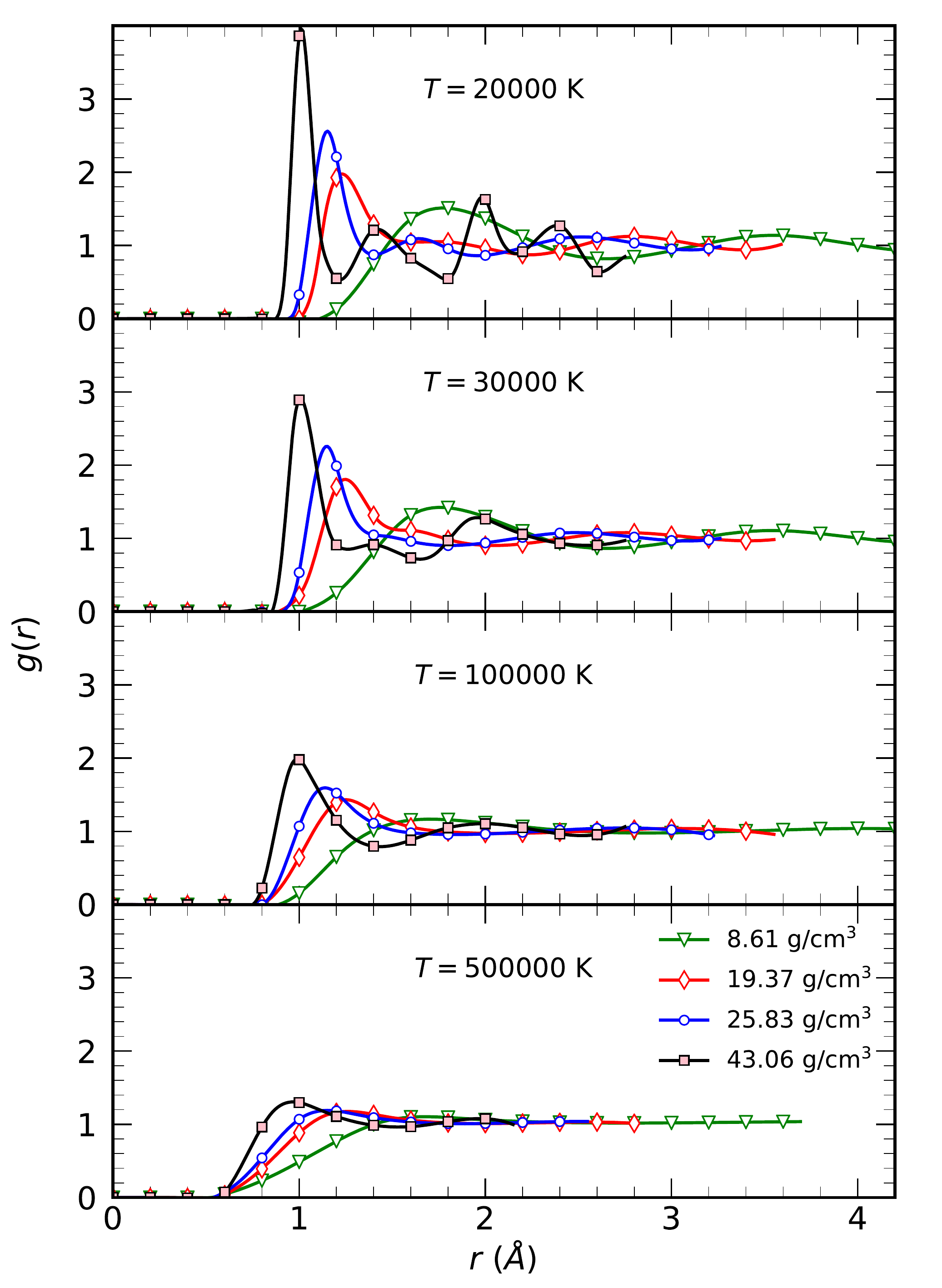}
    \caption{ Nuclear pair correlation functions computed with DFT-MD simulations at different temperatures and densities. Simulations with 32 atoms were used at $5 \times 10^5$ K while 64 atoms were used for lower temperatures. All simulations are liquid except for one at T=20,000 K and 43.06 \gcc.
  \label{fig:g(r)}
  }
\end{figure}

In Fig.~\ref{fig:g(r)}, we compare the different $g(r)$ functions at selected temperature and density conditions. As temperature increases and density decreases, the particle motion becomes less correlated and the liquid gradually loses its structure as the nuclei become more homogeneously distributed. At close range, a strong repulsion persists at all conditions, which is the result of Coulomb forces and Pauli exclusion. These two effects cause the fluid to freeze into an amorphous solid at the lowest temperature (20\,000 K) and highest density (43.04~\gcc) under consideration. The black curve in the top panel of Fig.~\ref{fig:g(r)} shows a number of additional peaks that are typical of amorphous samples.~\cite{Gutierrez2010,Kalkan2018,Drewitt2020} The position of the first peak in the $g(r)$ function in Fig.~\ref{fig:g(r)} does not change much with temperature. So, the average nearest-neighbor distance between Mg nuclei is always about 1~\AA.

As density decreases, the height of the first peak is reduced. The peak broadens and shift towards larger distances. This means that the nearest-neighbor distance increases, as expected, and that the separation between atoms covers a wider range of distances. The second and third peaks indicate the average positions of 2nd and 3rd nearest neighbors. These peaks are smoothed out with increasing temperature until there is no signature left at approximately 500\,000 K. Besides the strong short-range repulsion, there is litte structure left in the liquid at this temperature.   Only at 43.06~\gcc, the first peak is still visible. For distances $r\geqslant 1.4$~\AA, no correlation effects are present. Correlations between Mg nuclei up to 5$\times10^5$ K have also been observed in MgO~\cite{Soubiran2019} where the average nearest-neighbor distance between Mg nuclei is also approximately 1~\AA. However, the correlations are slightly stronger than in pure Mg due to the presence of oxygen nuclei. Similar correlations effects have been observed in simulations of MgSiO$_3$~\cite{GonzalezMilitzer2019} where the first peak in Mg-Mg pair correlation function can still be identified at 32.08~\gcc and 250\,000 K, which is consistent with the changes in the electronic structure that we discussed earlier.

%%%%%%%%%%%%%%%%%%%%%%%%%%%%%%%%%%%%%%%%%%%%%%%%%%%%%%%%%%%%%%%%%%%%%%%%%%%%%%%%%%%%%%%%%%%%%%%%%%%
\subsection{Shock Hugoniot Curves}
\label{hug}

The EOS can be used to infer the conditions reached by a material when subjected to dynamical shock compression. Assuming thermodynamic equilibrium is reached in experiments, the measured shock and particle velocity can be converted into pressure, density, and energy through the Rankine-Hugoniot equations.~\cite{Hugoniot1887,Hugoniot1889,Ze66} The energy conservation equation,
\begin{equation}
(E-E_0) + \frac{1}{2} (P+P_0)(V-V_0) = 0,
\label{eq:hug}
\end{equation}
is particularly convenient to derive the shock Hugoniot curve with theoretical methods.
Here, $E_0$, $V_0$, and $P_0$ represent the initial conditions of energy, volume, and pressure, respectively. 
$E$, $V$, and $P$ are the final conditions after the material behind the shock front has reached a equilibrium state. 
The shock Hugoniot curves of many materials have been measured up to megabar, and in some cases gigabar, pressures.~\cite{Root2010,Bolis2016,Root2018,Fratanduono2018}
Even at extreme conditions,~\cite{Wang2010,Mattsson2014,ZhangBN2019,ZhangCH2018,Soubiran2019,GonzalezMilitzer2019} predictions from {\it ab initio} simulations have been validated. 

\begin{figure}
    \centering
    \includegraphics[width=\columnwidth]{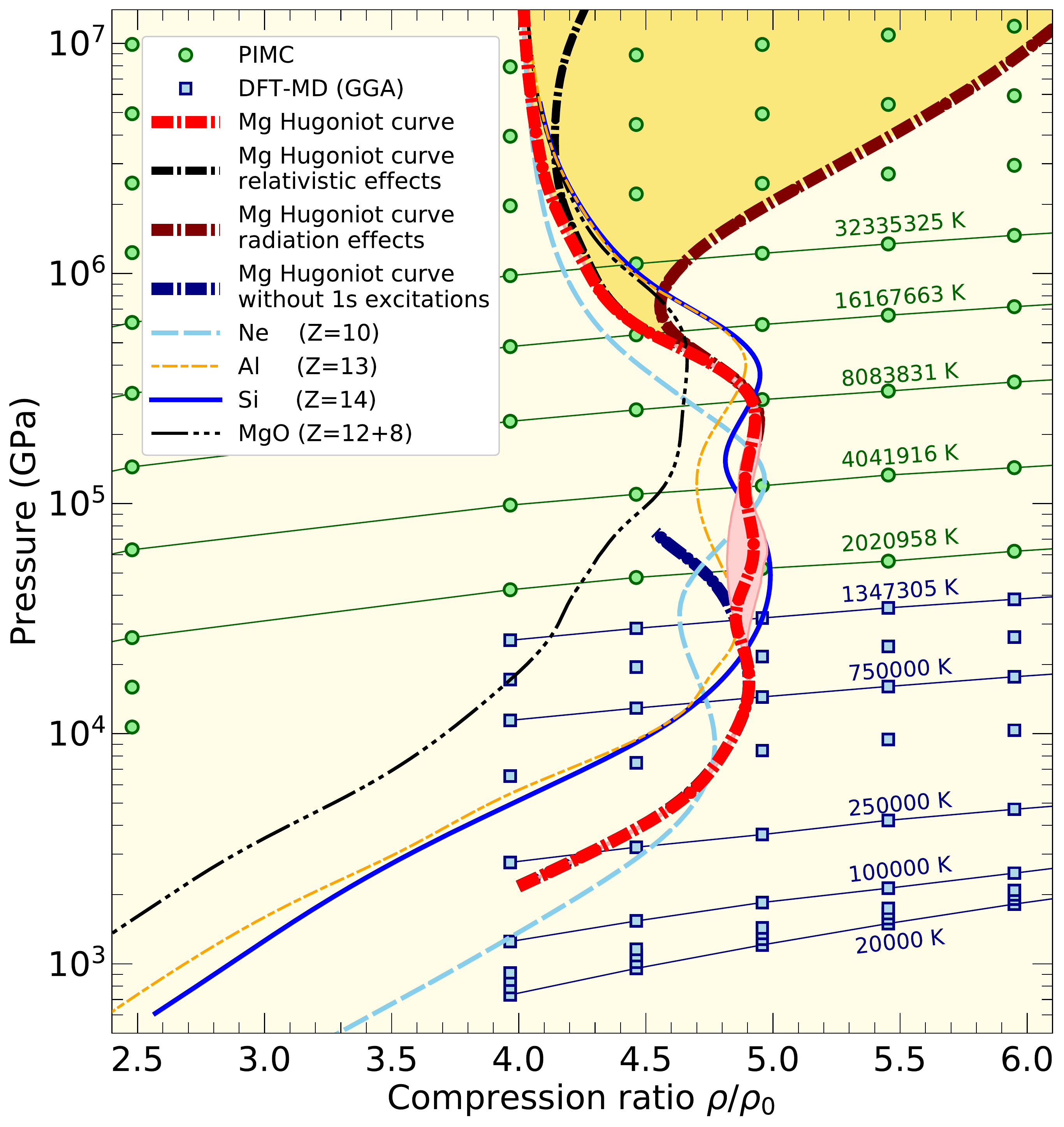}
    \caption{The Mg shock Hugoniot curves with and without relativistic and radiation effects are compared with the Hugoniot curves of %carbon~\cite{Driver2012,Benedict2014},
    %oxygen~\cite{Driver2015b},
    neon,~\cite{Driver2015}
    aluminum,~\cite{Driver2018} and
    silicon.~\cite{MilitzerDriver2015,Hu2016}
    The compression ratio is given with respect to the initial density of $\rho_0=1.73686577$~\gcc. One finds a broad temperature interval from 250,000 to 1.6 $\times 10^7$ K where the compression ratio of Mg exceeds 4.5, which can be attributed to the ionization of the K and L shell electrons. Without the excitation of K shell electrons the compression ratio decreases for temperatures above $1.3 \times 10^6$ K (blue dot-dashed line). The pink shaded region shows the uncertainties of the principal Hugniot curve, which is largest in the region where we switch between PIMC (circles) and DFT-MD (squares) EOS points. The horizontal lines show several isotherms. }
    \label{fig:HugoniotComp}
\end{figure}

\begin{figure}
    \centering
    \includegraphics[width=\columnwidth]{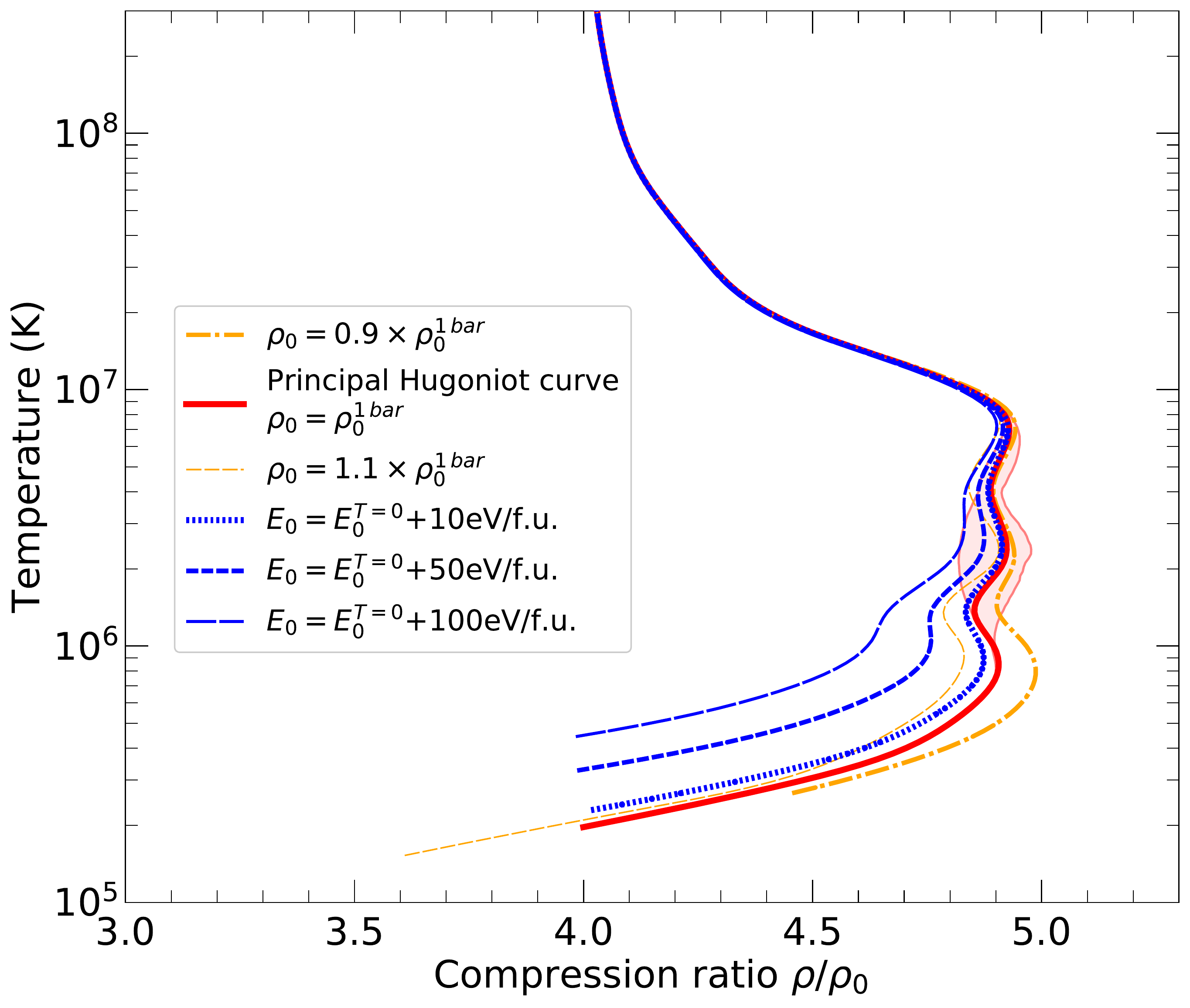}
    \caption{Shock Hugoniot curves with preheat and precompression. In the Hugoniot calculations, the initial density $\rho_0^{\rm 1\,bar}= 1.73686577$ \gcc~was modified to simulate the effects of static precompression. Similarly, the initial internal energy, $E_0^{T=0}$, was modified to simulate the effects of preheat. The shaded region illustrates the uncertainties of the principal Hugoniot curve from Fig.~\ref{fig:HugoniotComp}.}
    \label{fig:pre}
\end{figure}

We solve the Eq.~\eqref{eq:hug} for $T$ and $V$ using a double-spline interpolation of the computed $E(\rho,T)$ and $P(\rho,T)$ in our EOS table (see supplementary material). We start from the ambient density of solid hcp magnesium, $\rho_0=1.73686577 $~\gcc~($V_0=23.236914$ \AA$^3$/atom) and $P_0=P^{\rm PBE}(\rho_0) \approx 0$ as initial conditions. Depending on whether we compute $E(\rho,T)$ with the LDA or PBE functional,  two slightly different initial energies, $E_0^{\rm LDA} = -199.722498$ and $E_0^{\rm PBE} = -200.011011$  Ha/atom, are used. This is a reasonable choice in order to minimize the error that arises from choosing a particular DFT functional. The difference between these two $E_0$ values is small compared the $\sim 10^4$ Ha/atom that the internal energy changes along the shock Hugoniot curve in the temperature interval that we study here. 
When we use PIMC values for $E$, we combine them with $E_0^{\rm PBE}$ because this approach has worked well in Ref.~\onlinecite{ZhangSodium2017} and ~\onlinecite{Soubiran2019}. The resulting shock Hugoniot curve has been added to Figs.~\ref{fig:Tvsrho}, \ref{fig:PvsT}, ~\ref{fig:HugoniotComp}, and \ref{fig:pre}.

In Fig.~\ref{fig:HugoniotComp}, we show the shock Hugoniot curve as a function of the compression ratio, which spans across a wide range of pressures. We find a single broader region of high compression. From $5\times10^5$ K (8000 GPa) to $1\times 10^7$ K (370,000 GPa) the compression exceeds 4.8-fold the initial density. The maximum compression is approximately $4.9\,\rho_0$. Already at 200$\,$000$\,$K and 2,200 GPa, the compression ratio exceeds 4.0, which is the asymptotic value for a non-relativistic ideal gas. The high compression ratio in our Mg shock Hugoniot curve is the result of excitations of internal degrees of freedom,~\cite{Mi06} which increase the internal energy term in Eq.~\eqref{eq:hug}. Consequently, the second term in this equation becomes more negative, which reduces the volume $V$ and thus increases the compression ratio. The compression maximum is the result of L shell ionization that dominates the lower temperature regime ($8 \times 10^5$ K and 16,000 GPa) and K shell ionization effects that primarily occur around $7 \times 10^6$ K and 230,000 GPa. The shock Hugoniot curves of neon,~\cite{Driver2015}
    aluminum,~\cite{Driver2018} and
    silicon~\cite{MilitzerDriver2015,Hu2016} in Fig.~\ref{fig:HugoniotComp} show two well-separated compression maxima for the L and K shell ionization and a minimum in between. We do not see such a minimum in our Mg Hugoniot curve. Instead, we find a very small third compression maximum, but that is within the error bars of our Hugoniot curve computation. We derived these error bars by including two effects. First we propagated the 1-$\sigma$ error bars in the computed pressures and energies and second, we included the changes that resulted from removing all EOS points at either 1.3 $\times 10^6$ K or 2.0 $\times 10^6$ K because we switch between PIMC and DFT-MD results at these temperatures.

In Fig.~\ref{fig:HugoniotComp}, the upper maximum compression ratio of $\rho/\rho_0=4.9$ corresponds to a density of $\rho=8.51$~\gcc, which is equivalent to 2 $\times \rho^*$ in Fig.~\ref{fig:N(r)}. At this density, most of the K shell ionization occurs in the temperature interval from 4 to 8 $\times 10^6$ K, which are precisely the conditions of the Hugoniot curve compression maximum.
At temperatures higher than $16\times10^6$ K, radiation effects become important and substantially increase the compression ratio predicted by the Rankine-Hugoniot equations, allowing compressions beyond 6-fold. The difference is highlighted by the shaded area in Fig.~\ref{fig:HugoniotComp}. Radiation effects have been included by considering an ideal black body correction to our EOS using $P_{\text{rad}}=(4\sigma/3c)T^4$ and $E_{\text{rad}}= 3VP_{\text{rad}}$ where $\sigma$ is the Stefan-Boltzmann constant and $c$ is the speed of light in vacuum. The energy correction drives the increase in compression as we have seen in case of the K and L shell ionizations. We also studied the relativistic effects of the free electrons but they only become relevant for temperatures above $32\times10^6$ K and do not change the Hugoniot curve as much as radiation effects do. 

%{\color{red}CAN THE FIRST PEAK BE ATTRIBUTED TO L-shell IONIZATION?}, as the maxima seen %in simulations of pure nitrogen and oxygen~\cite{DriverNitrogen2016,Driver2018}. 
%We conclude that L shells of the Mg nuclei are ionized gradually, as it occurs in dense %carbon and boron materials~\cite{ZhangCH2018,Zhang2018,ZhangBN2019}. This interpretation %is consistent with the electronic structure observed in Fig. \ref{fig:DOS} where the %oxygen and magnesium electronic states are strongly hybridized at high compression. 

%%%%%%%%%%%%%%%%%%%%%%%%%%%%%%%%%%%%%%%%%%%%%%%%%%%%%%%%%%%%%%%%%%%%%%%%%%%

In Fig.~\ref{fig:pre}, we study how preheat and a change in the initial density affect the Hugoniot curve. Both modifications do not affect in any significant way the upper compression maximum that is dominated by the ionization of K shell electrons. The lower L shell compression maximum is affected, however. The higher the initial density the smaller is the peak compression ratio because particles interact more strongly, which increases the pressure and thus reduces the compression ratio.~\cite{Mi06,Mi09} In experiments, a higher initial density may be achieved with static precompression in diamond anvil cells.~\cite{Militzer2007,Jeanloz2007} A lower initial density, may be realized by heating the material or when Mg is part of a compound that has a lower overall density. 

In Fig.~\ref{fig:pre}, we also study the effect of preheat that we may occur in shock experiments when the laser drive generates x-rays that, despite shielding, heat the sample before the shock reaches it. The effects of radiative preheat were studied in detail with hydrocode simulation by Nilsen {\it et al.}~\cite{Nilsen2020} Here we performed only a simplified analysis where we simulate the preheat effect by increasing the initial internal energy $E_0$ by different amounts. An increase of 10 eV/atom leads to a moderate reduction in shock compression only. Fig.~\ref{fig:pre}, shows that an increase of 50 eV per atom reduces the L shell compression maximum significantly. However, for 100 eV/atom, this maximum is reduced to only a shoulder in the Hugoniot curve.

%%%%%%%%%%%%%%%%%%%%%%%%%%%%%%%%%%%%%%%%%%%%%%%%%%%%%%%%%%%%%%%%%%%%%%%%%%%
\begin{figure}
    \centering
    \includegraphics[width=\columnwidth]{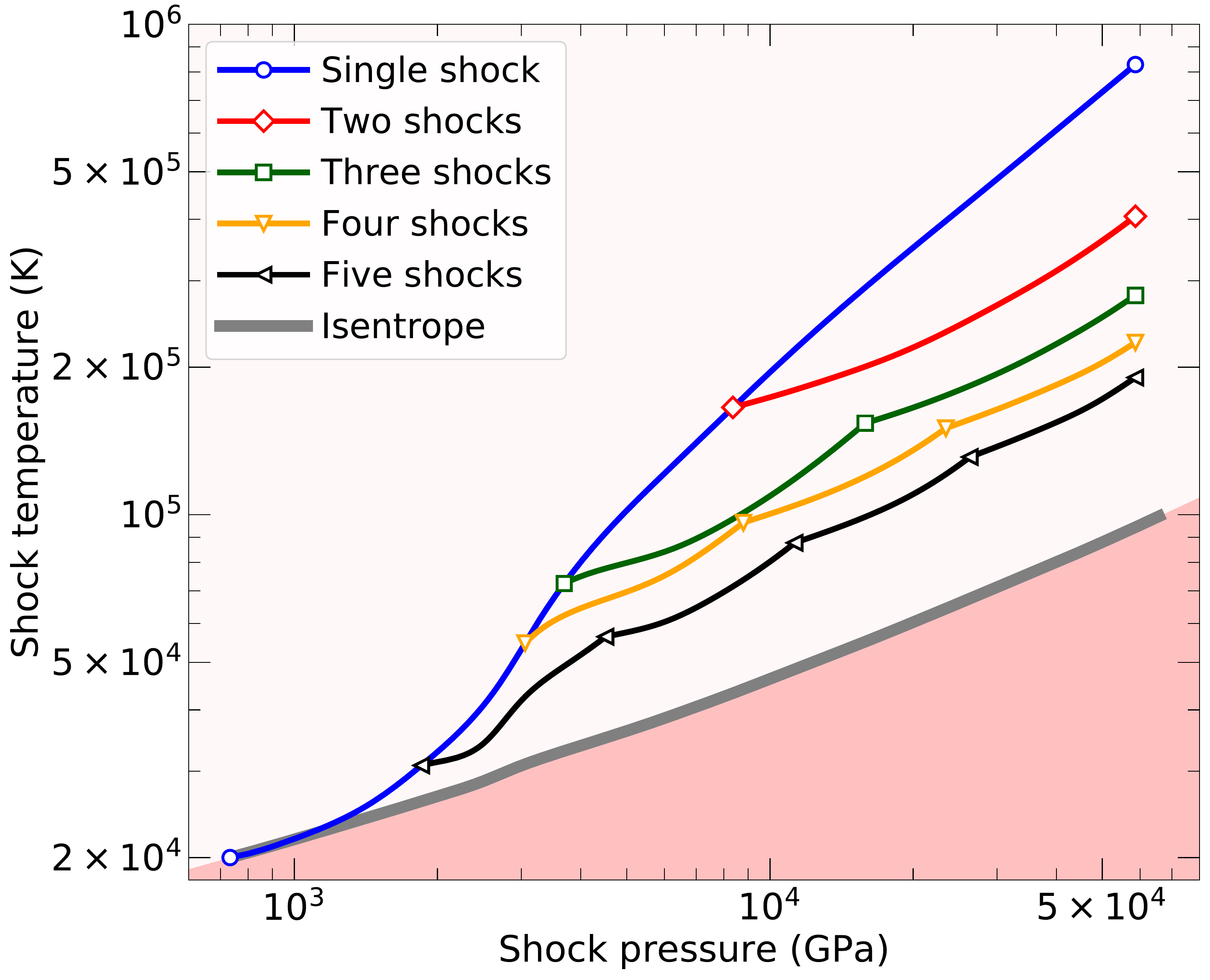}
    \caption{The amount of shock heating is reduced if multiple shocks are used to compress a material rather than just one. Here we plot the temperature-pressure stages for experiments with different numbers of shocks. The more shocks are employed, the closer the results are to an isentrope (thick grey line). The initial conditions were $T=20\,000\,$K and $P=733.2\,$GPa. The final pressure was kept at 58$\,$656$\,$GPa.}
    \label{fig:MultiShock}
\end{figure}

Isentropic compression can be thought of as the limit of infinite number of consecutive small shocks. Less and less heat is generated when the compression path is broken up into more and more shocks. In Fig.~\ref{fig:MultiShock}, we compare an isentrope with various multi-shock Hugoniot curves. All curves start from 20$\,$000$\,$K, twice the ambient density, and 733 GPa. The isentrope~\cite{MH08b} was derived from our EOS table using $\frac{dT}{dV}|_S = -T
{\frac{dP}{dT}|_V}/{ \frac{dE}{dT}|_V}$. For weak shocks, the Hugoniot curve does not deviate very much from an isentrope. For strong shocks, a substantial amount of shock heating occurs. The resulting single-shock Hugoniot curves are thus much hotter than an isentrope assuming both temperatures are compared for the same final pressure. The difference in temperature depends significantly on the final pressure.  To reach a large final pressure with a single shock, a substantial contribution to the pressure must come from the thermal pressure because final shock density cannot exceed 4.9 times the initial density (Fig.~\ref{fig:HugoniotComp}). 
The purpose of Fig.~\ref{fig:MultiShock} is to determine how much shock heating occurs if the shock is broken up into $N=2$--$5$ steps. In these multi-shock calculations, we successively solve Eq.~\ref{eq:hug} to connect the intermediate shock states. In order to obtain the lowest possible shock temperature for a given number of shocks, we keep the final shock pressure fixed while we carefully adjust the temperatures of the intermediate shocks until we determined the global minimum of the final shock temperature with sufficient accuracy.

As expected, the resulting multi-shock Hugoniot curves converge to an isentrope if the number of shocks is increased. For strong shocks, such as $P_{\rm final}/P_{\rm initial} \approx 80$, we find that the temperature of single-shock is 8.3 times higher than the corresponding temperature on the isentrope.
The final shock temperature can be reduced to 4.1 times the value in the isentrope if broken up into two.
If three, four or five shocks are employed, the final shock temperature can, respectively, be reduced to 2.8, 2.3, and 1.9 times the isentropic value. These are substantial reductions compared to the single-shock temperatures.

%%%%%%%%%%%%%%%%%%%%%%%%%%%%%%%%%%%%%%%%%%%%%%%%%%%%%%%%%%%%%%%%%%%%%%%%%%%%%%%%%%%%%%%%%%%%%%%%%%%
\section{Conclusion}

With PIMC and DFT-MD computer simulations, we have constructed a consistent EOS table for magnesium over a wide temperature-density range that bridges the WDM and plasma regimes. Our results provide the first detailed characterization of K shell ionization in magnesium. The ionization of the L shell gradually increases the compression along the principal shock Hugoniot curve until it reaches K shell ionization, where the compression ratio reaches a maximum of 4.9, which is as high as the maximum compression ratio of pure silicon. 

We find good agreement between results from PIMC and DFT-MD simulations, which provides evidence that the combination of these two different formulations of quantum mechanics can be used to accurately describe WDM. The precision of first-principles computer simulations will guide the design of inertial confinement fusion (ICF) experiments under conditions where the K and L shell electrons are gradually ionized, which is challenging to predict accurately with analytical EOS models.

The analysis of the density of states of magnesium at high pressures and temperatures led us to conclude that the 2s and 2p bands merge, as they do in magnesium oxide, but the band gap with the conduction band remains for all the conditions explored in this work. This band gap changes significantly with temperature. It decreases as temperature increases, until it reaches a minimum around $250\,000$ K, where an increasing number of electrons start populating the conduction band. Above this temperature, the band gap considerably increases with temperature as the atoms become more ionized and the liquid becomes less structured.

Finally, we found interesting features in the shock Hugoniot curve that can be attributed to ionization of electronic shells. The effects of preheating and precompression have significant impact in the predicted shock temperatures, but do not significantly change the shape of the curve, unless the preheating is considerably high. We observed three compression maxima that occur between $5\times10^5$ and $10^7$ K, which correspond to pressures between $8000$ and $370\,$000 K GPa. Although the highest temperature peak is certainly correlated with the K shell ionization and the lowest temperature peak to the end of L shell ionization, we do not find a physical mechanism that explains the intermediate peak. We cannot rule out the possibility of interpolation or uncertainty effects in this temperature region, as it corresponds to the boundary between our PIMC and DFT-MD data. Nevertheless, our prediction of a maximum compression ratio of $\rho/\rho_0=4.9$ is robust. More experiments are required in order to explore the different ionization regimes predicted by our calculations.

%From the present work it is clear that mixing magnesium and oxygen under extreme conditions has non-ideal effects on the electronic structure and thus the ionization processes. It will make it challenging for other theoretical methods to approximate the properties of Mg by combining EOS tables of Mg and O without fully taking into account their interaction effects. But it also demonstrates that elements with a relatively close atomic number may be strongly influenced by one another. This could have some significance for ICF experiments and for stellar interior opacities where the radiative properties and thus the ionization stage is of essence.

\section*{Supplementary Material}
See supplementary material for the complete EOS table of Mg pressures and internal energies at density-temperature conditions simulated in this work.

%%%%%%%%%%%%%%%%%%%%%%%%%%%%%%%%%%%%%%%%%%%%%%%%%%%%%%%%%%%%%%%%%%%%%%%%%%%%%%%%%%%%%%%%%%%%%%%%%%%
\begin{acknowledgments}
  This work was in part supported by the National Science
  Foundation-Department of Energy (DOE) partnership for plasma science   and engineering (grant DE-SC0016248), by the DOE-National Nuclear Security Administration (grant DE-NA0003842), and the University of California Laboratory Fees Research Program (grant LFR-17-449059). 
F.S. was in part supported by the European Union through a Marie Sk\l odowska-Curie action (grant 750901). F.G.-C. acknowledges support from the CONICYT Postdoctoral fellowship (grant 74160058). Computational support was provided by the Blue Waters sustained-petascale computing project (NSF ACI 1640776) and the National Energy Research Scientific Computing Center (NERSC).
\end{acknowledgments}

%%%%%%%%%%%%%%%%%%%%%%%%%%%%%%%%%%%%%%%%%%%%%%%%%%%%%%%%%%%%%%%%%%%%%%%%%%%%%%%%%%%%%%%%%%%%%%%%%%%

%%%%%%%%%%%%%%%%%%%%%%%%%%%%%%%%%%%%%%%%%%%%%%%%%%%%%%%%%%%%%%%%%%%%%%%%%%%%%%%%%%%%%%%%%%%%%%%%%%%
\section*{Data Availability}
The data that supports the findings of this study are available within the article and its supplementary material.
%%%%%%%%%%%%%%%%%%%%%%%%%%%%%%%%%%%%%%%%%%%%%%%%%%%%%%%%%%%%%%%%%%%%%%%%%%%%%%%%%%%%%%%%%%%%%%%%%%%

%\bibliography{bibliography}% Produces the bibliography via BibTeX.

\begin{thebibliography}{103}%
\makeatletter
\providecommand \@ifxundefined [1]{%
 \@ifx{#1\undefined}
}%
\providecommand \@ifnum [1]{%
 \ifnum #1\expandafter \@firstoftwo
 \else \expandafter \@secondoftwo
 \fi
}%
\providecommand \@ifx [1]{%
 \ifx #1\expandafter \@firstoftwo
 \else \expandafter \@secondoftwo
 \fi
}%
\providecommand \natexlab [1]{#1}%
\providecommand \enquote  [1]{``#1''}%
\providecommand \bibnamefont  [1]{#1}%
\providecommand \bibfnamefont [1]{#1}%
\providecommand \citenamefont [1]{#1}%
\providecommand \href@noop [0]{\@secondoftwo}%
\providecommand \href [0]{\begingroup \@sanitize@url \@href}%
\providecommand \@href[1]{\@@startlink{#1}\@@href}%
\providecommand \@@href[1]{\endgroup#1\@@endlink}%
\providecommand \@sanitize@url [0]{\catcode `\\12\catcode `\$12\catcode
  `\&12\catcode `\#12\catcode `\^12\catcode `\_12\catcode `\%12\relax}%
\providecommand \@@startlink[1]{}%
\providecommand \@@endlink[0]{}%
\providecommand \url  [0]{\begingroup\@sanitize@url \@url }%
\providecommand \@url [1]{\endgroup\@href {#1}{\urlprefix }}%
\providecommand \urlprefix  [0]{URL }%
\providecommand \Eprint [0]{\href }%
\providecommand \doibase [0]{http://dx.doi.org/}%
\providecommand \selectlanguage [0]{\@gobble}%
\providecommand \bibinfo  [0]{\@secondoftwo}%
\providecommand \bibfield  [0]{\@secondoftwo}%
\providecommand \translation [1]{[#1]}%
\providecommand \BibitemOpen [0]{}%
\providecommand \bibitemStop [0]{}%
\providecommand \bibitemNoStop [0]{.\EOS\space}%
\providecommand \EOS [0]{\spacefactor3000\relax}%
\providecommand \BibitemShut  [1]{\csname bibitem#1\endcsname}%
\let\auto@bib@innerbib\@empty
%</preamble>
\bibitem [{\citenamefont {Ebeling}\ \emph {et~al.}(1991)\citenamefont
  {Ebeling}, \citenamefont {Foerster}, \citenamefont {Fortov}, \citenamefont
  {Gryaznov},\ and\ \citenamefont {Polishchuk}}]{Ebeling1991}%
  \BibitemOpen
  \bibfield  {author} {\bibinfo {author} {\bibfnamefont {W.}~\bibnamefont
  {Ebeling}}, \bibinfo {author} {\bibfnamefont {A.}~\bibnamefont {Foerster}},
  \bibinfo {author} {\bibfnamefont {V.}~\bibnamefont {Fortov}}, \bibinfo
  {author} {\bibfnamefont {V.}~\bibnamefont {Gryaznov}}, \ and\ \bibinfo
  {author} {\bibfnamefont {A.}~\bibnamefont {Polishchuk}},\ }\href@noop {}
  {\emph {\bibinfo {title} {Thermophysical properties of hot dense plasmas}}},\
  Vol.~\bibinfo {volume} {25}\ (\bibinfo  {publisher} {B.G. Teubner
  Verlagsgesellschaft},\ \bibinfo {year} {1991})\BibitemShut {NoStop}%
\bibitem [{\citenamefont {Zhang}\ \emph
  {et~al.}(2018{\natexlab{a}})\citenamefont {Zhang}, \citenamefont {Militzer},
  \citenamefont {Gregor}, \citenamefont {Caspersen}, \citenamefont {Yang},
  \citenamefont {Gaffney}, \citenamefont {Ogitsu}, \citenamefont {Swift},
  \citenamefont {Lazicki}, \citenamefont {Erskine}, \citenamefont {London},
  \citenamefont {Celliers}, \citenamefont {Nilsen}, \citenamefont {Sterne},\
  and\ \citenamefont {Whitley}}]{Zhang2018}%
  \BibitemOpen
  \bibfield  {author} {\bibinfo {author} {\bibfnamefont {S.}~\bibnamefont
  {Zhang}}, \bibinfo {author} {\bibfnamefont {B.}~\bibnamefont {Militzer}},
  \bibinfo {author} {\bibfnamefont {M.~C.}\ \bibnamefont {Gregor}}, \bibinfo
  {author} {\bibfnamefont {K.}~\bibnamefont {Caspersen}}, \bibinfo {author}
  {\bibfnamefont {L.~H.}\ \bibnamefont {Yang}}, \bibinfo {author}
  {\bibfnamefont {J.}~\bibnamefont {Gaffney}}, \bibinfo {author} {\bibfnamefont
  {T.}~\bibnamefont {Ogitsu}}, \bibinfo {author} {\bibfnamefont
  {D.}~\bibnamefont {Swift}}, \bibinfo {author} {\bibfnamefont
  {A.}~\bibnamefont {Lazicki}}, \bibinfo {author} {\bibfnamefont
  {D.}~\bibnamefont {Erskine}}, \bibinfo {author} {\bibfnamefont {R.~A.}\
  \bibnamefont {London}}, \bibinfo {author} {\bibfnamefont {P.~M.}\
  \bibnamefont {Celliers}}, \bibinfo {author} {\bibfnamefont {J.}~\bibnamefont
  {Nilsen}}, \bibinfo {author} {\bibfnamefont {P.~A.}\ \bibnamefont {Sterne}},
  \ and\ \bibinfo {author} {\bibfnamefont {H.~D.}\ \bibnamefont {Whitley}},\
  }\bibfield  {title} {\enquote {\bibinfo {title} {{Theoretical and
  experimental investigation of the equation of state of boron plasmas}},}\
  }\href@noop {} {\bibfield  {journal} {\bibinfo  {journal} {Phys. Rev. E}\
  }\textbf {\bibinfo {volume} {98}},\ \bibinfo {pages} {023205} (\bibinfo
  {year} {2018}{\natexlab{a}})}\BibitemShut {NoStop}%
\bibitem [{\citenamefont {Betti}\ and\ \citenamefont
  {Hurricane}(2016)}]{Betti2016}%
  \BibitemOpen
  \bibfield  {author} {\bibinfo {author} {\bibfnamefont {R.}~\bibnamefont
  {Betti}}\ and\ \bibinfo {author} {\bibfnamefont {O.}~\bibnamefont
  {Hurricane}},\ }\bibfield  {title} {\enquote {\bibinfo {title}
  {Inertial-confinement fusion with lasers},}\ }\href@noop {} {\bibfield
  {journal} {\bibinfo  {journal} {Nature Physics}\ }\textbf {\bibinfo {volume}
  {12}},\ \bibinfo {pages} {435} (\bibinfo {year} {2016})}\BibitemShut
  {NoStop}%
\bibitem [{\citenamefont {Seidl}\ \emph {et~al.}(2009)\citenamefont {Seidl},
  \citenamefont {Anders}, \citenamefont {Bieniosek}, \citenamefont {Barnard},
  \citenamefont {Calanog}, \citenamefont {Chen}, \citenamefont {Cohen},
  \citenamefont {Coleman}, \citenamefont {Dorf}, \citenamefont {Gilson} \emph
  {et~al.}}]{Seidl2009}%
  \BibitemOpen
  \bibfield  {author} {\bibinfo {author} {\bibfnamefont {P.}~\bibnamefont
  {Seidl}}, \bibinfo {author} {\bibfnamefont {A.}~\bibnamefont {Anders}},
  \bibinfo {author} {\bibfnamefont {F.}~\bibnamefont {Bieniosek}}, \bibinfo
  {author} {\bibfnamefont {J.}~\bibnamefont {Barnard}}, \bibinfo {author}
  {\bibfnamefont {J.}~\bibnamefont {Calanog}}, \bibinfo {author} {\bibfnamefont
  {A.}~\bibnamefont {Chen}}, \bibinfo {author} {\bibfnamefont {R.}~\bibnamefont
  {Cohen}}, \bibinfo {author} {\bibfnamefont {J.}~\bibnamefont {Coleman}},
  \bibinfo {author} {\bibfnamefont {M.}~\bibnamefont {Dorf}}, \bibinfo {author}
  {\bibfnamefont {E.}~\bibnamefont {Gilson}},  \emph {et~al.},\ }\bibfield
  {title} {\enquote {\bibinfo {title} {Progress in beam focusing and
  compression for warm-dense matter experiments},}\ }\href@noop {} {\bibfield
  {journal} {\bibinfo  {journal} {Nuclear Instruments and Methods in Physics
  Research Section A: Accelerators, Spectrometers, Detectors and Associated
  Equipment}\ }\textbf {\bibinfo {volume} {606}},\ \bibinfo {pages} {75--82}
  (\bibinfo {year} {2009})}\BibitemShut {NoStop}%
\bibitem [{\citenamefont {Miyanishi}\ \emph {et~al.}(2015)\citenamefont
  {Miyanishi}, \citenamefont {Tange}, \citenamefont {Ozaki}, \citenamefont
  {Kimura}, \citenamefont {Sano}, \citenamefont {Sakawa}, \citenamefont
  {Tsuchiya},\ and\ \citenamefont {Kodama}}]{Miyanishi2015}%
  \BibitemOpen
  \bibfield  {author} {\bibinfo {author} {\bibfnamefont {K.}~\bibnamefont
  {Miyanishi}}, \bibinfo {author} {\bibfnamefont {Y.}~\bibnamefont {Tange}},
  \bibinfo {author} {\bibfnamefont {N.}~\bibnamefont {Ozaki}}, \bibinfo
  {author} {\bibfnamefont {T.}~\bibnamefont {Kimura}}, \bibinfo {author}
  {\bibfnamefont {T.}~\bibnamefont {Sano}}, \bibinfo {author} {\bibfnamefont
  {Y.}~\bibnamefont {Sakawa}}, \bibinfo {author} {\bibfnamefont
  {T.}~\bibnamefont {Tsuchiya}}, \ and\ \bibinfo {author} {\bibfnamefont
  {R.}~\bibnamefont {Kodama}},\ }\bibfield  {title} {\enquote {\bibinfo {title}
  {Laser-shock compression of magnesium oxide in the warm-dense-matter
  regime},}\ }\href@noop {} {\bibfield  {journal} {\bibinfo  {journal} {Phys.
  Rev. E}\ }\textbf {\bibinfo {volume} {92}},\ \bibinfo {pages} {023103}
  (\bibinfo {year} {2015})}\BibitemShut {NoStop}%
\bibitem [{\citenamefont {Hammel}\ \emph {et~al.}(2010)\citenamefont {Hammel},
  \citenamefont {Haan}, \citenamefont {Clark}, \citenamefont {Edwards},
  \citenamefont {Langer}, \citenamefont {Marinak}, \citenamefont {Patel},
  \citenamefont {Salmonson},\ and\ \citenamefont {Scott}}]{Hammel2010}%
  \BibitemOpen
  \bibfield  {author} {\bibinfo {author} {\bibfnamefont {B.}~\bibnamefont
  {Hammel}}, \bibinfo {author} {\bibfnamefont {S.}~\bibnamefont {Haan}},
  \bibinfo {author} {\bibfnamefont {D.}~\bibnamefont {Clark}}, \bibinfo
  {author} {\bibfnamefont {M.}~\bibnamefont {Edwards}}, \bibinfo {author}
  {\bibfnamefont {S.}~\bibnamefont {Langer}}, \bibinfo {author} {\bibfnamefont
  {M.}~\bibnamefont {Marinak}}, \bibinfo {author} {\bibfnamefont
  {M.}~\bibnamefont {Patel}}, \bibinfo {author} {\bibfnamefont
  {J.}~\bibnamefont {Salmonson}}, \ and\ \bibinfo {author} {\bibfnamefont
  {H.}~\bibnamefont {Scott}},\ }\bibfield  {title} {\enquote {\bibinfo {title}
  {High-mode rayleigh-taylor growth in nif ignition capsules},}\ }\href@noop {}
  {\bibfield  {journal} {\bibinfo  {journal} {High Energy Density Physics}\
  }\textbf {\bibinfo {volume} {6}},\ \bibinfo {pages} {171 -- 178} (\bibinfo
  {year} {2010})},\ \bibinfo {note} {iCHED 2009 - 2nd International Conference
  on High Energy Density Physics}\BibitemShut {NoStop}%
\bibitem [{\citenamefont {Millot}\ \emph {et~al.}(2015)\citenamefont {Millot},
  \citenamefont {Dubrovinskaia}, \citenamefont {{\v{C}}ernok}, \citenamefont
  {Blaha}, \citenamefont {Dubrovinsky}, \citenamefont {Braun}, \citenamefont
  {Celliers}, \citenamefont {Collins}, \citenamefont {Eggert},\ and\
  \citenamefont {Jeanloz}}]{Millot2015}%
  \BibitemOpen
  \bibfield  {author} {\bibinfo {author} {\bibfnamefont {M.}~\bibnamefont
  {Millot}}, \bibinfo {author} {\bibfnamefont {N.~a.}\ \bibnamefont
  {Dubrovinskaia}}, \bibinfo {author} {\bibfnamefont {A.}~\bibnamefont
  {{\v{C}}ernok}}, \bibinfo {author} {\bibfnamefont {S.}~\bibnamefont {Blaha}},
  \bibinfo {author} {\bibfnamefont {L.}~\bibnamefont {Dubrovinsky}}, \bibinfo
  {author} {\bibfnamefont {D.}~\bibnamefont {Braun}}, \bibinfo {author}
  {\bibfnamefont {P.}~\bibnamefont {Celliers}}, \bibinfo {author}
  {\bibfnamefont {G.}~\bibnamefont {Collins}}, \bibinfo {author} {\bibfnamefont
  {J.}~\bibnamefont {Eggert}}, \ and\ \bibinfo {author} {\bibfnamefont
  {R.}~\bibnamefont {Jeanloz}},\ }\bibfield  {title} {\enquote {\bibinfo
  {title} {Shock compression of stishovite and melting of silica at planetary
  interior conditions},}\ }\href@noop {} {\bibfield  {journal} {\bibinfo
  {journal} {Science}\ }\textbf {\bibinfo {volume} {347}},\ \bibinfo {pages}
  {418--420} (\bibinfo {year} {2015})}\BibitemShut {NoStop}%
\bibitem [{\citenamefont {Kirsch}\ \emph {et~al.}(2019)\citenamefont {Kirsch},
  \citenamefont {Ali}, \citenamefont {Fratanduono}, \citenamefont {Kraus},
  \citenamefont {Braun}, \citenamefont {Fernandez-Pa{\~n}ella}, \citenamefont
  {Smith}, \citenamefont {McNaney},\ and\ \citenamefont {Eggert}}]{Kirsch2019}%
  \BibitemOpen
  \bibfield  {author} {\bibinfo {author} {\bibfnamefont {L.}~\bibnamefont
  {Kirsch}}, \bibinfo {author} {\bibfnamefont {S.}~\bibnamefont {Ali}},
  \bibinfo {author} {\bibfnamefont {D.}~\bibnamefont {Fratanduono}}, \bibinfo
  {author} {\bibfnamefont {R.}~\bibnamefont {Kraus}}, \bibinfo {author}
  {\bibfnamefont {D.}~\bibnamefont {Braun}}, \bibinfo {author} {\bibfnamefont
  {A.}~\bibnamefont {Fernandez-Pa{\~n}ella}}, \bibinfo {author} {\bibfnamefont
  {R.}~\bibnamefont {Smith}}, \bibinfo {author} {\bibfnamefont
  {J.}~\bibnamefont {McNaney}}, \ and\ \bibinfo {author} {\bibfnamefont
  {J.}~\bibnamefont {Eggert}},\ }\bibfield  {title} {\enquote {\bibinfo {title}
  {Refractive index of lithium fluoride to 900 gigapascal and implications for
  dynamic equation of state measurements},}\ }\href@noop {} {\bibfield
  {journal} {\bibinfo  {journal} {Journal of Applied Physics}\ }\textbf
  {\bibinfo {volume} {125}},\ \bibinfo {pages} {175901} (\bibinfo {year}
  {2019})}\BibitemShut {NoStop}%
\bibitem [{\citenamefont {Cotelo}\ \emph {et~al.}(2011)\citenamefont {Cotelo},
  \citenamefont {Velarde}, \citenamefont {de~La~Varga},\ and\ \citenamefont
  {Garc{\'\i}a-Fern{\'a}ndez}}]{Cotelo2011}%
  \BibitemOpen
  \bibfield  {author} {\bibinfo {author} {\bibfnamefont {M.}~\bibnamefont
  {Cotelo}}, \bibinfo {author} {\bibfnamefont {P.}~\bibnamefont {Velarde}},
  \bibinfo {author} {\bibfnamefont {A.}~\bibnamefont {de~La~Varga}}, \ and\
  \bibinfo {author} {\bibfnamefont {C.}~\bibnamefont
  {Garc{\'\i}a-Fern{\'a}ndez}},\ }\bibfield  {title} {\enquote {\bibinfo
  {title} {Equation of state for laboratory astrophysics applications},}\
  }\href@noop {} {\bibfield  {journal} {\bibinfo  {journal} {Astrophysics and
  Space Science}\ }\textbf {\bibinfo {volume} {336}},\ \bibinfo {pages}
  {53--59} (\bibinfo {year} {2011})}\BibitemShut {NoStop}%
\bibitem [{\citenamefont {Chabrier}, \citenamefont {Douchin},\ and\
  \citenamefont {Potekhin}(2002)}]{Chabrier2002}%
  \BibitemOpen
  \bibfield  {author} {\bibinfo {author} {\bibfnamefont {G.}~\bibnamefont
  {Chabrier}}, \bibinfo {author} {\bibfnamefont {F.}~\bibnamefont {Douchin}}, \
  and\ \bibinfo {author} {\bibfnamefont {A.}~\bibnamefont {Potekhin}},\
  }\bibfield  {title} {\enquote {\bibinfo {title} {Dense astrophysical
  plasmas},}\ }\href@noop {} {\bibfield  {journal} {\bibinfo  {journal}
  {Journal of Physics: Condensed Matter}\ }\textbf {\bibinfo {volume} {14}},\
  \bibinfo {pages} {9133} (\bibinfo {year} {2002})}\BibitemShut {NoStop}%
\bibitem [{Exo()}]{ExoplanetArchive}%
  \BibitemOpen
  \href {http://exoplanet.eu/} {\enquote {\bibinfo {title} {{The Extrasolar
  Planets Encyclopedia http://exoplanet.eu/}},}\ }\BibitemShut {NoStop}%
\bibitem [{\citenamefont {Guillot}(1999)}]{Guillot1999}%
  \BibitemOpen
  \bibfield  {author} {\bibinfo {author} {\bibfnamefont {T.}~\bibnamefont
  {Guillot}},\ }\bibfield  {title} {\enquote {\bibinfo {title} {{Interiors of
  Giant Planets Inside and Outside the Solar System}},}\ }\href {\doibase
  10.1126/science.286.5437.72} {\bibfield  {journal} {\bibinfo  {journal}
  {Science}\ }\textbf {\bibinfo {volume} {286}},\ \bibinfo {pages} {72--77}
  (\bibinfo {year} {1999})}\BibitemShut {NoStop}%
\bibitem [{\citenamefont {Militzer}\ \emph {et~al.}(2016)\citenamefont
  {Militzer}, \citenamefont {Soubiran}, \citenamefont {Wahl},\ and\
  \citenamefont {Hubbard}}]{Militzer2016b}%
  \BibitemOpen
  \bibfield  {author} {\bibinfo {author} {\bibfnamefont {B.}~\bibnamefont
  {Militzer}}, \bibinfo {author} {\bibfnamefont {F.}~\bibnamefont {Soubiran}},
  \bibinfo {author} {\bibfnamefont {S.~M.}\ \bibnamefont {Wahl}}, \ and\
  \bibinfo {author} {\bibfnamefont {W.}~\bibnamefont {Hubbard}},\ }\bibfield
  {title} {\enquote {\bibinfo {title} {{Understanding Jupiter's interior}},}\
  }\href {\doibase 10.1002/2016JE005080} {\bibfield  {journal} {\bibinfo
  {journal} {Journal of Geophysical Research: Planets}\ }\textbf {\bibinfo
  {volume} {121}},\ \bibinfo {pages} {1552--1572} (\bibinfo {year} {2016})},\
  \Eprint {http://arxiv.org/abs/1608.02685} {1608.02685} \BibitemShut {NoStop}%
\bibitem [{\citenamefont {Baraffe}\ \emph {et~al.}(2014)\citenamefont
  {Baraffe}, \citenamefont {Chabrier}, \citenamefont {Fortney},\ and\
  \citenamefont {Sotin}}]{Baraffe2014}%
  \BibitemOpen
  \bibfield  {author} {\bibinfo {author} {\bibfnamefont {I.}~\bibnamefont
  {Baraffe}}, \bibinfo {author} {\bibfnamefont {G.}~\bibnamefont {Chabrier}},
  \bibinfo {author} {\bibfnamefont {J.}~\bibnamefont {Fortney}}, \ and\
  \bibinfo {author} {\bibfnamefont {C.}~\bibnamefont {Sotin}},\ }\bibfield
  {title} {\enquote {\bibinfo {title} {{Planetary Internal Structures}},}\ }in\
  \href {\doibase 10.2458/azu_uapress_9780816531240-ch033} {\emph {\bibinfo
  {booktitle} {Protostars and Planets VI}}}\ (\bibinfo  {publisher} {University
  of Arizona Press},\ \bibinfo {year} {2014})\ \Eprint
  {http://arxiv.org/abs/arXiv:1401.4738v1} {arXiv:1401.4738v1} \BibitemShut
  {NoStop}%
\bibitem [{\citenamefont {Valencia}\ \emph {et~al.}(2010)\citenamefont
  {Valencia}, \citenamefont {Ikoma}, \citenamefont {Guillot},\ and\
  \citenamefont {Nettelmann}}]{Valencia2010}%
  \BibitemOpen
  \bibfield  {author} {\bibinfo {author} {\bibfnamefont {D.}~\bibnamefont
  {Valencia}}, \bibinfo {author} {\bibfnamefont {M.}~\bibnamefont {Ikoma}},
  \bibinfo {author} {\bibfnamefont {T.}~\bibnamefont {Guillot}}, \ and\
  \bibinfo {author} {\bibfnamefont {N.}~\bibnamefont {Nettelmann}},\ }\bibfield
   {title} {\enquote {\bibinfo {title} {{Composition and fate of short-period
  super-Earths}},}\ }\href
  {http://www.aanda.org/articles/aa/abs/2010/08/aa12839-09/aa12839-09.html
  http://arxiv.org/abs/0907.3067
  http://www.aanda.org/10.1051/0004-6361/200912839} {\bibfield  {journal}
  {\bibinfo  {journal} {Astronomy {\&} Astrophysics}\ }\textbf {\bibinfo
  {volume} {516}},\ \bibinfo {pages} {A20} (\bibinfo {year}
  {2010})}\BibitemShut {NoStop}%
\bibitem [{\citenamefont {Bolis}\ \emph {et~al.}(2016)\citenamefont {Bolis},
  \citenamefont {Morard}, \citenamefont {Vinci}, \citenamefont {Ravasio},
  \citenamefont {Bambrink}, \citenamefont {Guarguaglini}, \citenamefont
  {Koenig}, \citenamefont {Musella}, \citenamefont {Remus}, \citenamefont
  {Bouchet}, \citenamefont {Ozaki}, \citenamefont {Miyanishi}, \citenamefont
  {Sekine}, \citenamefont {Sakawa}, \citenamefont {Sano}, \citenamefont
  {Kodama}, \citenamefont {Guyot},\ and\ \citenamefont
  {Benuzzi-Mounaix}}]{Bolis2016}%
  \BibitemOpen
  \bibfield  {author} {\bibinfo {author} {\bibfnamefont {R.~M.}\ \bibnamefont
  {Bolis}}, \bibinfo {author} {\bibfnamefont {G.}~\bibnamefont {Morard}},
  \bibinfo {author} {\bibfnamefont {T.}~\bibnamefont {Vinci}}, \bibinfo
  {author} {\bibfnamefont {A.}~\bibnamefont {Ravasio}}, \bibinfo {author}
  {\bibfnamefont {E.}~\bibnamefont {Bambrink}}, \bibinfo {author}
  {\bibfnamefont {M.}~\bibnamefont {Guarguaglini}}, \bibinfo {author}
  {\bibfnamefont {M.}~\bibnamefont {Koenig}}, \bibinfo {author} {\bibfnamefont
  {R.}~\bibnamefont {Musella}}, \bibinfo {author} {\bibfnamefont
  {F.}~\bibnamefont {Remus}}, \bibinfo {author} {\bibfnamefont
  {J.}~\bibnamefont {Bouchet}}, \bibinfo {author} {\bibfnamefont
  {N.}~\bibnamefont {Ozaki}}, \bibinfo {author} {\bibfnamefont
  {K.}~\bibnamefont {Miyanishi}}, \bibinfo {author} {\bibfnamefont
  {T.}~\bibnamefont {Sekine}}, \bibinfo {author} {\bibfnamefont
  {Y.}~\bibnamefont {Sakawa}}, \bibinfo {author} {\bibfnamefont
  {T.}~\bibnamefont {Sano}}, \bibinfo {author} {\bibfnamefont {R.}~\bibnamefont
  {Kodama}}, \bibinfo {author} {\bibfnamefont {F.}~\bibnamefont {Guyot}}, \
  and\ \bibinfo {author} {\bibfnamefont {A.}~\bibnamefont {Benuzzi-Mounaix}},\
  }\bibfield  {title} {\enquote {\bibinfo {title} {{Decaying shock studies of
  phase transitions in MgO-SiO$_2$ systems: Implications for the super-Earths'
  interiors}},}\ }\href@noop {} {\bibfield  {journal} {\bibinfo  {journal}
  {Geophysical Research Letters}\ }\textbf {\bibinfo {volume} {43}},\ \bibinfo
  {pages} {9475--9483} (\bibinfo {year} {2016})}\BibitemShut {NoStop}%
\bibitem [{\citenamefont {Musella}, \citenamefont {Mazevet},\ and\
  \citenamefont {Guyot}(2019)}]{Musella2019}%
  \BibitemOpen
  \bibfield  {author} {\bibinfo {author} {\bibfnamefont {R.}~\bibnamefont
  {Musella}}, \bibinfo {author} {\bibfnamefont {S.}~\bibnamefont {Mazevet}}, \
  and\ \bibinfo {author} {\bibfnamefont {F.}~\bibnamefont {Guyot}},\ }\bibfield
   {title} {\enquote {\bibinfo {title} {{Physical properties of MgO at deep
  planetary conditions}},}\ }\href {\doibase 10.1103/PhysRevB.99.064110}
  {\bibfield  {journal} {\bibinfo  {journal} {Physical Review B}\ }\textbf
  {\bibinfo {volume} {99}},\ \bibinfo {pages} {064110} (\bibinfo {year}
  {2019})},\ \Eprint {http://arxiv.org/abs/1805.12439} {arXiv:1805.12439}
  \BibitemShut {NoStop}%
\bibitem [{\citenamefont {Gonz{\'{a}}lez-Cataldo}\ \emph
  {et~al.}(2020)\citenamefont {Gonz{\'{a}}lez-Cataldo}, \citenamefont
  {Soubiran}, \citenamefont {Peterson},\ and\ \citenamefont
  {Militzer}}]{GonzalezMilitzer2019}%
  \BibitemOpen
  \bibfield  {author} {\bibinfo {author} {\bibfnamefont {F.}~\bibnamefont
  {Gonz{\'{a}}lez-Cataldo}}, \bibinfo {author} {\bibfnamefont {F.}~\bibnamefont
  {Soubiran}}, \bibinfo {author} {\bibfnamefont {H.}~\bibnamefont {Peterson}},
  \ and\ \bibinfo {author} {\bibfnamefont {B.}~\bibnamefont {Militzer}},\
  }\bibfield  {title} {\enquote {\bibinfo {title} {{Path integral Monte Carlo
  and density functional molecular dynamics simulations of warm dense
  MgSiO$_3$}},}\ }\href {\doibase 10.1103/PhysRevB.101.024107} {\bibfield
  {journal} {\bibinfo  {journal} {Physical Review B}\ }\textbf {\bibinfo
  {volume} {101}},\ \bibinfo {pages} {024107} (\bibinfo {year}
  {2020})}\BibitemShut {NoStop}%
\bibitem [{\citenamefont {Soubiran}\ \emph {et~al.}(2019)\citenamefont
  {Soubiran}, \citenamefont {Gonz{\'{a}}lez-Cataldo}, \citenamefont {Driver},
  \citenamefont {Zhang},\ and\ \citenamefont {Militzer}}]{Soubiran2019}%
  \BibitemOpen
  \bibfield  {author} {\bibinfo {author} {\bibfnamefont {F.}~\bibnamefont
  {Soubiran}}, \bibinfo {author} {\bibfnamefont {F.}~\bibnamefont
  {Gonz{\'{a}}lez-Cataldo}}, \bibinfo {author} {\bibfnamefont {K.~P.}\
  \bibnamefont {Driver}}, \bibinfo {author} {\bibfnamefont {S.}~\bibnamefont
  {Zhang}}, \ and\ \bibinfo {author} {\bibfnamefont {B.}~\bibnamefont
  {Militzer}},\ }\bibfield  {title} {\enquote {\bibinfo {title} {{Magnesium
  oxide at extreme temperatures and pressures studied with first-principles
  simulations}},}\ }\href {\doibase 10.1063/1.5126624} {\bibfield  {journal}
  {\bibinfo  {journal} {The Journal of Chemical Physics}\ }\textbf {\bibinfo
  {volume} {151}},\ \bibinfo {pages} {214104} (\bibinfo {year}
  {2019})}\BibitemShut {NoStop}%
\bibitem [{\citenamefont {Gonz{\'{a}}lez-Cataldo}\ and\ \citenamefont
  {Militzer}(2020)}]{GonzalezMilitzer2020}%
  \BibitemOpen
  \bibfield  {author} {\bibinfo {author} {\bibfnamefont {F.}~\bibnamefont
  {Gonz{\'{a}}lez-Cataldo}}\ and\ \bibinfo {author} {\bibfnamefont
  {B.}~\bibnamefont {Militzer}},\ }\bibfield  {title} {\enquote {\bibinfo
  {title} {{Thermal and Pressure Ionization in Warm, Dense MgSiO$_3$ Studied
  with First-Principles Computer Simulations}},}\ }\href@noop {} {\bibfield
  {journal} {\bibinfo  {journal} {AIP Conference Proceedings}\ } (\bibinfo
  {year} {2020})}\BibitemShut {NoStop}%
\bibitem [{\citenamefont {McWilliams}\ \emph {et~al.}(2012)\citenamefont
  {McWilliams}, \citenamefont {Spaulding}, \citenamefont {Eggert},
  \citenamefont {Celliers}, \citenamefont {Hicks}, \citenamefont {Smith},
  \citenamefont {Collins},\ and\ \citenamefont {Jeanloz}}]{Mcwilliams2012}%
  \BibitemOpen
  \bibfield  {author} {\bibinfo {author} {\bibfnamefont {R.~S.}\ \bibnamefont
  {McWilliams}}, \bibinfo {author} {\bibfnamefont {D.~K.}\ \bibnamefont
  {Spaulding}}, \bibinfo {author} {\bibfnamefont {J.~H.}\ \bibnamefont
  {Eggert}}, \bibinfo {author} {\bibfnamefont {P.~M.}\ \bibnamefont
  {Celliers}}, \bibinfo {author} {\bibfnamefont {D.~G.}\ \bibnamefont {Hicks}},
  \bibinfo {author} {\bibfnamefont {R.~F.}\ \bibnamefont {Smith}}, \bibinfo
  {author} {\bibfnamefont {G.~W.}\ \bibnamefont {Collins}}, \ and\ \bibinfo
  {author} {\bibfnamefont {R.}~\bibnamefont {Jeanloz}},\ }\bibfield  {title}
  {\enquote {\bibinfo {title} {Phase transformations and metallization of
  magnesium oxide at high pressure and temperature},}\ }\href@noop {}
  {\bibfield  {journal} {\bibinfo  {journal} {Science}\ }\textbf {\bibinfo
  {volume} {338}},\ \bibinfo {pages} {1330--1333} (\bibinfo {year}
  {2012})}\BibitemShut {NoStop}%
\bibitem [{\citenamefont {Hicks}\ \emph {et~al.}(2006)\citenamefont {Hicks},
  \citenamefont {Boehly}, \citenamefont {Eggert}, \citenamefont {Miller},
  \citenamefont {Celliers},\ and\ \citenamefont {Collins}}]{Hicks2006}%
  \BibitemOpen
  \bibfield  {author} {\bibinfo {author} {\bibfnamefont {D.~G.}\ \bibnamefont
  {Hicks}}, \bibinfo {author} {\bibfnamefont {T.~R.}\ \bibnamefont {Boehly}},
  \bibinfo {author} {\bibfnamefont {J.~H.}\ \bibnamefont {Eggert}}, \bibinfo
  {author} {\bibfnamefont {J.~E.}\ \bibnamefont {Miller}}, \bibinfo {author}
  {\bibfnamefont {P.~M.}\ \bibnamefont {Celliers}}, \ and\ \bibinfo {author}
  {\bibfnamefont {G.~W.}\ \bibnamefont {Collins}},\ }\bibfield  {title}
  {\enquote {\bibinfo {title} {{Dissociation of Liquid Silica at High Pressures
  and Temperatures}},}\ }\href
  {https://link.aps.org/doi/10.1103/PhysRevLett.97.025502} {\bibfield
  {journal} {\bibinfo  {journal} {Phys. Rev. Lett.}\ }\textbf {\bibinfo
  {volume} {97}},\ \bibinfo {pages} {025502} (\bibinfo {year}
  {2006})}\BibitemShut {NoStop}%
\bibitem [{\citenamefont {Soubiran}\ and\ \citenamefont
  {Militzer}(2018)}]{Soubiran2018}%
  \BibitemOpen
  \bibfield  {author} {\bibinfo {author} {\bibfnamefont {F.}~\bibnamefont
  {Soubiran}}\ and\ \bibinfo {author} {\bibfnamefont {B.}~\bibnamefont
  {Militzer}},\ }\bibfield  {title} {\enquote {\bibinfo {title} {{Electrical
  conductivity and magnetic dynamos in magma oceans of Super-Earths}},}\
  }\href@noop {} {\bibfield  {journal} {\bibinfo  {journal} {Nature
  Communications}\ }\textbf {\bibinfo {volume} {9}},\ \bibinfo {pages} {3883}
  (\bibinfo {year} {2018})}\BibitemShut {NoStop}%
\bibitem [{\citenamefont {Stixrude}, \citenamefont {Scipioni},\ and\
  \citenamefont {Desjarlais}(2020)}]{Stixrude2020}%
  \BibitemOpen
  \bibfield  {author} {\bibinfo {author} {\bibfnamefont {L.}~\bibnamefont
  {Stixrude}}, \bibinfo {author} {\bibfnamefont {R.}~\bibnamefont {Scipioni}},
  \ and\ \bibinfo {author} {\bibfnamefont {M.~P.}\ \bibnamefont {Desjarlais}},\
  }\bibfield  {title} {\enquote {\bibinfo {title} {{A silicate dynamo in the
  early Earth}},}\ }\href {\doibase 10.1038/s41467-020-14773-4} {\bibfield
  {journal} {\bibinfo  {journal} {Nature Communications}\ }\textbf {\bibinfo
  {volume} {11}},\ \bibinfo {pages} {935} (\bibinfo {year} {2020})}\BibitemShut
  {NoStop}%
\bibitem [{\citenamefont {Stinton}\ \emph {et~al.}(2014)\citenamefont
  {Stinton}, \citenamefont {MacLeod}, \citenamefont {Cynn}, \citenamefont
  {Errandonea}, \citenamefont {Evans}, \citenamefont {Proctor}, \citenamefont
  {Meng},\ and\ \citenamefont {McMahon}}]{Stinton2014}%
  \BibitemOpen
  \bibfield  {author} {\bibinfo {author} {\bibfnamefont {G.~W.}\ \bibnamefont
  {Stinton}}, \bibinfo {author} {\bibfnamefont {S.~G.}\ \bibnamefont
  {MacLeod}}, \bibinfo {author} {\bibfnamefont {H.}~\bibnamefont {Cynn}},
  \bibinfo {author} {\bibfnamefont {D.}~\bibnamefont {Errandonea}}, \bibinfo
  {author} {\bibfnamefont {W.~J.}\ \bibnamefont {Evans}}, \bibinfo {author}
  {\bibfnamefont {J.~E.}\ \bibnamefont {Proctor}}, \bibinfo {author}
  {\bibfnamefont {Y.}~\bibnamefont {Meng}}, \ and\ \bibinfo {author}
  {\bibfnamefont {M.~I.}\ \bibnamefont {McMahon}},\ }\bibfield  {title}
  {\enquote {\bibinfo {title} {Equation of state and
  high-pressure/high-temperature phase diagram of magnesium},}\ }\href
  {\doibase 10.1103/PhysRevB.90.134105} {\bibfield  {journal} {\bibinfo
  {journal} {Phys. Rev. B}\ }\textbf {\bibinfo {volume} {90}},\ \bibinfo
  {pages} {134105} (\bibinfo {year} {2014})}\BibitemShut {NoStop}%
\bibitem [{\citenamefont {Urtiew}\ and\ \citenamefont
  {Grover}(1977)}]{Urtiew1977}%
  \BibitemOpen
  \bibfield  {author} {\bibinfo {author} {\bibfnamefont {P.~A.}\ \bibnamefont
  {Urtiew}}\ and\ \bibinfo {author} {\bibfnamefont {R.}~\bibnamefont
  {Grover}},\ }\bibfield  {title} {\enquote {\bibinfo {title} {The melting
  temperature of magnesium under shock loading},}\ }\href@noop {} {\bibfield
  {journal} {\bibinfo  {journal} {Journal of Applied Physics}\ }\textbf
  {\bibinfo {volume} {48}},\ \bibinfo {pages} {1122--1126} (\bibinfo {year}
  {1977})}\BibitemShut {NoStop}%
\bibitem [{\citenamefont {Errandonea}(2010)}]{Errandonea2010}%
  \BibitemOpen
  \bibfield  {author} {\bibinfo {author} {\bibfnamefont {D.}~\bibnamefont
  {Errandonea}},\ }\bibfield  {title} {\enquote {\bibinfo {title} {{The melting
  curve of ten metals up to 12 GPa and 1600 K}},}\ }\href@noop {} {\bibfield
  {journal} {\bibinfo  {journal} {Journal of Applied Physics}\ }\textbf
  {\bibinfo {volume} {108}},\ \bibinfo {pages} {033517} (\bibinfo {year}
  {2010})}\BibitemShut {NoStop}%
\bibitem [{\citenamefont {Errandonea}, \citenamefont {Boehler},\ and\
  \citenamefont {Ross}(2001)}]{Errandonea2001}%
  \BibitemOpen
  \bibfield  {author} {\bibinfo {author} {\bibfnamefont {D.}~\bibnamefont
  {Errandonea}}, \bibinfo {author} {\bibfnamefont {R.}~\bibnamefont {Boehler}},
  \ and\ \bibinfo {author} {\bibfnamefont {M.}~\bibnamefont {Ross}},\
  }\bibfield  {title} {\enquote {\bibinfo {title} {{Melting of the
  alkaline-earth metals to 80 GPa}},}\ }\href {\doibase
  10.1103/PhysRevB.65.012108} {\bibfield  {journal} {\bibinfo  {journal} {Phys.
  Rev. B}\ }\textbf {\bibinfo {volume} {65}},\ \bibinfo {pages} {012108}
  (\bibinfo {year} {2001})}\BibitemShut {NoStop}%
\bibitem [{\citenamefont {Qiang}, \citenamefont {Fu-Qian},\ and\ \citenamefont
  {Xin-Zhu}(2002)}]{Qiang2002}%
  \BibitemOpen
  \bibfield  {author} {\bibinfo {author} {\bibfnamefont {W.}~\bibnamefont
  {Qiang}}, \bibinfo {author} {\bibfnamefont {J.}~\bibnamefont {Fu-Qian}}, \
  and\ \bibinfo {author} {\bibfnamefont {L.}~\bibnamefont {Xin-Zhu}},\
  }\bibfield  {title} {\enquote {\bibinfo {title} {Behaviour of gr{\"u}neisen
  parameter at high pressure and temperature inferred from shock compression
  data},}\ }\href@noop {} {\bibfield  {journal} {\bibinfo  {journal} {Chinese
  Physics Letters}\ }\textbf {\bibinfo {volume} {19}},\ \bibinfo {pages} {528}
  (\bibinfo {year} {2002})}\BibitemShut {NoStop}%
\bibitem [{\citenamefont {Beason}, \citenamefont {Mandal},\ and\ \citenamefont
  {Jensen}(2020)}]{Beason2020}%
  \BibitemOpen
  \bibfield  {author} {\bibinfo {author} {\bibfnamefont {M.~T.}\ \bibnamefont
  {Beason}}, \bibinfo {author} {\bibfnamefont {A.}~\bibnamefont {Mandal}}, \
  and\ \bibinfo {author} {\bibfnamefont {B.~J.}\ \bibnamefont {Jensen}},\
  }\bibfield  {title} {\enquote {\bibinfo {title} {{Direct observation of the
  hcp-bcc phase transition and melting along the principal Hugoniot of Mg}},}\
  }\href {https://doi.org/10.1103/PhysRevB.101.024110
  https://link.aps.org/doi/10.1103/PhysRevB.101.024110} {\bibfield  {journal}
  {\bibinfo  {journal} {Physical Review B}\ }\textbf {\bibinfo {volume}
  {101}},\ \bibinfo {pages} {024110} (\bibinfo {year} {2020})}\BibitemShut
  {NoStop}%
\bibitem [{\citenamefont {Sin'ko}\ and\ \citenamefont
  {Smirnov}(2009)}]{Sinko2009}%
  \BibitemOpen
  \bibfield  {author} {\bibinfo {author} {\bibfnamefont {G.}~\bibnamefont
  {Sin'ko}}\ and\ \bibinfo {author} {\bibfnamefont {N.}~\bibnamefont
  {Smirnov}},\ }\bibfield  {title} {\enquote {\bibinfo {title} {Ab initio
  calculations for the elastic properties of magnesium under pressure},}\
  }\href@noop {} {\bibfield  {journal} {\bibinfo  {journal} {Physical Review
  B}\ }\textbf {\bibinfo {volume} {80}},\ \bibinfo {pages} {104113} (\bibinfo
  {year} {2009})}\BibitemShut {NoStop}%
\bibitem [{\citenamefont {Greeff}\ and\ \citenamefont
  {Moriarty}(1999)}]{Greeff1999}%
  \BibitemOpen
  \bibfield  {author} {\bibinfo {author} {\bibfnamefont {C.}~\bibnamefont
  {Greeff}}\ and\ \bibinfo {author} {\bibfnamefont {J.~A.}\ \bibnamefont
  {Moriarty}},\ }\bibfield  {title} {\enquote {\bibinfo {title} {Ab initio
  thermoelasticity of magnesium},}\ }\href@noop {} {\bibfield  {journal}
  {\bibinfo  {journal} {Physical Review B}\ }\textbf {\bibinfo {volume} {59}},\
  \bibinfo {pages} {3427} (\bibinfo {year} {1999})}\BibitemShut {NoStop}%
\bibitem [{\citenamefont {Khishchenko}(2004)}]{Khishchenko2004}%
  \BibitemOpen
  \bibfield  {author} {\bibinfo {author} {\bibfnamefont {K.}~\bibnamefont
  {Khishchenko}},\ }\bibfield  {title} {\enquote {\bibinfo {title} {The
  equation of state for magnesium at high pressures},}\ }\href@noop {}
  {\bibfield  {journal} {\bibinfo  {journal} {Technical physics letters}\
  }\textbf {\bibinfo {volume} {30}},\ \bibinfo {pages} {829--831} (\bibinfo
  {year} {2004})}\BibitemShut {NoStop}%
\bibitem [{\citenamefont {Lomonosov}\ \emph {et~al.}(2002)\citenamefont
  {Lomonosov}, \citenamefont {Fortov}, \citenamefont {Khishchenko},\ and\
  \citenamefont {Levashov}}]{Lomonosov2002}%
  \BibitemOpen
  \bibfield  {author} {\bibinfo {author} {\bibfnamefont {I.}~\bibnamefont
  {Lomonosov}}, \bibinfo {author} {\bibfnamefont {V.}~\bibnamefont {Fortov}},
  \bibinfo {author} {\bibfnamefont {K.}~\bibnamefont {Khishchenko}}, \ and\
  \bibinfo {author} {\bibfnamefont {P.}~\bibnamefont {Levashov}},\ }\bibfield
  {title} {\enquote {\bibinfo {title} {Phase diagrams and thermodynamic
  properties of metals at high pressures, high temperatures},}\ \ }(\bibinfo
  {organization} {American Institute of Physics},\ \bibinfo {year} {2002})\
  pp.\ \bibinfo {pages} {111--114}\BibitemShut {NoStop}%
\bibitem [{\citenamefont {Hong}\ and\ \citenamefont {van~de
  Walle}(2019)}]{Hong2019}%
  \BibitemOpen
  \bibfield  {author} {\bibinfo {author} {\bibfnamefont {Q.-J.}\ \bibnamefont
  {Hong}}\ and\ \bibinfo {author} {\bibfnamefont {A.}~\bibnamefont {van~de
  Walle}},\ }\bibfield  {title} {\enquote {\bibinfo {title} {Reentrant melting
  of sodium, magnesium, and aluminum: General trend},}\ }\href {\doibase
  10.1103/PhysRevB.100.140102} {\bibfield  {journal} {\bibinfo  {journal}
  {Phys. Rev. B}\ }\textbf {\bibinfo {volume} {100}},\ \bibinfo {pages}
  {140102} (\bibinfo {year} {2019})}\BibitemShut {NoStop}%
\bibitem [{\citenamefont {Mehta}, \citenamefont {Price},\ and\ \citenamefont
  {Alf{\`e}}(2006)}]{Mehta2006}%
  \BibitemOpen
  \bibfield  {author} {\bibinfo {author} {\bibfnamefont {S.}~\bibnamefont
  {Mehta}}, \bibinfo {author} {\bibfnamefont {G.}~\bibnamefont {Price}}, \ and\
  \bibinfo {author} {\bibfnamefont {D.}~\bibnamefont {Alf{\`e}}},\ }\bibfield
  {title} {\enquote {\bibinfo {title} {{Ab initio thermodynamics and phase
  diagram of solid magnesium: A comparison of the LDA and GGA}},}\ }\href@noop
  {} {\bibfield  {journal} {\bibinfo  {journal} {The Journal of chemical
  physics}\ }\textbf {\bibinfo {volume} {125}},\ \bibinfo {pages} {194507}
  (\bibinfo {year} {2006})}\BibitemShut {NoStop}%
\bibitem [{\citenamefont {Militzer}(2006)}]{Mi06}%
  \BibitemOpen
  \bibfield  {author} {\bibinfo {author} {\bibfnamefont {B.}~\bibnamefont
  {Militzer}},\ }\bibfield  {title} {\enquote {\bibinfo {title} {First
  principles calculations of shock compressed fluid helium},}\ }\href@noop {}
  {\bibfield  {journal} {\bibinfo  {journal} {Phys. Rev. Lett.}\ }\textbf
  {\bibinfo {volume} {97}},\ \bibinfo {pages} {175501} (\bibinfo {year}
  {2006})}\BibitemShut {NoStop}%
\bibitem [{\citenamefont {Benedict}\ \emph {et~al.}(2014)\citenamefont
  {Benedict}, \citenamefont {Driver}, \citenamefont {Hamel}, \citenamefont
  {Militzer}, \citenamefont {Qi}, \citenamefont {Correa}, \citenamefont
  {Saul},\ and\ \citenamefont {Schwegler}}]{Benedict2014}%
  \BibitemOpen
  \bibfield  {author} {\bibinfo {author} {\bibfnamefont {L.~X.}\ \bibnamefont
  {Benedict}}, \bibinfo {author} {\bibfnamefont {K.~P.}\ \bibnamefont
  {Driver}}, \bibinfo {author} {\bibfnamefont {S.}~\bibnamefont {Hamel}},
  \bibinfo {author} {\bibfnamefont {B.}~\bibnamefont {Militzer}}, \bibinfo
  {author} {\bibfnamefont {T.}~\bibnamefont {Qi}}, \bibinfo {author}
  {\bibfnamefont {A.~A.}\ \bibnamefont {Correa}}, \bibinfo {author}
  {\bibfnamefont {A.}~\bibnamefont {Saul}}, \ and\ \bibinfo {author}
  {\bibfnamefont {E.}~\bibnamefont {Schwegler}},\ }\bibfield  {title} {\enquote
  {\bibinfo {title} {A multiphase equation of state for carbon addressing high
  pressures and temperatures},}\ }\href@noop {} {\bibfield  {journal} {\bibinfo
   {journal} {Phys. Rev. B}\ }\textbf {\bibinfo {volume} {89}},\ \bibinfo
  {pages} {224109} (\bibinfo {year} {2014})}\BibitemShut {NoStop}%
\bibitem [{\citenamefont {Driver}\ and\ \citenamefont
  {Militzer}(2015)}]{Driver2015}%
  \BibitemOpen
  \bibfield  {author} {\bibinfo {author} {\bibfnamefont {K.~P.}\ \bibnamefont
  {Driver}}\ and\ \bibinfo {author} {\bibfnamefont {B.}~\bibnamefont
  {Militzer}},\ }\bibfield  {title} {\enquote {\bibinfo {title}
  {{First-principles simulations and shock Hugoniot calculations of warm dense
  neon}},}\ }\href@noop {} {\bibfield  {journal} {\bibinfo  {journal} {Phys.
  Rev. B}\ }\textbf {\bibinfo {volume} {91}},\ \bibinfo {pages} {045103}
  (\bibinfo {year} {2015})}\BibitemShut {NoStop}%
\bibitem [{\citenamefont {Hu}\ \emph {et~al.}(2016)\citenamefont {Hu},
  \citenamefont {Militzer}, \citenamefont {Collins}, \citenamefont {Driver},\
  and\ \citenamefont {Kress}}]{Hu2016}%
  \BibitemOpen
  \bibfield  {author} {\bibinfo {author} {\bibfnamefont {S.~X.}\ \bibnamefont
  {Hu}}, \bibinfo {author} {\bibfnamefont {B.}~\bibnamefont {Militzer}},
  \bibinfo {author} {\bibfnamefont {L.~A.}\ \bibnamefont {Collins}}, \bibinfo
  {author} {\bibfnamefont {K.~P.}\ \bibnamefont {Driver}}, \ and\ \bibinfo
  {author} {\bibfnamefont {J.~D.}\ \bibnamefont {Kress}},\ }\bibfield  {title}
  {\enquote {\bibinfo {title} {First-principles prediction of the softening of
  the silicon shock hugoniot curve},}\ }\href@noop {} {\bibfield  {journal}
  {\bibinfo  {journal} {Phys. Rev. B}\ }\textbf {\bibinfo {volume} {94}},\
  \bibinfo {pages} {094109} (\bibinfo {year} {2016})}\BibitemShut {NoStop}%
\bibitem [{\citenamefont {Zhang}\ \emph {et~al.}(2019)\citenamefont {Zhang},
  \citenamefont {Lazicki}, \citenamefont {Militzer}, \citenamefont {Yang},
  \citenamefont {Caspersen}, \citenamefont {Gaffney}, \citenamefont {D\"ane},
  \citenamefont {Pask}, \citenamefont {Johnson}, \citenamefont {Sharma},
  \citenamefont {Suryanarayana}, \citenamefont {Johnson}, \citenamefont
  {Smirnov}, \citenamefont {Sterne}, \citenamefont {Erskine}, \citenamefont
  {London}, \citenamefont {Coppari}, \citenamefont {Swift}, \citenamefont
  {Nilsen}, \citenamefont {Nelson},\ and\ \citenamefont
  {Whitley}}]{ZhangBN2019}%
  \BibitemOpen
  \bibfield  {author} {\bibinfo {author} {\bibfnamefont {S.}~\bibnamefont
  {Zhang}}, \bibinfo {author} {\bibfnamefont {A.}~\bibnamefont {Lazicki}},
  \bibinfo {author} {\bibfnamefont {B.}~\bibnamefont {Militzer}}, \bibinfo
  {author} {\bibfnamefont {L.~H.}\ \bibnamefont {Yang}}, \bibinfo {author}
  {\bibfnamefont {K.}~\bibnamefont {Caspersen}}, \bibinfo {author}
  {\bibfnamefont {J.~A.}\ \bibnamefont {Gaffney}}, \bibinfo {author}
  {\bibfnamefont {M.~W.}\ \bibnamefont {D\"ane}}, \bibinfo {author}
  {\bibfnamefont {J.~E.}\ \bibnamefont {Pask}}, \bibinfo {author}
  {\bibfnamefont {W.~R.}\ \bibnamefont {Johnson}}, \bibinfo {author}
  {\bibfnamefont {A.}~\bibnamefont {Sharma}}, \bibinfo {author} {\bibfnamefont
  {P.}~\bibnamefont {Suryanarayana}}, \bibinfo {author} {\bibfnamefont {D.~D.}\
  \bibnamefont {Johnson}}, \bibinfo {author} {\bibfnamefont {A.~V.}\
  \bibnamefont {Smirnov}}, \bibinfo {author} {\bibfnamefont {P.~A.}\
  \bibnamefont {Sterne}}, \bibinfo {author} {\bibfnamefont {D.}~\bibnamefont
  {Erskine}}, \bibinfo {author} {\bibfnamefont {R.~A.}\ \bibnamefont {London}},
  \bibinfo {author} {\bibfnamefont {F.}~\bibnamefont {Coppari}}, \bibinfo
  {author} {\bibfnamefont {D.}~\bibnamefont {Swift}}, \bibinfo {author}
  {\bibfnamefont {J.}~\bibnamefont {Nilsen}}, \bibinfo {author} {\bibfnamefont
  {A.~J.}\ \bibnamefont {Nelson}}, \ and\ \bibinfo {author} {\bibfnamefont
  {H.~D.}\ \bibnamefont {Whitley}},\ }\bibfield  {title} {\enquote {\bibinfo
  {title} {Equation of state of boron nitride combining computation, modeling,
  and experiment},}\ }\href {\doibase 10.1103/PhysRevB.99.165103} {\bibfield
  {journal} {\bibinfo  {journal} {Phys. Rev. B}\ }\textbf {\bibinfo {volume}
  {99}},\ \bibinfo {pages} {165103} (\bibinfo {year} {2019})}\BibitemShut
  {NoStop}%
\bibitem [{\citenamefont {Ceperley}(1995)}]{Ce95}%
  \BibitemOpen
  \bibfield  {author} {\bibinfo {author} {\bibfnamefont {D.~M.}\ \bibnamefont
  {Ceperley}},\ }\bibfield  {title} {\enquote {\bibinfo {title} {Path integrals
  in the theory of condensed helium},}\ }\href {\doibase
  10.1103/RevModPhys.67.279} {\bibfield  {journal} {\bibinfo  {journal} {Rev.
  Mod. Phys.}\ }\textbf {\bibinfo {volume} {67}},\ \bibinfo {pages} {279--355}
  (\bibinfo {year} {1995})}\BibitemShut {NoStop}%
\bibitem [{\citenamefont {Ceperley}(1996)}]{Ce96}%
  \BibitemOpen
  \bibfield  {author} {\bibinfo {author} {\bibfnamefont {D.}~\bibnamefont
  {Ceperley}},\ }\bibfield  {title} {\enquote {\bibinfo {title} {Monte carlo
  and molecular dynamics of condensed matter systems},}\ \ }(\bibinfo
  {publisher} {Editrice Compositori, Bologna, Italy},\ \bibinfo {year} {1996})\
  p.\ \bibinfo {pages} {443}\BibitemShut {NoStop}%
\bibitem [{\citenamefont {Militzer}\ and\ \citenamefont
  {Driver}(2015)}]{MilitzerDriver2015}%
  \BibitemOpen
  \bibfield  {author} {\bibinfo {author} {\bibfnamefont {B.}~\bibnamefont
  {Militzer}}\ and\ \bibinfo {author} {\bibfnamefont {K.~P.}\ \bibnamefont
  {Driver}},\ }\bibfield  {title} {\enquote {\bibinfo {title} {{Development of
  Path Integral Monte Carlo Simulations with Localized Nodal Surfaces for
  Second-Row Elements}},}\ }\href@noop {} {\bibfield  {journal} {\bibinfo
  {journal} {Phys. Rev. Lett.}\ }\textbf {\bibinfo {volume} {115}},\ \bibinfo
  {pages} {176403} (\bibinfo {year} {2015})}\BibitemShut {NoStop}%
\bibitem [{\citenamefont {Zhang}\ \emph
  {et~al.}(2017{\natexlab{a}})\citenamefont {Zhang}, \citenamefont {Driver},
  \citenamefont {Soubiran},\ and\ \citenamefont {Militzer}}]{ZhangSodium2017}%
  \BibitemOpen
  \bibfield  {author} {\bibinfo {author} {\bibfnamefont {S.}~\bibnamefont
  {Zhang}}, \bibinfo {author} {\bibfnamefont {K.~P.}\ \bibnamefont {Driver}},
  \bibinfo {author} {\bibfnamefont {F.}~\bibnamefont {Soubiran}}, \ and\
  \bibinfo {author} {\bibfnamefont {B.}~\bibnamefont {Militzer}},\ }\bibfield
  {title} {\enquote {\bibinfo {title} {{Equation of state and shock compression
  of warm dense sodium—A first-principles study}},}\ }\href@noop {}
  {\bibfield  {journal} {\bibinfo  {journal} {J. Chem. Phys.}\ }\textbf
  {\bibinfo {volume} {146}},\ \bibinfo {pages} {074505} (\bibinfo {year}
  {2017}{\natexlab{a}})}\BibitemShut {NoStop}%
\bibitem [{\citenamefont {Driver}, \citenamefont {Soubiran},\ and\
  \citenamefont {Militzer}(2018)}]{Driver2018}%
  \BibitemOpen
  \bibfield  {author} {\bibinfo {author} {\bibfnamefont {K.~P.}\ \bibnamefont
  {Driver}}, \bibinfo {author} {\bibfnamefont {F.}~\bibnamefont {Soubiran}}, \
  and\ \bibinfo {author} {\bibfnamefont {B.}~\bibnamefont {Militzer}},\
  }\bibfield  {title} {\enquote {\bibinfo {title} {{Path integral Monte Carlo
  simulations of warm dense aluminum}},}\ }\href@noop {} {\bibfield  {journal}
  {\bibinfo  {journal} {Phys. Rev. E}\ }\textbf {\bibinfo {volume} {97}},\
  \bibinfo {pages} {063207} (\bibinfo {year} {2018})}\BibitemShut {NoStop}%
\bibitem [{\citenamefont {Militzer}(2009{\natexlab{a}})}]{Mi09}%
  \BibitemOpen
  \bibfield  {author} {\bibinfo {author} {\bibfnamefont {B.}~\bibnamefont
  {Militzer}},\ }\bibfield  {title} {\enquote {\bibinfo {title} {Path integral
  monte carlo and density functional molecular dynamics simulations of hot,
  dense helium},}\ }\href@noop {} {\bibfield  {journal} {\bibinfo  {journal}
  {Phys. Rev. B}\ }\textbf {\bibinfo {volume} {79}},\ \bibinfo {pages} {155105}
  (\bibinfo {year} {2009}{\natexlab{a}})}\BibitemShut {NoStop}%
\bibitem [{\citenamefont {Zhang}\ \emph
  {et~al.}(2018{\natexlab{b}})\citenamefont {Zhang}, \citenamefont {Militzer},
  \citenamefont {Benedict}, \citenamefont {Soubiran}, \citenamefont {Sterne},\
  and\ \citenamefont {Driver}}]{ZhangCH2018}%
  \BibitemOpen
  \bibfield  {author} {\bibinfo {author} {\bibfnamefont {S.}~\bibnamefont
  {Zhang}}, \bibinfo {author} {\bibfnamefont {B.}~\bibnamefont {Militzer}},
  \bibinfo {author} {\bibfnamefont {L.~X.}\ \bibnamefont {Benedict}}, \bibinfo
  {author} {\bibfnamefont {F.}~\bibnamefont {Soubiran}}, \bibinfo {author}
  {\bibfnamefont {P.~A.}\ \bibnamefont {Sterne}}, \ and\ \bibinfo {author}
  {\bibfnamefont {K.~P.}\ \bibnamefont {Driver}},\ }\bibfield  {title}
  {\enquote {\bibinfo {title} {{Path integral Monte Carlo simulations of dense
  carbon-hydrogen plasmas}},}\ }\href@noop {} {\bibfield  {journal} {\bibinfo
  {journal} {J. Chem. Phys.}\ }\textbf {\bibinfo {volume} {148}},\ \bibinfo
  {pages} {102318} (\bibinfo {year} {2018}{\natexlab{b}})}\BibitemShut
  {NoStop}%
\bibitem [{\citenamefont {Hohenberg}\ and\ \citenamefont
  {Kohn}(1964)}]{Hohenberg1964}%
  \BibitemOpen
  \bibfield  {author} {\bibinfo {author} {\bibfnamefont {P.}~\bibnamefont
  {Hohenberg}}\ and\ \bibinfo {author} {\bibfnamefont {W.}~\bibnamefont
  {Kohn}},\ }\bibfield  {title} {\enquote {\bibinfo {title} {Inhomogeneous
  electron gas},}\ }\href@noop {} {\bibfield  {journal} {\bibinfo  {journal}
  {Phys. Rev.}\ }\textbf {\bibinfo {volume} {136}},\ \bibinfo {pages}
  {B864--B871} (\bibinfo {year} {1964})}\BibitemShut {NoStop}%
\bibitem [{\citenamefont {Kohn}\ and\ \citenamefont {Sham}(1965)}]{Kohn1965}%
  \BibitemOpen
  \bibfield  {author} {\bibinfo {author} {\bibfnamefont {W.}~\bibnamefont
  {Kohn}}\ and\ \bibinfo {author} {\bibfnamefont {L.~J.}\ \bibnamefont
  {Sham}},\ }\bibfield  {title} {\enquote {\bibinfo {title} {Self-consistent
  equations including exchange and correlation effects},}\ }\href@noop {}
  {\bibfield  {journal} {\bibinfo  {journal} {Phys. Rev.}\ }\textbf {\bibinfo
  {volume} {140}},\ \bibinfo {pages} {A1133--A1138} (\bibinfo {year}
  {1965})}\BibitemShut {NoStop}%
\bibitem [{\citenamefont {Mermin}(1965)}]{Mermin1965}%
  \BibitemOpen
  \bibfield  {author} {\bibinfo {author} {\bibfnamefont {N.~D.}\ \bibnamefont
  {Mermin}},\ }\bibfield  {title} {\enquote {\bibinfo {title} {Thermal
  properties of the inhomogeneous electron gas},}\ }\href@noop {} {\bibfield
  {journal} {\bibinfo  {journal} {Phys. Rev.}\ }\textbf {\bibinfo {volume}
  {137}},\ \bibinfo {pages} {A1441--A1443} (\bibinfo {year}
  {1965})}\BibitemShut {NoStop}%
\bibitem [{\citenamefont {Root}\ \emph {et~al.}(2010)\citenamefont {Root},
  \citenamefont {Magyar}, \citenamefont {Carpenter}, \citenamefont {Hanson},\
  and\ \citenamefont {Mattsson}}]{Root2010}%
  \BibitemOpen
  \bibfield  {author} {\bibinfo {author} {\bibfnamefont {S.}~\bibnamefont
  {Root}}, \bibinfo {author} {\bibfnamefont {R.~J.}\ \bibnamefont {Magyar}},
  \bibinfo {author} {\bibfnamefont {J.~H.}\ \bibnamefont {Carpenter}}, \bibinfo
  {author} {\bibfnamefont {D.~L.}\ \bibnamefont {Hanson}}, \ and\ \bibinfo
  {author} {\bibfnamefont {T.~R.}\ \bibnamefont {Mattsson}},\ }\bibfield
  {title} {\enquote {\bibinfo {title} {{Shock Compression of a Fifth Period
  Element: Liquid Xenon to 840 GPa}},}\ }\href@noop {} {\bibfield  {journal}
  {\bibinfo  {journal} {Phys. Rev. Lett.}\ }\textbf {\bibinfo {volume} {105}},\
  \bibinfo {pages} {085501} (\bibinfo {year} {2010})}\BibitemShut {NoStop}%
\bibitem [{\citenamefont {Wang}\ \emph {et~al.}(2010)\citenamefont {Wang},
  \citenamefont {Tian}, \citenamefont {Wang}, \citenamefont {Cui},
  \citenamefont {Liu},\ and\ \citenamefont {Zou}}]{Wang2010}%
  \BibitemOpen
  \bibfield  {author} {\bibinfo {author} {\bibfnamefont {X.}~\bibnamefont
  {Wang}}, \bibinfo {author} {\bibfnamefont {F.}~\bibnamefont {Tian}}, \bibinfo
  {author} {\bibfnamefont {L.}~\bibnamefont {Wang}}, \bibinfo {author}
  {\bibfnamefont {T.}~\bibnamefont {Cui}}, \bibinfo {author} {\bibfnamefont
  {B.}~\bibnamefont {Liu}}, \ and\ \bibinfo {author} {\bibfnamefont
  {G.}~\bibnamefont {Zou}},\ }\bibfield  {title} {\enquote {\bibinfo {title}
  {Structural stability of polymeric nitrogen: A first-principles
  investigation},}\ }\href@noop {} {\bibfield  {journal} {\bibinfo  {journal}
  {J. Chem. Phys.}\ }\textbf {\bibinfo {volume} {132}},\ \bibinfo {pages}
  {024502} (\bibinfo {year} {2010})}\BibitemShut {NoStop}%
\bibitem [{\citenamefont {Mattsson}\ \emph {et~al.}(2014)\citenamefont
  {Mattsson}, \citenamefont {Root}, \citenamefont {Mattsson}, \citenamefont
  {Shulenburger}, \citenamefont {Magyar},\ and\ \citenamefont
  {Flicker}}]{Mattsson2014}%
  \BibitemOpen
  \bibfield  {author} {\bibinfo {author} {\bibfnamefont {T.~R.}\ \bibnamefont
  {Mattsson}}, \bibinfo {author} {\bibfnamefont {S.}~\bibnamefont {Root}},
  \bibinfo {author} {\bibfnamefont {A.~E.}\ \bibnamefont {Mattsson}}, \bibinfo
  {author} {\bibfnamefont {L.}~\bibnamefont {Shulenburger}}, \bibinfo {author}
  {\bibfnamefont {R.~J.}\ \bibnamefont {Magyar}}, \ and\ \bibinfo {author}
  {\bibfnamefont {D.~G.}\ \bibnamefont {Flicker}},\ }\bibfield  {title}
  {\enquote {\bibinfo {title} {{Validating density-functional theory
  simulations at high energy-density conditions with liquid krypton shock
  experiments to 850 GPa on Sandia's Z machine}},}\ }\href@noop {} {\bibfield
  {journal} {\bibinfo  {journal} {Phys. Rev. B}\ }\textbf {\bibinfo {volume}
  {90}},\ \bibinfo {pages} {184105} (\bibinfo {year} {2014})}\BibitemShut
  {NoStop}%
\bibitem [{\citenamefont {Karasiev}, \citenamefont {Calder\'{\i}n},\ and\
  \citenamefont {Trickey}(2016)}]{Karasiev2016}%
  \BibitemOpen
  \bibfield  {author} {\bibinfo {author} {\bibfnamefont {V.~V.}\ \bibnamefont
  {Karasiev}}, \bibinfo {author} {\bibfnamefont {L.}~\bibnamefont
  {Calder\'{\i}n}}, \ and\ \bibinfo {author} {\bibfnamefont {S.~B.}\
  \bibnamefont {Trickey}},\ }\bibfield  {title} {\enquote {\bibinfo {title}
  {Importance of finite-temperature exchange correlation for warm dense matter
  calculations},}\ }\href@noop {} {\bibfield  {journal} {\bibinfo  {journal}
  {Phys. Rev. E}\ }\textbf {\bibinfo {volume} {93}},\ \bibinfo {pages} {063207}
  (\bibinfo {year} {2016})}\BibitemShut {NoStop}%
\bibitem [{\citenamefont {Pollock}\ and\ \citenamefont
  {Ceperley}(1984)}]{PC84}%
  \BibitemOpen
  \bibfield  {author} {\bibinfo {author} {\bibfnamefont {E.~L.}\ \bibnamefont
  {Pollock}}\ and\ \bibinfo {author} {\bibfnamefont {D.~M.}\ \bibnamefont
  {Ceperley}},\ }\bibfield  {title} {\enquote {\bibinfo {title} {Simulation of
  quantum many-body systems by path-integral methods},}\ }\href {\doibase
  10.1103/PhysRevB.30.2555} {\bibfield  {journal} {\bibinfo  {journal} {Phys.
  Rev. B}\ }\textbf {\bibinfo {volume} {30}},\ \bibinfo {pages} {2555--2568}
  (\bibinfo {year} {1984})}\BibitemShut {NoStop}%
\bibitem [{\citenamefont {Ceperley}(1991)}]{Ce91}%
  \BibitemOpen
  \bibfield  {author} {\bibinfo {author} {\bibfnamefont {D.~M.}\ \bibnamefont
  {Ceperley}},\ }\bibfield  {title} {\enquote {\bibinfo {title} {{Fermion
  nodes}},}\ }\href@noop {} {\bibfield  {journal} {\bibinfo  {journal} {Journal
  of Statistical Physics}\ }\textbf {\bibinfo {volume} {63}},\ \bibinfo {pages}
  {1237--1267} (\bibinfo {year} {1991})}\BibitemShut {NoStop}%
\bibitem [{\citenamefont {Ceperley}(1992)}]{Ce92}%
  \BibitemOpen
  \bibfield  {author} {\bibinfo {author} {\bibfnamefont {D.~M.}\ \bibnamefont
  {Ceperley}},\ }\bibfield  {title} {\enquote {\bibinfo {title} {Path-integral
  calculations of normal liquid $^{3}\mathrm{He}$},}\ }\href {\doibase
  10.1103/PhysRevLett.69.331} {\bibfield  {journal} {\bibinfo  {journal} {Phys.
  Rev. Lett.}\ }\textbf {\bibinfo {volume} {69}},\ \bibinfo {pages} {331--334}
  (\bibinfo {year} {1992})}\BibitemShut {NoStop}%
\bibitem [{\citenamefont {Pierleoni}\ \emph {et~al.}(1994)\citenamefont
  {Pierleoni}, \citenamefont {Ceperley}, \citenamefont {Bernu},\ and\
  \citenamefont {Magro}}]{PC94}%
  \BibitemOpen
  \bibfield  {author} {\bibinfo {author} {\bibfnamefont {C.}~\bibnamefont
  {Pierleoni}}, \bibinfo {author} {\bibfnamefont {D.~M.}\ \bibnamefont
  {Ceperley}}, \bibinfo {author} {\bibfnamefont {B.}~\bibnamefont {Bernu}}, \
  and\ \bibinfo {author} {\bibfnamefont {W.~R.}\ \bibnamefont {Magro}},\
  }\bibfield  {title} {\enquote {\bibinfo {title} {Equation of state of the
  hydrogen plasma by path integral monte carlo simulation},}\ }\href@noop {}
  {\bibfield  {journal} {\bibinfo  {journal} {Phys. Rev. Lett.}\ }\textbf
  {\bibinfo {volume} {73}},\ \bibinfo {pages} {2145--2149} (\bibinfo {year}
  {1994})}\BibitemShut {NoStop}%
\bibitem [{\citenamefont {Magro}\ \emph {et~al.}(1996)\citenamefont {Magro},
  \citenamefont {Ceperley}, \citenamefont {Pierleoni},\ and\ \citenamefont
  {Bernu}}]{Ma96}%
  \BibitemOpen
  \bibfield  {author} {\bibinfo {author} {\bibfnamefont {W.~R.}\ \bibnamefont
  {Magro}}, \bibinfo {author} {\bibfnamefont {D.~M.}\ \bibnamefont {Ceperley}},
  \bibinfo {author} {\bibfnamefont {C.}~\bibnamefont {Pierleoni}}, \ and\
  \bibinfo {author} {\bibfnamefont {B.}~\bibnamefont {Bernu}},\ }\bibfield
  {title} {\enquote {\bibinfo {title} {Molecular dissociation in hot, dense
  hydrogen},}\ }\href {\doibase 10.1103/PhysRevLett.76.1240} {\bibfield
  {journal} {\bibinfo  {journal} {Phys. Rev. Lett.}\ }\textbf {\bibinfo
  {volume} {76}},\ \bibinfo {pages} {1240--1243} (\bibinfo {year}
  {1996})}\BibitemShut {NoStop}%
\bibitem [{\citenamefont {Militzer}, \citenamefont {Magro},\ and\ \citenamefont
  {Ceperley}(1999)}]{Mi99}%
  \BibitemOpen
  \bibfield  {author} {\bibinfo {author} {\bibfnamefont {B.}~\bibnamefont
  {Militzer}}, \bibinfo {author} {\bibfnamefont {W.}~\bibnamefont {Magro}}, \
  and\ \bibinfo {author} {\bibfnamefont {D.}~\bibnamefont {Ceperley}},\
  }\bibfield  {title} {\enquote {\bibinfo {title} {Characterization of the
  state of hydrogen at high temperature and density},}\ }\href@noop {}
  {\bibfield  {journal} {\bibinfo  {journal} {Contributions to Plasma Physics}\
  }\textbf {\bibinfo {volume} {39}},\ \bibinfo {pages} {151--154} (\bibinfo
  {year} {1999})}\BibitemShut {NoStop}%
\bibitem [{\citenamefont {Militzer}(2000)}]{MilitzerThesis}%
  \BibitemOpen
  \bibfield  {author} {\bibinfo {author} {\bibfnamefont {B.}~\bibnamefont
  {Militzer}},\ }\emph {\bibinfo {title} {Path Integral Monte Carlo Simulations
  of Hot Dense Hydrogen}},\ \href@noop {} {Ph.D. thesis},\ \bibinfo  {school}
  {University of Illinois at Urbana-Champaign} (\bibinfo {year}
  {2000})\BibitemShut {NoStop}%
\bibitem [{\citenamefont {Militzer}\ and\ \citenamefont
  {Ceperley}(2000)}]{MC00}%
  \BibitemOpen
  \bibfield  {author} {\bibinfo {author} {\bibfnamefont {B.}~\bibnamefont
  {Militzer}}\ and\ \bibinfo {author} {\bibfnamefont {D.~M.}\ \bibnamefont
  {Ceperley}},\ }\bibfield  {title} {\enquote {\bibinfo {title} {{Path Integral
  Monte Carlo Calculation of the Deuterium Hugoniot}},}\ }\href@noop {}
  {\bibfield  {journal} {\bibinfo  {journal} {Phys. Rev. Lett.}\ }\textbf
  {\bibinfo {volume} {85}},\ \bibinfo {pages} {1890--1893} (\bibinfo {year}
  {2000})}\BibitemShut {NoStop}%
\bibitem [{\citenamefont {Militzer}\ and\ \citenamefont
  {Ceperley}(2001)}]{MC01}%
  \BibitemOpen
  \bibfield  {author} {\bibinfo {author} {\bibfnamefont {B.}~\bibnamefont
  {Militzer}}\ and\ \bibinfo {author} {\bibfnamefont {D.~M.}\ \bibnamefont
  {Ceperley}},\ }\bibfield  {title} {\enquote {\bibinfo {title} {Path integral
  monte carlo simulation of the low-density hydrogen plasma},}\ }\href@noop {}
  {\bibfield  {journal} {\bibinfo  {journal} {Phys. Rev. E}\ }\textbf {\bibinfo
  {volume} {63}},\ \bibinfo {pages} {066404} (\bibinfo {year}
  {2001})}\BibitemShut {NoStop}%
\bibitem [{\citenamefont {Militzer}\ \emph {et~al.}(2001)\citenamefont
  {Militzer}, \citenamefont {Ceperley}, \citenamefont {Kress}, \citenamefont
  {Johnson}, \citenamefont {Collins},\ and\ \citenamefont {Mazevet}}]{Mi01}%
  \BibitemOpen
  \bibfield  {author} {\bibinfo {author} {\bibfnamefont {B.}~\bibnamefont
  {Militzer}}, \bibinfo {author} {\bibfnamefont {D.~M.}\ \bibnamefont
  {Ceperley}}, \bibinfo {author} {\bibfnamefont {J.~D.}\ \bibnamefont {Kress}},
  \bibinfo {author} {\bibfnamefont {J.~D.}\ \bibnamefont {Johnson}}, \bibinfo
  {author} {\bibfnamefont {L.~A.}\ \bibnamefont {Collins}}, \ and\ \bibinfo
  {author} {\bibfnamefont {S.}~\bibnamefont {Mazevet}},\ }\bibfield  {title}
  {\enquote {\bibinfo {title} {Calculation of a deuterium double shock hugoniot
  from ab initio simulations},}\ }\href@noop {} {\bibfield  {journal} {\bibinfo
   {journal} {Phys. Rev. Lett.}\ }\textbf {\bibinfo {volume} {87}},\ \bibinfo
  {pages} {275502} (\bibinfo {year} {2001})}\BibitemShut {NoStop}%
\bibitem [{\citenamefont {Militzer}(2009{\natexlab{b}})}]{Mi09b}%
  \BibitemOpen
  \bibfield  {author} {\bibinfo {author} {\bibfnamefont {B.}~\bibnamefont
  {Militzer}},\ }\bibfield  {title} {\enquote {\bibinfo {title} {Correlations
  in hot dense helium},}\ }\href@noop {} {\bibfield  {journal} {\bibinfo
  {journal} {Journal of Physics A: Mathematical and Theoretical}\ }\textbf
  {\bibinfo {volume} {42}},\ \bibinfo {pages} {214001} (\bibinfo {year}
  {2009}{\natexlab{b}})}\BibitemShut {NoStop}%
\bibitem [{\citenamefont {Militzer}(2005)}]{Mi05}%
  \BibitemOpen
  \bibfield  {author} {\bibinfo {author} {\bibfnamefont {B.}~\bibnamefont
  {Militzer}},\ }\bibfield  {title} {\enquote {\bibinfo {title}
  {Hydrogen--helium mixtures at high pressure},}\ }\href@noop {} {\bibfield
  {journal} {\bibinfo  {journal} {Journal of Low Temperature Physics}\ }\textbf
  {\bibinfo {volume} {139}},\ \bibinfo {pages} {739--752} (\bibinfo {year}
  {2005})}\BibitemShut {NoStop}%
\bibitem [{\citenamefont {Jones}\ and\ \citenamefont {Ceperley}(1996)}]{JC96}%
  \BibitemOpen
  \bibfield  {author} {\bibinfo {author} {\bibfnamefont {M.~D.}\ \bibnamefont
  {Jones}}\ and\ \bibinfo {author} {\bibfnamefont {D.~M.}\ \bibnamefont
  {Ceperley}},\ }\bibfield  {title} {\enquote {\bibinfo {title}
  {Crystallization of the one-component plasma at finite temperature},}\
  }\href@noop {} {\bibfield  {journal} {\bibinfo  {journal} {Phys. Rev. Lett.}\
  }\textbf {\bibinfo {volume} {76}},\ \bibinfo {pages} {4572--4575} (\bibinfo
  {year} {1996})}\BibitemShut {NoStop}%
\bibitem [{\citenamefont {Pollock}\ and\ \citenamefont
  {Militzer}(2004)}]{MP04}%
  \BibitemOpen
  \bibfield  {author} {\bibinfo {author} {\bibfnamefont {E.~L.}\ \bibnamefont
  {Pollock}}\ and\ \bibinfo {author} {\bibfnamefont {B.}~\bibnamefont
  {Militzer}},\ }\bibfield  {title} {\enquote {\bibinfo {title} {Dense plasma
  effects on nuclear reaction rates},}\ }\href@noop {} {\bibfield  {journal}
  {\bibinfo  {journal} {Phys. Rev. Lett.}\ }\textbf {\bibinfo {volume} {92}},\
  \bibinfo {pages} {021101} (\bibinfo {year} {2004})}\BibitemShut {NoStop}%
\bibitem [{\citenamefont {Militzer}\ and\ \citenamefont
  {Pollock}(2005)}]{MP05}%
  \BibitemOpen
  \bibfield  {author} {\bibinfo {author} {\bibfnamefont {B.}~\bibnamefont
  {Militzer}}\ and\ \bibinfo {author} {\bibfnamefont {E.~L.}\ \bibnamefont
  {Pollock}},\ }\bibfield  {title} {\enquote {\bibinfo {title} {Equilibrium
  contact probabilities in dense plasmas},}\ }\href@noop {} {\bibfield
  {journal} {\bibinfo  {journal} {Phys. Rev. B}\ }\textbf {\bibinfo {volume}
  {71}},\ \bibinfo {pages} {134303} (\bibinfo {year} {2005})}\BibitemShut
  {NoStop}%
\bibitem [{\citenamefont {Driver}\ and\ \citenamefont
  {Militzer}(2016)}]{DriverNitrogen2016}%
  \BibitemOpen
  \bibfield  {author} {\bibinfo {author} {\bibfnamefont {K.~P.}\ \bibnamefont
  {Driver}}\ and\ \bibinfo {author} {\bibfnamefont {B.}~\bibnamefont
  {Militzer}},\ }\bibfield  {title} {\enquote {\bibinfo {title}
  {First-principles equation of state calculations of warm dense nitrogen},}\
  }\href {\doibase 10.1103/PhysRevB.93.064101} {\bibfield  {journal} {\bibinfo
  {journal} {Phys. Rev. B}\ }\textbf {\bibinfo {volume} {93}},\ \bibinfo
  {pages} {064101} (\bibinfo {year} {2016})}\BibitemShut {NoStop}%
\bibitem [{\citenamefont {Driver}\ and\ \citenamefont
  {Militzer}(2017)}]{Driver2017}%
  \BibitemOpen
  \bibfield  {author} {\bibinfo {author} {\bibfnamefont {K.~P.}\ \bibnamefont
  {Driver}}\ and\ \bibinfo {author} {\bibfnamefont {B.}~\bibnamefont
  {Militzer}},\ }\bibfield  {title} {\enquote {\bibinfo {title}
  {{First-principles simulations of warm dense lithium fluoride}},}\
  }\href@noop {} {\bibfield  {journal} {\bibinfo  {journal} {Phys. Rev. E}\
  }\textbf {\bibinfo {volume} {95}},\ \bibinfo {pages} {043205} (\bibinfo
  {year} {2017})}\BibitemShut {NoStop}%
\bibitem [{\citenamefont {Zhang}\ \emph
  {et~al.}(2017{\natexlab{b}})\citenamefont {Zhang}, \citenamefont {Driver},
  \citenamefont {Soubiran},\ and\ \citenamefont {Militzer}}]{ZhangCH2017}%
  \BibitemOpen
  \bibfield  {author} {\bibinfo {author} {\bibfnamefont {S.}~\bibnamefont
  {Zhang}}, \bibinfo {author} {\bibfnamefont {K.~P.}\ \bibnamefont {Driver}},
  \bibinfo {author} {\bibfnamefont {F.}~\bibnamefont {Soubiran}}, \ and\
  \bibinfo {author} {\bibfnamefont {B.}~\bibnamefont {Militzer}},\ }\bibfield
  {title} {\enquote {\bibinfo {title} {{First-principles equation of state and
  shock compression predictions of warm dense hydrocarbons}},}\ }\href@noop {}
  {\bibfield  {journal} {\bibinfo  {journal} {Phys. Rev. E}\ }\textbf {\bibinfo
  {volume} {96}},\ \bibinfo {pages} {013204} (\bibinfo {year}
  {2017}{\natexlab{b}})}\BibitemShut {NoStop}%
\bibitem [{\citenamefont {Natoli}\ and\ \citenamefont {Ceperley}(1995)}]{Na95}%
  \BibitemOpen
  \bibfield  {author} {\bibinfo {author} {\bibfnamefont {V.}~\bibnamefont
  {Natoli}}\ and\ \bibinfo {author} {\bibfnamefont {D.~M.}\ \bibnamefont
  {Ceperley}},\ }\bibfield  {title} {\enquote {\bibinfo {title} {An optimized
  method for treating long-range potentials},}\ }\href@noop {} {\bibfield
  {journal} {\bibinfo  {journal} {Journal of Computational Physics}\ }\textbf
  {\bibinfo {volume} {117}},\ \bibinfo {pages} {171--178} (\bibinfo {year}
  {1995})}\BibitemShut {NoStop}%
\bibitem [{\citenamefont {Militzer}(2016{\natexlab{a}})}]{BM2016}%
  \BibitemOpen
  \bibfield  {author} {\bibinfo {author} {\bibfnamefont {B.}~\bibnamefont
  {Militzer}},\ }\bibfield  {title} {\enquote {\bibinfo {title} {Computation of
  the high temperature coulomb density matrix in periodic boundary
  conditions},}\ }\href@noop {} {\bibfield  {journal} {\bibinfo  {journal}
  {Comp. Phys. Comm.}\ }\textbf {\bibinfo {volume} {204}},\ \bibinfo {pages}
  {88} (\bibinfo {year} {2016}{\natexlab{a}})}\BibitemShut {NoStop}%
\bibitem [{\citenamefont {Militzer}, \citenamefont {Pollock},\ and\
  \citenamefont {Ceperley}(2019)}]{Militzer2019}%
  \BibitemOpen
  \bibfield  {author} {\bibinfo {author} {\bibfnamefont {B.}~\bibnamefont
  {Militzer}}, \bibinfo {author} {\bibfnamefont {E.}~\bibnamefont {Pollock}}, \
  and\ \bibinfo {author} {\bibfnamefont {D.}~\bibnamefont {Ceperley}},\
  }\bibfield  {title} {\enquote {\bibinfo {title} {{Path integral Monte Carlo
  calculation of the momentum distribution of the homogeneous electron gas at
  finite temperature}},}\ }\href {https://doi.org/10.1016/j.hedp.2018.12.004
  https://linkinghub.elsevier.com/retrieve/pii/S1574181818300995} {\bibfield
  {journal} {\bibinfo  {journal} {High Energy Density Physics}\ }\textbf
  {\bibinfo {volume} {30}},\ \bibinfo {pages} {13--20} (\bibinfo {year}
  {2019})}\BibitemShut {NoStop}%
\bibitem [{\citenamefont {Militzer}(2016{\natexlab{b}})}]{Militzer2016}%
  \BibitemOpen
  \bibfield  {author} {\bibinfo {author} {\bibfnamefont {B.}~\bibnamefont
  {Militzer}},\ }\bibfield  {title} {\enquote {\bibinfo {title} {{Computation
  of the high temperature Coulomb density matrix in periodic boundary
  conditions}},}\ }\href {\doibase 10.1016/j.cpc.2016.03.011} {\bibfield
  {journal} {\bibinfo  {journal} {Computer Physics Communications}\ }\textbf
  {\bibinfo {volume} {204}},\ \bibinfo {pages} {88--96} (\bibinfo {year}
  {2016}{\natexlab{b}})}\BibitemShut {NoStop}%
\bibitem [{\citenamefont {Driver}\ and\ \citenamefont
  {Militzer}(2012)}]{Driver2012}%
  \BibitemOpen
  \bibfield  {author} {\bibinfo {author} {\bibfnamefont {K.~P.}\ \bibnamefont
  {Driver}}\ and\ \bibinfo {author} {\bibfnamefont {B.}~\bibnamefont
  {Militzer}},\ }\bibfield  {title} {\enquote {\bibinfo {title} {{All-Electron
  Path Integral Monte Carlo Simulations of Warm Dense Matter: Application to
  Water and Carbon Plasmas}},}\ }\href@noop {} {\bibfield  {journal} {\bibinfo
  {journal} {Phys. Rev. Lett.}\ }\textbf {\bibinfo {volume} {108}},\ \bibinfo
  {pages} {115502} (\bibinfo {year} {2012})}\BibitemShut {NoStop}%
\bibitem [{\citenamefont {Kresse}\ and\ \citenamefont
  {Joubert}(1999)}]{Kresse1999}%
  \BibitemOpen
  \bibfield  {author} {\bibinfo {author} {\bibfnamefont {G.}~\bibnamefont
  {Kresse}}\ and\ \bibinfo {author} {\bibfnamefont {D.}~\bibnamefont
  {Joubert}},\ }\bibfield  {title} {\enquote {\bibinfo {title} {{From ultrasoft
  pseudopotentials to the projector augmented-wave method}},}\ }\href@noop {}
  {\bibfield  {journal} {\bibinfo  {journal} {Phys. Rev. B}\ }\textbf {\bibinfo
  {volume} {59}},\ \bibinfo {pages} {1758--1775} (\bibinfo {year}
  {1999})}\BibitemShut {NoStop}%
\bibitem [{\citenamefont {Driver}\ \emph {et~al.}(2015)\citenamefont {Driver},
  \citenamefont {Soubiran}, \citenamefont {Zhang},\ and\ \citenamefont
  {Militzer}}]{Driver2015b}%
  \BibitemOpen
  \bibfield  {author} {\bibinfo {author} {\bibfnamefont {K.~P.}\ \bibnamefont
  {Driver}}, \bibinfo {author} {\bibfnamefont {F.}~\bibnamefont {Soubiran}},
  \bibinfo {author} {\bibfnamefont {S.}~\bibnamefont {Zhang}}, \ and\ \bibinfo
  {author} {\bibfnamefont {B.}~\bibnamefont {Militzer}},\ }\bibfield  {title}
  {\enquote {\bibinfo {title} {{First-principles equation of state and
  electronic properties of warm dense oxygen}},}\ }\href@noop {} {\bibfield
  {journal} {\bibinfo  {journal} {J. Chem. Phys.}\ }\textbf {\bibinfo {volume}
  {143}},\ \bibinfo {pages} {164507} (\bibinfo {year} {2015})}\BibitemShut
  {NoStop}%
\bibitem [{\citenamefont {Nos{\'{e}}}(1984)}]{Nose1984}%
  \BibitemOpen
  \bibfield  {author} {\bibinfo {author} {\bibfnamefont {S.}~\bibnamefont
  {Nos{\'{e}}}},\ }\bibfield  {title} {\enquote {\bibinfo {title} {{A unified
  formulation of the constant temperature molecular dynamics methods}},}\
  }\href {\doibase 10.1063/1.447334} {\bibfield  {journal} {\bibinfo  {journal}
  {J. Chem. Phys.}\ }\textbf {\bibinfo {volume} {81}},\ \bibinfo {pages}
  {511--519} (\bibinfo {year} {1984})}\BibitemShut {NoStop}%
\bibitem [{\citenamefont {Nos{\'{e}}}(1991)}]{Nose1991}%
  \BibitemOpen
  \bibfield  {author} {\bibinfo {author} {\bibfnamefont {S.}~\bibnamefont
  {Nos{\'{e}}}},\ }\bibfield  {title} {\enquote {\bibinfo {title} {{Constant
  Temperature Molecular Dynamics Methods}},}\ }\href@noop {} {\bibfield
  {journal} {\bibinfo  {journal} {Prog. Theor. Phys. Suppl.}\ }\textbf
  {\bibinfo {volume} {103}},\ \bibinfo {pages} {1} (\bibinfo {year}
  {1991})}\BibitemShut {NoStop}%
\bibitem [{\citenamefont {Bl{\"{o}}chl}(1994)}]{Blochl1994}%
  \BibitemOpen
  \bibfield  {author} {\bibinfo {author} {\bibfnamefont {P.~E.}\ \bibnamefont
  {Bl{\"{o}}chl}},\ }\bibfield  {title} {\enquote {\bibinfo {title} {{Projector
  augmented-wave method}},}\ }\href {\doibase 10.1103/PhysRevB.50.17953}
  {\bibfield  {journal} {\bibinfo  {journal} {Phys. Rev. B}\ }\textbf {\bibinfo
  {volume} {50}},\ \bibinfo {pages} {17953--17979} (\bibinfo {year}
  {1994})}\BibitemShut {NoStop}%
\bibitem [{\citenamefont {Perdew}, \citenamefont {Burke},\ and\ \citenamefont
  {Ernzerhof}(1996)}]{PBE}%
  \BibitemOpen
  \bibfield  {author} {\bibinfo {author} {\bibfnamefont {J.~P.}\ \bibnamefont
  {Perdew}}, \bibinfo {author} {\bibfnamefont {K.}~\bibnamefont {Burke}}, \
  and\ \bibinfo {author} {\bibfnamefont {M.}~\bibnamefont {Ernzerhof}},\
  }\bibfield  {title} {\enquote {\bibinfo {title} {{Generalized Gradient
  Approximation Made Simple}},}\ }\href@noop {} {\bibfield  {journal} {\bibinfo
   {journal} {Phys. Rev. Lett.}\ }\textbf {\bibinfo {volume} {77}},\ \bibinfo
  {pages} {3865--3868} (\bibinfo {year} {1996})}\BibitemShut {NoStop}%
\bibitem [{OPI()}]{OPIUM}%
  \BibitemOpen
  \href@noop {} {}\bibinfo {howpublished} {For OPIUM pseudopotential generation
  programs, see \url{http://opium.sourceforge.net}.}\BibitemShut {Stop}%
\bibitem [{\citenamefont {Debye}\ and\ \citenamefont
  {H{\"u}ckel}(1923)}]{Debye1923}%
  \BibitemOpen
  \bibfield  {author} {\bibinfo {author} {\bibfnamefont {P.}~\bibnamefont
  {Debye}}\ and\ \bibinfo {author} {\bibfnamefont {E.}~\bibnamefont
  {H{\"u}ckel}},\ }\bibfield  {title} {\enquote {\bibinfo {title} {Zur theorie
  der elektrolyte},}\ }\href@noop {} {\bibfield  {journal} {\bibinfo  {journal}
  {Phys. Z}\ }\textbf {\bibinfo {volume} {24}},\ \bibinfo {pages} {185}
  (\bibinfo {year} {1923})}\BibitemShut {NoStop}%
\bibitem [{\citenamefont {Allen}\ and\ \citenamefont {Tildesley}(1987)}]{AT87}%
  \BibitemOpen
  \bibfield  {author} {\bibinfo {author} {\bibfnamefont {M.}~\bibnamefont
  {Allen}}\ and\ \bibinfo {author} {\bibfnamefont {D.}~\bibnamefont
  {Tildesley}},\ }\href@noop {} {\emph {\bibinfo {title} {Computer Simulation
  of Liquids}}}\ (\bibinfo  {publisher} {Oxford University Press},\ \bibinfo
  {address} {New York},\ \bibinfo {year} {1987})\BibitemShut {NoStop}%
\bibitem [{\citenamefont {Vinko}, \citenamefont {Ciricosta},\ and\
  \citenamefont {Wark}(2014)}]{Vinko2014}%
  \BibitemOpen
  \bibfield  {author} {\bibinfo {author} {\bibfnamefont {S.}~\bibnamefont
  {Vinko}}, \bibinfo {author} {\bibfnamefont {O.}~\bibnamefont {Ciricosta}}, \
  and\ \bibinfo {author} {\bibfnamefont {J.}~\bibnamefont {Wark}},\ }\bibfield
  {title} {\enquote {\bibinfo {title} {Density functional theory calculations
  of continuum lowering in strongly coupled plasmas},}\ }\href@noop {}
  {\bibfield  {journal} {\bibinfo  {journal} {Nature communications}\ }\textbf
  {\bibinfo {volume} {5}},\ \bibinfo {pages} {3533} (\bibinfo {year}
  {2014})}\BibitemShut {NoStop}%
\bibitem [{\citenamefont {Lin}\ \emph {et~al.}(2017)\citenamefont {Lin},
  \citenamefont {R\"opke}, \citenamefont {Kraeft},\ and\ \citenamefont
  {Reinholz}}]{Lin2017}%
  \BibitemOpen
  \bibfield  {author} {\bibinfo {author} {\bibfnamefont {C.}~\bibnamefont
  {Lin}}, \bibinfo {author} {\bibfnamefont {G.}~\bibnamefont {R\"opke}},
  \bibinfo {author} {\bibfnamefont {W.-D.}\ \bibnamefont {Kraeft}}, \ and\
  \bibinfo {author} {\bibfnamefont {H.}~\bibnamefont {Reinholz}},\ }\bibfield
  {title} {\enquote {\bibinfo {title} {Ionization-potential depression and
  dynamical structure factor in dense plasmas},}\ }\href {\doibase
  10.1103/PhysRevE.96.013202} {\bibfield  {journal} {\bibinfo  {journal} {Phys.
  Rev. E}\ }\textbf {\bibinfo {volume} {96}},\ \bibinfo {pages} {013202}
  (\bibinfo {year} {2017})}\BibitemShut {NoStop}%
\bibitem [{GAM()}]{GAMESS}%
  \BibitemOpen
  \href@noop {} {}\bibinfo {howpublished} {General Atomic and Molecular
  Electronic Structure System (GAMESS). Visit
  \url{http://www.msg.ameslab.gov/gamess/} for more information.}\BibitemShut
  {Stop}%
\bibitem [{\citenamefont {Driver}\ \emph {et~al.}(2017)\citenamefont {Driver},
  \citenamefont {Soubiran}, \citenamefont {Zhang},\ and\ \citenamefont
  {Militzer}}]{Driver2017b}%
  \BibitemOpen
  \bibfield  {author} {\bibinfo {author} {\bibfnamefont {K.~P.}\ \bibnamefont
  {Driver}}, \bibinfo {author} {\bibfnamefont {F.}~\bibnamefont {Soubiran}},
  \bibinfo {author} {\bibfnamefont {S.}~\bibnamefont {Zhang}}, \ and\ \bibinfo
  {author} {\bibfnamefont {B.}~\bibnamefont {Militzer}},\ }\bibfield  {title}
  {\enquote {\bibinfo {title} {{Comparison of path integral Monte Carlo
  simulations of helium, carbon, nitrogen, oxygen, water, neon, and silicon
  plasmas}},}\ }\href@noop {} {\bibfield  {journal} {\bibinfo  {journal} {High
  Energy Density Physics}\ }\textbf {\bibinfo {volume} {23}},\ \bibinfo {pages}
  {81--89} (\bibinfo {year} {2017})}\BibitemShut {NoStop}%
\bibitem [{\citenamefont {Guti{\'{e}}rrez}\ \emph {et~al.}(2010)\citenamefont
  {Guti{\'{e}}rrez}, \citenamefont {Men{\'{e}}ndez-Proupin}, \citenamefont
  {Loyola}, \citenamefont {Peralta},\ and\ \citenamefont
  {Davis}}]{Gutierrez2010}%
  \BibitemOpen
  \bibfield  {author} {\bibinfo {author} {\bibfnamefont {G.}~\bibnamefont
  {Guti{\'{e}}rrez}}, \bibinfo {author} {\bibfnamefont {E.}~\bibnamefont
  {Men{\'{e}}ndez-Proupin}}, \bibinfo {author} {\bibfnamefont {C.}~\bibnamefont
  {Loyola}}, \bibinfo {author} {\bibfnamefont {J.}~\bibnamefont {Peralta}}, \
  and\ \bibinfo {author} {\bibfnamefont {S.}~\bibnamefont {Davis}},\ }\bibfield
   {title} {\enquote {\bibinfo {title} {{Computer simulation study of amorphous
  compounds: structural and vibrational properties}},}\ }\href {\doibase
  10.1007/s10853-010-4579-0} {\bibfield  {journal} {\bibinfo  {journal}
  {Journal of Materials Science}\ }\textbf {\bibinfo {volume} {45}},\ \bibinfo
  {pages} {5124--5134} (\bibinfo {year} {2010})}\BibitemShut {NoStop}%
\bibitem [{\citenamefont {Kalkan}\ \emph {et~al.}(2018)\citenamefont {Kalkan},
  \citenamefont {Godwal}, \citenamefont {Raju},\ and\ \citenamefont
  {Jeanloz}}]{Kalkan2018}%
  \BibitemOpen
  \bibfield  {author} {\bibinfo {author} {\bibfnamefont {B.}~\bibnamefont
  {Kalkan}}, \bibinfo {author} {\bibfnamefont {B.}~\bibnamefont {Godwal}},
  \bibinfo {author} {\bibfnamefont {S.~V.}\ \bibnamefont {Raju}}, \ and\
  \bibinfo {author} {\bibfnamefont {R.}~\bibnamefont {Jeanloz}},\ }\bibfield
  {title} {\enquote {\bibinfo {title} {{Local structure of molten AuGa2 under
  pressure: Evidence for coordination change and planetary implications}},}\
  }\href {\doibase 10.1038/s41598-018-25297-9} {\bibfield  {journal} {\bibinfo
  {journal} {Scientific Reports}\ }\textbf {\bibinfo {volume} {8}},\ \bibinfo
  {pages} {6844} (\bibinfo {year} {2018})}\BibitemShut {NoStop}%
\bibitem [{\citenamefont {Drewitt}\ \emph {et~al.}(2020)\citenamefont
  {Drewitt}, \citenamefont {Turci}, \citenamefont {Heinen}, \citenamefont
  {Macleod}, \citenamefont {Qin}, \citenamefont {Kleppe},\ and\ \citenamefont
  {Lord}}]{Drewitt2020}%
  \BibitemOpen
  \bibfield  {author} {\bibinfo {author} {\bibfnamefont {J.~W.~E.}\
  \bibnamefont {Drewitt}}, \bibinfo {author} {\bibfnamefont {F.}~\bibnamefont
  {Turci}}, \bibinfo {author} {\bibfnamefont {B.~J.}\ \bibnamefont {Heinen}},
  \bibinfo {author} {\bibfnamefont {S.~G.}\ \bibnamefont {Macleod}}, \bibinfo
  {author} {\bibfnamefont {F.}~\bibnamefont {Qin}}, \bibinfo {author}
  {\bibfnamefont {A.~K.}\ \bibnamefont {Kleppe}}, \ and\ \bibinfo {author}
  {\bibfnamefont {O.~T.}\ \bibnamefont {Lord}},\ }\bibfield  {title} {\enquote
  {\bibinfo {title} {{Structural Ordering in Liquid Gallium under Extreme
  Conditions}},}\ }\href {\doibase 10.1103/PhysRevLett.124.145501} {\bibfield
  {journal} {\bibinfo  {journal} {Physical Review Letters}\ }\textbf {\bibinfo
  {volume} {124}},\ \bibinfo {pages} {145501} (\bibinfo {year}
  {2020})}\BibitemShut {NoStop}%
\bibitem [{\citenamefont {Hugoniot}(1887)}]{Hugoniot1887}%
  \BibitemOpen
  \bibfield  {author} {\bibinfo {author} {\bibfnamefont {H.}~\bibnamefont
  {Hugoniot}},\ }\bibfield  {title} {\enquote {\bibinfo {title} {Memoir on the
  propagation of movements in bodies, especially perfect gases (first part)},}\
  }\href@noop {} {\bibfield  {journal} {\bibinfo  {journal} {J. de l’Ecole
  Polytechnique}\ }\textbf {\bibinfo {volume} {57}},\ \bibinfo {pages} {3--97}
  (\bibinfo {year} {1887})}\BibitemShut {NoStop}%
\bibitem [{\citenamefont {Hugoniot}(1889)}]{Hugoniot1889}%
  \BibitemOpen
  \bibfield  {author} {\bibinfo {author} {\bibfnamefont {H.}~\bibnamefont
  {Hugoniot}},\ }\bibfield  {title} {\enquote {\bibinfo {title} {Memoir on the
  propagation of movements in bodies, especially perfect gases (second
  part)},}\ }\href@noop {} {\bibfield  {journal} {\bibinfo  {journal} {J. de
  l’Ecole Polytechnique}\ }\textbf {\bibinfo {volume} {58}},\ \bibinfo
  {pages} {1--125} (\bibinfo {year} {1889})}\BibitemShut {NoStop}%
\bibitem [{\citenamefont {Zeldovich}\ and\ \citenamefont
  {Raizer}(1968)}]{Ze66}%
  \BibitemOpen
  \bibfield  {author} {\bibinfo {author} {\bibfnamefont {Y.~B.}\ \bibnamefont
  {Zeldovich}}\ and\ \bibinfo {author} {\bibfnamefont {Y.~P.}\ \bibnamefont
  {Raizer}},\ }\href@noop {} {\emph {\bibinfo {title} {Elements of Gasdynamics
  and the Classical Theory of Shock Waves}}}\ (\bibinfo  {publisher} {Academic
  Press},\ \bibinfo {address} {New York},\ \bibinfo {year} {1968})\BibitemShut
  {NoStop}%
\bibitem [{\citenamefont {Root}\ \emph {et~al.}(2018)\citenamefont {Root},
  \citenamefont {Townsend}, \citenamefont {Davies}, \citenamefont {Lemke},
  \citenamefont {Bliss}, \citenamefont {Fratanduono}, \citenamefont {Kraus},
  \citenamefont {Millot}, \citenamefont {Spaulding}, \citenamefont
  {Shulenburger}, \citenamefont {Stewart},\ and\ \citenamefont
  {Jacobsen}}]{Root2018}%
  \BibitemOpen
  \bibfield  {author} {\bibinfo {author} {\bibfnamefont {S.}~\bibnamefont
  {Root}}, \bibinfo {author} {\bibfnamefont {J.~P.}\ \bibnamefont {Townsend}},
  \bibinfo {author} {\bibfnamefont {E.}~\bibnamefont {Davies}}, \bibinfo
  {author} {\bibfnamefont {R.~W.}\ \bibnamefont {Lemke}}, \bibinfo {author}
  {\bibfnamefont {D.~E.}\ \bibnamefont {Bliss}}, \bibinfo {author}
  {\bibfnamefont {D.~E.}\ \bibnamefont {Fratanduono}}, \bibinfo {author}
  {\bibfnamefont {R.~G.}\ \bibnamefont {Kraus}}, \bibinfo {author}
  {\bibfnamefont {M.}~\bibnamefont {Millot}}, \bibinfo {author} {\bibfnamefont
  {D.~K.}\ \bibnamefont {Spaulding}}, \bibinfo {author} {\bibfnamefont
  {L.}~\bibnamefont {Shulenburger}}, \bibinfo {author} {\bibfnamefont {S.~T.}\
  \bibnamefont {Stewart}}, \ and\ \bibinfo {author} {\bibfnamefont {S.~B.}\
  \bibnamefont {Jacobsen}},\ }\bibfield  {title} {\enquote {\bibinfo {title}
  {{The Principal Hugoniot of Forsterite to 950 GPa}},}\ }\href@noop {}
  {\bibfield  {journal} {\bibinfo  {journal} {Geophysical Research Letters}\
  }\textbf {\bibinfo {volume} {45}},\ \bibinfo {pages} {3865--3872} (\bibinfo
  {year} {2018})}\BibitemShut {NoStop}%
\bibitem [{\citenamefont {Fratanduono}\ \emph {et~al.}(2018)\citenamefont
  {Fratanduono}, \citenamefont {Millot}, \citenamefont {Kraus}, \citenamefont
  {Spaulding}, \citenamefont {Collins}, \citenamefont {Celliers},\ and\
  \citenamefont {Eggert}}]{Fratanduono2018}%
  \BibitemOpen
  \bibfield  {author} {\bibinfo {author} {\bibfnamefont {D.~E.}\ \bibnamefont
  {Fratanduono}}, \bibinfo {author} {\bibfnamefont {M.}~\bibnamefont {Millot}},
  \bibinfo {author} {\bibfnamefont {R.~G.}\ \bibnamefont {Kraus}}, \bibinfo
  {author} {\bibfnamefont {D.~K.}\ \bibnamefont {Spaulding}}, \bibinfo {author}
  {\bibfnamefont {G.~W.}\ \bibnamefont {Collins}}, \bibinfo {author}
  {\bibfnamefont {P.~M.}\ \bibnamefont {Celliers}}, \ and\ \bibinfo {author}
  {\bibfnamefont {J.~H.}\ \bibnamefont {Eggert}},\ }\bibfield  {title}
  {\enquote {\bibinfo {title} {{Thermodynamic properties of MgSiO$_3$ at
  super-Earth mantle conditions}},}\ }\href
  {https://link.aps.org/doi/10.1103/PhysRevB.97.214105} {\bibfield  {journal}
  {\bibinfo  {journal} {Phys. Rev. B}\ }\textbf {\bibinfo {volume} {97}},\
  \bibinfo {pages} {214105} (\bibinfo {year} {2018})}\BibitemShut {NoStop}%
\bibitem [{\citenamefont {Militzer}\ \emph {et~al.}(2007)\citenamefont
  {Militzer}, \citenamefont {Hubbard}, \citenamefont {Elert}, \citenamefont
  {Furnish}, \citenamefont {Chau}, \citenamefont {Holmes},\ and\ \citenamefont
  {Nguyen}}]{Militzer2007}%
  \BibitemOpen
  \bibfield  {author} {\bibinfo {author} {\bibfnamefont {B.}~\bibnamefont
  {Militzer}}, \bibinfo {author} {\bibfnamefont {W.}~\bibnamefont {Hubbard}},
  \bibinfo {author} {\bibfnamefont {M.}~\bibnamefont {Elert}}, \bibinfo
  {author} {\bibfnamefont {M.~D.}\ \bibnamefont {Furnish}}, \bibinfo {author}
  {\bibfnamefont {R.}~\bibnamefont {Chau}}, \bibinfo {author} {\bibfnamefont
  {N.}~\bibnamefont {Holmes}}, \ and\ \bibinfo {author} {\bibfnamefont
  {J.}~\bibnamefont {Nguyen}},\ }\bibfield  {title} {\enquote {\bibinfo {title}
  {{Implications of Shock Wave Experiments With Precompressed Materials for
  Giant Planet Interiors}},}\ }\href {\doibase 10.1063/1.2832986} {\bibfield
  {journal} {\bibinfo  {journal} {AIP Conference Proceedings}\ }\textbf
  {\bibinfo {volume} {955}},\ \bibinfo {pages} {1395--1398} (\bibinfo {year}
  {2007})}\BibitemShut {NoStop}%
\bibitem [{\citenamefont {Jeanloz}\ \emph {et~al.}(2007)\citenamefont
  {Jeanloz}, \citenamefont {Celliers}, \citenamefont {Collins}, \citenamefont
  {Eggert}, \citenamefont {Lee}, \citenamefont {McWilliams}, \citenamefont
  {Brygoo},\ and\ \citenamefont {Loubeyre}}]{Jeanloz2007}%
  \BibitemOpen
  \bibfield  {author} {\bibinfo {author} {\bibfnamefont {R.}~\bibnamefont
  {Jeanloz}}, \bibinfo {author} {\bibfnamefont {P.~M.}\ \bibnamefont
  {Celliers}}, \bibinfo {author} {\bibfnamefont {G.~W.}\ \bibnamefont
  {Collins}}, \bibinfo {author} {\bibfnamefont {J.~H.}\ \bibnamefont {Eggert}},
  \bibinfo {author} {\bibfnamefont {K.~K.~M.}\ \bibnamefont {Lee}}, \bibinfo
  {author} {\bibfnamefont {R.~S.}\ \bibnamefont {McWilliams}}, \bibinfo
  {author} {\bibfnamefont {S.}~\bibnamefont {Brygoo}}, \ and\ \bibinfo {author}
  {\bibfnamefont {P.}~\bibnamefont {Loubeyre}},\ }\bibfield  {title} {\enquote
  {\bibinfo {title} {{Achieving high-density states through shock-wave loading
  of precompressed samples.}}}\ }\href {\doibase 10.1073/pnas.0608170104}
  {\bibfield  {journal} {\bibinfo  {journal} {Proceedings of the National
  Academy of Sciences of the United States of America}\ }\textbf {\bibinfo
  {volume} {104}},\ \bibinfo {pages} {9172--9177} (\bibinfo {year}
  {2007})}\BibitemShut {NoStop}%
\bibitem [{\citenamefont {Nilsen}\ \emph {et~al.}(2020)\citenamefont {Nilsen},
  \citenamefont {Kritcher}, \citenamefont {Martin}, \citenamefont {Tipton},
  \citenamefont {Whitley}, \citenamefont {Swift}, \citenamefont
  {D{\"{o}}ppner}, \citenamefont {Bachmann}, \citenamefont {Lazicki},
  \citenamefont {Kostinski}, \citenamefont {Maddox}, \citenamefont {Collins},
  \citenamefont {Glenzer},\ and\ \citenamefont {Falcone}}]{Nilsen2020}%
  \BibitemOpen
  \bibfield  {author} {\bibinfo {author} {\bibfnamefont {J.}~\bibnamefont
  {Nilsen}}, \bibinfo {author} {\bibfnamefont {A.~L.}\ \bibnamefont
  {Kritcher}}, \bibinfo {author} {\bibfnamefont {M.~E.}\ \bibnamefont
  {Martin}}, \bibinfo {author} {\bibfnamefont {R.~E.}\ \bibnamefont {Tipton}},
  \bibinfo {author} {\bibfnamefont {H.~D.}\ \bibnamefont {Whitley}}, \bibinfo
  {author} {\bibfnamefont {D.~C.}\ \bibnamefont {Swift}}, \bibinfo {author}
  {\bibfnamefont {T.}~\bibnamefont {D{\"{o}}ppner}}, \bibinfo {author}
  {\bibfnamefont {B.~L.}\ \bibnamefont {Bachmann}}, \bibinfo {author}
  {\bibfnamefont {A.~E.}\ \bibnamefont {Lazicki}}, \bibinfo {author}
  {\bibfnamefont {N.~B.}\ \bibnamefont {Kostinski}}, \bibinfo {author}
  {\bibfnamefont {B.~R.}\ \bibnamefont {Maddox}}, \bibinfo {author}
  {\bibfnamefont {G.~W.}\ \bibnamefont {Collins}}, \bibinfo {author}
  {\bibfnamefont {S.~H.}\ \bibnamefont {Glenzer}}, \ and\ \bibinfo {author}
  {\bibfnamefont {R.~W.}\ \bibnamefont {Falcone}},\ }\bibfield  {title}
  {\enquote {\bibinfo {title} {{Understanding the effects of radiative preheat
  and self-emission from shock heating on equation of state measurement at 100s
  of Mbar using spherically converging shock waves in a NIF hohlraum}},}\
  }\href {\doibase 10.1063/1.5131748} {\bibfield  {journal} {\bibinfo
  {journal} {Matter and Radiation at Extremes}\ }\textbf {\bibinfo {volume}
  {5}},\ \bibinfo {pages} {018401} (\bibinfo {year} {2020})}\BibitemShut
  {NoStop}%
\bibitem [{\citenamefont {Militzer}\ and\ \citenamefont
  {Hubbard}(2009)}]{MH08b}%
  \BibitemOpen
  \bibfield  {author} {\bibinfo {author} {\bibfnamefont {B.}~\bibnamefont
  {Militzer}}\ and\ \bibinfo {author} {\bibfnamefont {W.~H.}\ \bibnamefont
  {Hubbard}},\ }\href@noop {} {\bibfield  {journal} {\bibinfo  {journal}
  {Astrophys. and Space Sci.}\ }\textbf {\bibinfo {volume} {322}},\ \bibinfo
  {pages} {129} (\bibinfo {year} {2009})}\BibitemShut {NoStop}%
\end{thebibliography}

\providecommand{\noopsort}[1]{}\providecommand{\singleletter}[1]{#1}%

\end{document}